\documentclass[%
 aip,
 amsmath,amssymb,
]{revtex4-1}
\usepackage{graphicx}% Include figure files
\usepackage{subfigure}
\usepackage{ulem}
\usepackage{dcolumn}% Align table columns on decimal point
\usepackage{bm}% bold math
\usepackage[utf8]{inputenc}
\usepackage[T1]{fontenc}
\usepackage{mathptmx}
\usepackage{xcolor}
\usepackage{soul}
\usepackage{hyperref}

\hypersetup{
    colorlinks=true,
    linkcolor=blue,
    filecolor=magenta,      citecolor=blue,
    urlcolor=cyan,
    pdftitle={Overleaf Example},
    pdfpagemode=FullScreen,
    }

\begin{document}

%\preprint{}

\title[]{Viscoelasticity and Rheological Hysteresis}
% Force line breaks with \\
\author{Shweta Sharma}
\author{V.~Shankar}%
 \email{vshankar@iitk.ac.in}
 %\altaffiliation[Author for correspondence]
 \author{Yogesh M. Joshi}
 \email{joshi@iitk.ac.in}
% \altaffiliation[Author for correspondence]
\affiliation{Department of Chemical Engineering, Indian Institute of Technology Kanpur, Kanpur 208016, India
%\\This line break forced with \textbackslash\textbackslash
}%


\date{\today}

\begin{abstract}
\begin{center}
    (Accepted for publication in \textit{Journal of Rheology})
\end{center}
 
Rheological characterization of complex fluids subjected to cyclic shear-rate sweep often exhibits hysteresis. Since both viscoelastic and thixotropic materials show hysteresis loops, it is important to understand distinguishing features (if any) in the same shown by either. Lately, there has been substantial work that attempts to relate the area enclosed by the hysteresis loop with the manner in which shear rate is varied in the cycle, in order to infer thixotropic parameters of a material. In this work, we use the nonlinear Giesekus model to study its response to the application of cyclic shear-rate sweep. We find that this model produces each type of ualitatively similar hysteresis loop that has hitherto been ascribed to thixotropic materials. We also show that the area of the hysteresis loop for a viscoelastic material as a function of sweep rate shows bell-shaped/bi-modal curves as has been observed for thixotropic materials. This study illustrates that caution needs to be exercised while attributing hysteresis loops and associated features observed in a material exclusively to thixotropy.  Another feature related to the hysteresis loop is the occurrence of shear banding instability. We find that viscoelastic hysteresis may not have any connection to shear banding instability.

\end{abstract}

%\pacs{}% insert suggested PACS numbers in braces on next line

\maketitle %\maketitle must follow title, authors, abstract and \pacs

\section{Introduction}

Viscoelasticity is concerned with the ability of a material to store elastic potential energy upon application of deformation field over timescales comparable to observation timescale. Correspondingly the time over which the stress relaxes is termed as relaxation time of a material. When a viscoelastic material is subjected to cyclic down-up (or up-down) shear rate ramp, either step wise or continuous, the transient stress induced in the material depends on the nature of a material, the deformation history, the value of instantaneous shear rate, and the step time (or the ramp rate). Consequently, depending upon specifics of the given material and the flow field, stress in a material at given shear rate in a forward path may differ from that associated with the reverse path leading to hysteresis. In the literature, the subject of hysteresis in viscoelastic materials has been studied in detail, both experimentally and theoretically. On an independent but related note, it has been well established that thixotropy, which is concerned with the increase in viscosity under no flow or weak flow conditions and decrease in viscosity under strong flow conditions, shows many behavioural similarities with viscoelasticity [\onlinecite{barnes1997thixotropy,mewis2009thixotropy,larson2019review,agarwal2021distinguishing}]. In particular, thixotropic materials also show hysteresis in cyclic shear rate sweep experiment. In the literature, certain features of hysteresis loops have been identified as discriminating signatures of thixotropy in a material. Specifically, the area enclosed by a hysteresis loop as a function of step time in a stepwise down-up ramp has been proposed to show some unique features for a thixotropic material. In our earlier study [\onlinecite{agarwal2021distinguishing}], we assessed four different experimental protocols that have been proposed to distinguish thixotropy and viscoelasticity. We found that stress relaxation and/or creep experiments at different waiting times lead to unique signatures for thixotropic vis-a-vis viscoelastic materials. In this work, we theoretically assess the hysteresis loop method by employing a standard nonlinear viscoelastic model and study to what extent a viscoelastic material demonstrates features of hysteresis considered to be unique for thixotropic materials and implications of the same.

Hysteresis in viscoelastic materials is observed because of the finite time required by the same to reach the steady state corresponding to the applied flow field, which is governed by the magnitude of their relaxation time. Consequently, if the sweep rate is faster than the relaxation time of the material, then due to incomplete relaxation of stress during down-sweep and incomplete buildup of stress during up-sweep, shear stress does not trace the same path in the down-sweep and up-sweep shear flow. Moreover, if the applied shear rate is in the non-linear region, then the hysteresis loop area increases due to presence of stress undershoot during down-sweep and stress overshoot during up-sweep shear flow.

\begin{figure}[htbp]
\centering
     \subfigure[]{
    \includegraphics[scale=0.05]{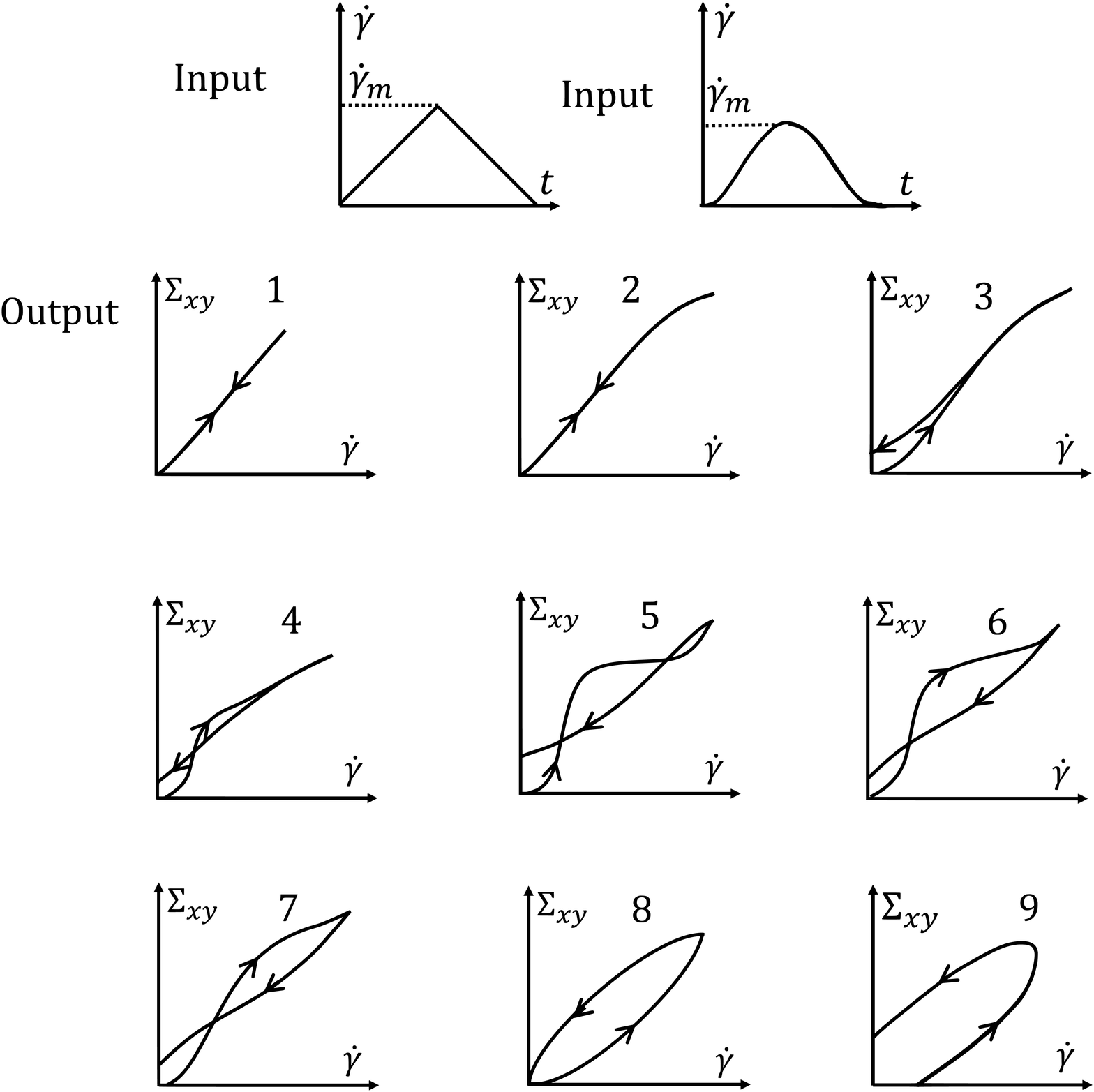}
    \label{marsh1}}
   \subfigure[]{
    \includegraphics[scale=0.05]{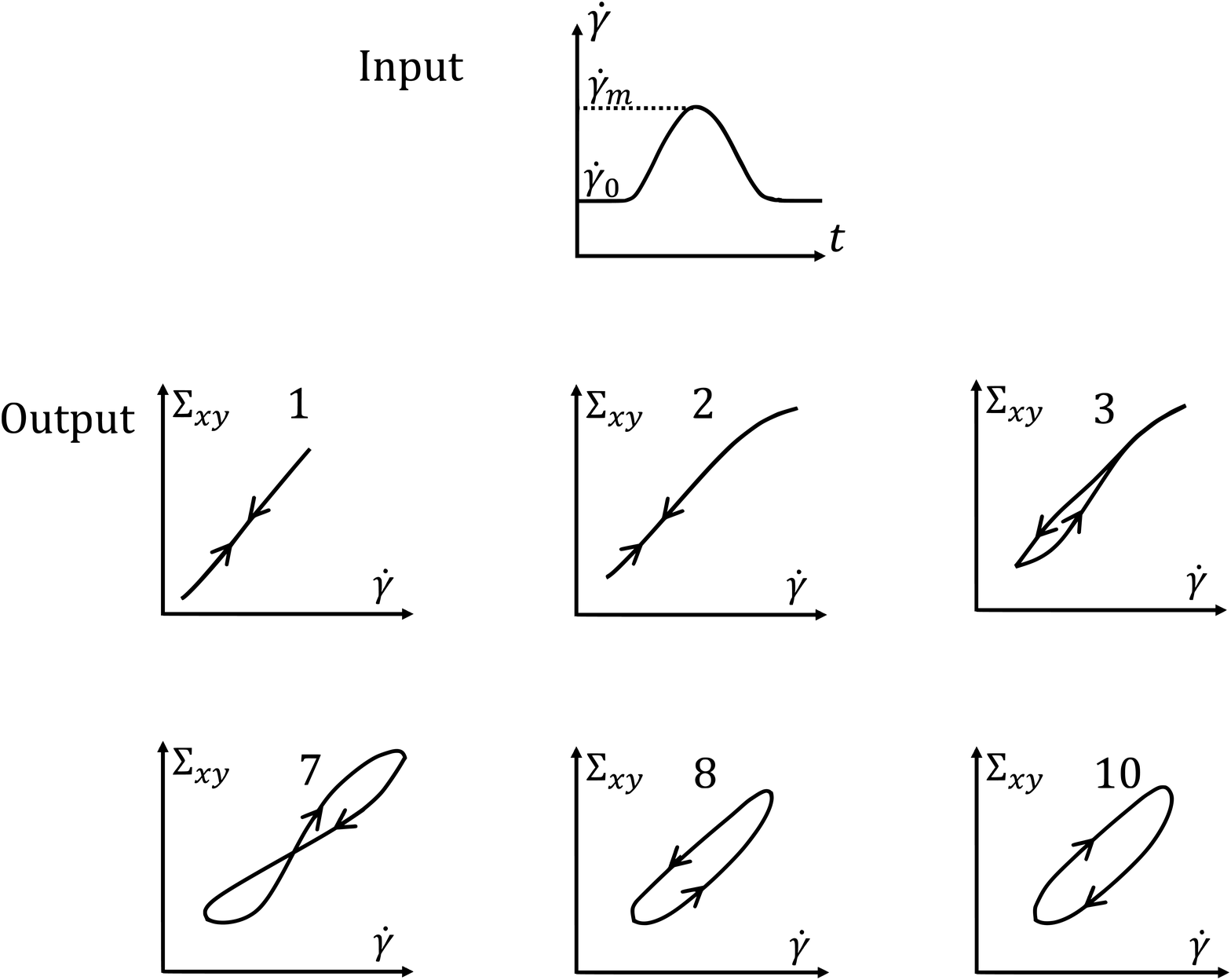}
    \label{marsh2}}
  
\caption{Different types of hysteresis loops proposed by Marsh [\onlinecite{marsh1968viscoelastic}] for a given input deformation flow field are shown in this figure. These loops can be obtained using a viscoelastic material for (a) triangular shear rate ramp up and ramp down flow or sinusoidal shear rate ramp up and ramp down, with a maximum shear rate, $\dot{\gamma }_m$ and (b) shifted sinusoidal shear rate ramp up and ramp down, presheared at $\dot{\gamma }_0$. These schematics are due to B. Duane Marsh, Transactions of the Society of Rheology 12:4, 489-510 (1968).  }
\label{fig:marsh_paper_figure}
\end{figure}

Viscoelastic hysteresis has been discussed in detail by Bird and Marsh [\onlinecite{bird1968viscoelastic}] and Marsh [\onlinecite{marsh1968viscoelastic}] using the viscoelastic model developed by Bird and Carreau [\onlinecite{bird1968nonlinear}]. They studied shear rate ramp flows with and without slow varying flow assumption ($De<1$). Marsh [\onlinecite{marsh1968viscoelastic}] proposed various kinds of viscoelastic hysteresis loops using the Bird-Carreau model that are redrawn and shown in Fig.\,\ref{fig:marsh_paper_figure}. In this figure, shear stress is plotted as a function of shear rate for different values of total time of experiment. Figure \ref{marsh1} shows loops that can be obtained using either triangular shear rate ramp up and ramp down flow or sinusoidal shear rate ramp up and ramp down. Figure \ref{marsh2} shows loops that can be obtained using shifted sinusoidal shear rate ramp up and ramp down. The type 1-9 curves in Fig.\,\ref{marsh1} and type 1-3, 7-9 curves in Fig.\,\ref{marsh2} are shown in the decreasing order of the value of total experiment time. The total experiment time is maximum for type 1 and it is minimum for type 9 (in Fig.\,\ref{marsh1}) and type 10 (in Fig.\,\ref{marsh2}) curves. The type 1 and 2 loops in Fig.\,\ref{marsh1} can be obtained for the case in which the the sweep rate is much greater than inverse of characteristic relaxation time of the fluid. The maximum shear rate for the type 1 loop is in the linear region and for the type 2 curve, the maximum shear rate is in the non-linear region. The type 3-9 loops can be obtained by varying the total experiment time in decreasing order. In these type of loops, it is shown that as the value of ramp time decreases, the area of hysteresis loop increases. The humps in these loops during up-sweep flow can be obtained due to stress overshoot and non-zero stress can be obtained at zero shear rate due to incomplete stress relaxation. The type 1-3 hysteresis loops in Fig.\,\ref{marsh2} for the case of shifted sinusoidal wave shear rate sweep flow are similar to loops presented in Fig.\,\ref{marsh1}. Type 7 loop shows that the hump in the loop is not as pronounced as type 6 loop in Fig.\,\ref{marsh1}. Type 10 hysteresis loop  shown in Fig.\,\ref{marsh2} can be obtained using a viscoelastic material in a shifted sinusoidal shear rate ramp up and ramp down flow field. Hysteresis loops obtained due to linear and non linear viscoleastic effects have also been explored in the literature [\onlinecite{greener1986response,rubio2008time}] using both Wagner and Maxwell model. To the best of our knowledge, the Maxwell model was first solved to show hysteresis loops due to linear viscoelastic effects in the classic monograph by Fredrickson [\onlinecite{fredrickson1964principles}].

Thixotropic materials are inherently out of thermodynamic equilibrium even in the stress free state, and therefore undergo time-dependent evolution of microstructure under quiescent conditions in order to attain the low energy states. As a result, viscosity of such materials continuously increases as a function of time. The microstructure developed in these materials, in principle, can be reversed or gradually broken down upon the application of deformation field of sufficiently high magnitude, which causes time dependent decrease in viscosity. During shear rate sweep experiments performed on such materials, the microstructure breaks down at high shear rates and as shear rate is decreased sequentially, it again starts to buildup. Furthermore, as shear rate is increased during the up-sweep shear flow, the developed microstructure requires more stress than during the down-sweep shear flow for the same value of shear rate. Therefore, shear stress does not trace the same path during the down-sweep and up-sweep shear flow for a thixotropic material. More detailed discussion on thixotropy, and hysteresis due to thixotropy is presented in the state of the art review articles [\onlinecite{larson2015constitutive,mewis1979thixotropy,mewis2009thixotropy,mujumdar2002transient,larson2019review,barnes1997thixotropy,armstrong2016dynamic}].

Hysteresis in complex fluids was first observed in paints by McMillen in 1932 [\onlinecite{mcmillen1932thixotropy}]. Since this study, many thixotropic and non-thixotropic complex materials in the literature have been reported to show hysteresis. Some of these materials are mineral oil [\onlinecite{weltmann1943thixotropic,hahn1959flow}], lithographic ink [\onlinecite{green1942high}], waxy potato starch [\onlinecite{krystyjan2016thixotropic}], cellulose nanocrystal suspensions [\onlinecite{fazilati2021thixotropy}], mud and cement pastes [\onlinecite{jeong2015thixotropic,perret1996thixotropic,fourmentin2015rheology}], sodium alginate solutions [\onlinecite{ma2014flow}], solder and adhesive pastes [\onlinecite{durairaj2004thixotropy}], ferrofluids [\onlinecite{li2018study}], sodium carboxymethylcellulose hydrogels [\onlinecite{ghica2016flow}], fluorinated guar gums [\onlinecite{wang2018self}], Carbopol microgels [\onlinecite{divoux2011stress}], microfibrillar cellulose water dispersions [\onlinecite{iotti2011rheological}], sodium polyacrylate Laponite solution [\onlinecite{labanda2005influence}], polystyrene solutions [\onlinecite{chen1992structural}], waxy crude oil [\onlinecite{mendes2015reversible}], acrylic emulsion paints [\onlinecite{baldewa2011thixotropy}], pulp fibre suspensions [\onlinecite{derakhshandeh2012thixotropy}], colloidal star gels [\onlinecite{holmes2004shear}], foams and emulsions [\onlinecite{da2002viscosity}], Ludox gels [\onlinecite{kurokawa2015avalanche}], etc. This list contains both non-thixotropic viscoelastic materials as well as thixotropic materials. Interestingly, the qualitative nature of the hysteresis loops observed in above materials, including those that are labelled as thixotropic, are similar to the hysteresis loops shown in Figure \ref{marsh2} obtained by Marsh [\onlinecite{marsh1968viscoelastic}] in the context of non-aging viscoelastic materials. It is therefore, imperative to detect characteristic features of hysteresis loops that distinguish viscoelasticity from thixotropy. Mewis [\onlinecite{mewis1979thixotropy}], Barnes [\onlinecite{barnes1997thixotropy}], and Mewis and Wagner [\onlinecite{mewis2009thixotropy}] considered hysteresis loop as a quick way to characterize thixotropy in a material and suggested this as an useful method for industrial applications. However, these review articles also noted that that hysteresis loop experiment is affected by both time and shear rate, and consequently, it is not an ideal way to identify a material as thixotropic. In these papers, the authors also sound a note of caution that hysteresis loops are common to both thixotropic and viscoelastic materials, but also mentioned that viscoelastic hysteresis might only be observed when the sweep rates are sufficiently high. Although, such cautionary advice in the literature is particularly focused on the viscoelastic materials that have a low relaxation time, very high relaxation time non-thixotropic viscoelastic materials may show hysteresis loops at sweep rates that are not sufficiently high (such sweep rates are commonly used in experiments to obtain flow curves as mentioned below). 

In the recent years, hysteresis in thixotropic and simple yield stress materials has been studied extensively. Divoux et al.\, [\onlinecite{divoux2013rheological}] experimentally studied rheological hysteresis using Laponite, carbon black, Mayonnaise, and Carbopol microgel. They performed shear rate down sweep followed by a shear rate up sweep experiment using all four samples and calculated the area of the resulting hysteresis loop. They also calculated the area of loop obtained due to inhomogeneity in the velocity profile during the shear rate sweep experiments. The area of hysteresis loop and area of inhomogeneous velocity profiles as a function of sweep time was found to show a bell shaped curve for Laponite and carbon black solution. The area of both loops showed a continuing decrease with sweep time for Mayonnaise and Carbopol microgel solutions. Divoux et al.\, related the value of sweep time at which a maximum loop area is observed to the characteristic timescale of the thixotropic material. The absence of peak in Carbopol and Mayonnaise was attributed to a negligible value (or of a significantly smaller order of magnitude) of thixotropic timescale. 

%, which was later termed as  timescale of the material [\onlinecite{mckinley2022mneymosymearxiv}]

Divoux et al.\, [\onlinecite{divoux2013rheological}] also showed results of the local velocity profile corresponding to the shear rate sweep experiments and found that hysteresis in these materials is interlinked with shear banding which was suggested to be triggered due to stress overshoot. The authors suggested a specific protocol to study hysteresis, wherein the material is first presheared at a very high shear rate so as to remove any prior history. After preshear, the next step is down-sweep from a very high shear rate to a very small value of shear rate and the shearing period $(\delta t)$ at each shear rate is fixed. The number of shear rates in each decade $(n)$ is also fixed. This protocol has been followed in most of the studies following the article by Divoux et al.\,[\onlinecite{divoux2013rheological}]. Puisto et. al.\,[\onlinecite{puisto2015dynamic}] studied a simple yield stress fluidity model using a homogeneous (parallel plate) and a concentric cylindrical geometry, and showed that these materials are also capable of showing a bell shaped curve. They also found that hysteresis and shear banding may not always be related in such fluids. Radhakrishnan et al.\, [\onlinecite{radhakrishnan2017understanding}] studied rheological hysteresis in detail using a simple fluidity model and the soft glassy rheology (SGR) model [\onlinecite{sollich1997rheology,sollich1998rheological}] to understand hysteresis in simple yield stress and viscosity bifurcating fluids. They compared their results with the experimental results of Divoux et al.\, [\onlinecite{divoux2013rheological}] and found good agreement between experimental and simulation results. Radhakrishanan et al.\, also emphasised on getting a closed hysteresis loop in simulation results so that these results can be compared with experiments. The authors attributed the lack of closed hysteresis loop to be the reason for getting a bell shaped curve for simple yield stress fluids by Puisto et al.\, [\onlinecite{puisto2015dynamic}]. Jamali et al.\, [\onlinecite{jamali2019multiscale}] studied hysteresis in soft glassy materials for three different lengthscale of material's structure- microscopic, mesoscopic and macroscopic scale. The authors employed dissipative particle dynamics and concluded that the soft glassy materials show hysteresis on all the three different scales. The plot of area of all the hysteresis loops with sweep time normalised by the characteristic timescale at each scale showed a bell-shaped master curve. This result is also in agreement with the results of Divoux et al.\, [\onlinecite{divoux2013rheological}] and Radhakrishnan et al.\, [\onlinecite{radhakrishnan2017understanding}].

The above discussion sets the context and provides the motivation for the present study. We address the following question - is it possible to identify a completely unknown material if it is viscoelastic or thixotropic purely on the basis of hysteresis loop? Larson [\onlinecite{larson2015constitutive}] in his seminal review noted that viscoelastic time scales range from nanoseconds to centuries. Bird and Marsh [\onlinecite{bird1968viscoelastic}] and Marsh [\onlinecite{marsh1968viscoelastic}] have already addressed this question, albeit partially, by clearly showing the various possible hysteresis loops in viscoleastic materials. In this study, we show that viscoelastic hysteresis loops proposed by Marsh [\onlinecite{marsh1968viscoelastic}] using a flow protocol of continuous ramp up and down shear flow can be obtained only for a low relaxation time viscoelastic material. If the relaxation time of viscoelastic materials is sufficiently high, then the type of hysteresis loops that can be observed will not remain same as of low relaxation time viscoelastic materials. In addition, more recently, the peculiar features of area of hysteresis loop as a function of sweep time (as discussed above) have also been reported for thixotropic materials. Therefore, an important question that remains to be addressed is what kind of signature viscoleastic materials would demonstrate for the protocol suggested by Divoux et al.\, [\onlinecite{divoux2013rheological}], wherein they reported a bell-shaped curve for the area of hysteresis loop when plotted against sweep time. To this end, in this work, we study the non-linear viscoelastic Giesekus model for the protocol of Divoux et al.\,, and analyse the obtained shear stress-strain rate hysteresis loops and compare these with the hysteresis loops obtained in literature for viscosity bifurcating, simple yield stress and other viscoelastic materials and models. 

The rest of this paper is organised as follows: we discuss the flow protocol, Giesekus model and the governing equations in section \ref{section_model}. We then discuss hysteresis in low and high relaxation time viscoelastic materials. Subsequently, we present features associated with area of viscoelastic hysteresis loops followed by a detailed analysis of the predicted behavior in section \ref{section_results}. Finally, at the end of this section, we discuss viscoelastic hysteresis in relation to shear banding in \ref{subsection_shear_banding}. The important conclusions of this study are discussed in section \ref{section_conclusions}. For the sake of comparison, we also investigate a simple inelastic thixotropic structural kinetic model [\onlinecite{goodeve1939general,larson2019review,moore1959rheology}] under the same shear rate sweep down-up protocol. The hysteresis behavior  of the same has been reported in the appendix section.

\section{Model and governing equations} \label{section_model}

We study shear rate sweep flow of a of a model viscoelastic fluid between two parallel plates. The plates are separated by a distance $H$ in the $y^*$ direction and are assumed to be of infinite length in $x^*$ and $z^*$ directions. We apply shear rate sweep in both increasing and decreasing order. For shearing the fluid at a fixed shear rate, we assume that both the plates are initially $(t^*=0)$ at rest and we shear the fluid by moving the top plate with a constant velocity $U$ in $x^*$ direction. In this case, the continuity equation gets satisfied by itself \( \underset{\sim}{\nabla} \cdot {{\underset{\sim}{u}}^{*}}=0 \), where \({\underset{\sim}{u}}^*\) is the velocity vector. We assume the shear flow to be inertialess and consequently, the simplified Cauchy momentum equation can be expressed as follows:

\begin{equation}\label{inertialess}
    \underset{\sim}{\nabla} \cdot {\underset{\approx}{\Sigma }^{*}}=0,
\end{equation}
where, \({\underset{\approx}{\Sigma }^{*}}\) is the total stress tensor. We consider the total stress to be a summation of viscoelastic stress (polymer contribution, ${\underset{\approx}{\sigma }^{*}}$) and a Newtonian  stress (solvent contribution) which can also accommodate the faster relaxing modes in a polymeric solution:

\begin{equation}\label{totalstress}
    {\underset{\approx}{\Sigma }^{*}}={\underset{\approx}{\sigma }^{*}}+{{\mu }_{s}}{\underset{\approx}{\dot{\gamma }}^{*},}
\end{equation}

where, ${\mu }_{s}$ is the viscosity of solvent in the polymeric solution and  ${\underset{\approx}{\dot{\gamma }}}^{*} = (\underset{\sim}{\nabla}{{\underset{\sim}{u}^{*}}} + (\underset{\sim}{\nabla}{{\underset{\sim}{u}^{*}}})^{T}) $, is the rate of strain tensor.

 The viscoelastic contibution to the total stress is governed by $\underset{\approx}{\sigma}^*$ according to the Giesekus model [\onlinecite{giesekus1982simple}]. The dimensional form of constitutive model is expressed as:
  \begin{equation} \label{gies_model}
     \underset{\approx}{\sigma}^* +{\tau}\overset{\nabla }{\mathop{\underset{\approx}{\sigma}^*}}=-\alpha\frac{\tau}{\eta_p}(\underset{\approx}{\sigma}^*\cdot\underset{\approx}{\sigma}^*)+\mu_{p}\underset{\approx}{\dot{\gamma}^*},
\end{equation}

where, $\alpha$ is the dimensionless mobility factor that accounts for anisotropic hydrodynamic drag on polymeric molecules and $\eta_p$ is the polymeric contribution to the zero shear viscosity. This model is a non-linear phenomenological model, which has been successful in predicting transient and steady state flow behaviour of polymeric and wormlike micellar solutions [\onlinecite{helgeson2009relating,khair2016large,vlassopoulos1995generalized}]. The advantage of using Giesekus model over Oldroyd-B model is its ability to fit shear thinning as well as first and second normal stress differences obtained in the experimental results of polymeric solution due to the presence of quadratic stress term.  

We non-dimensionalize the length scale by gap width $(H)$, velocity by top plate velocity $(U)$ and time scale by longest relaxation time of the polymeric solution $(\tau)$. We non-dimensionalize all the stress components by $\left(\displaystyle \frac{\eta_s+\eta_p}{\tau}\right)$. We represent dimensional terms with * superscript. The dimensionless numbers relevant to this study are the Weissenberg number $Wi = \displaystyle \frac{\tau U}{H}$ and the ratio of solvent viscosity to zero shear viscosity of the polymeric solution $\bar{\eta_s} = \displaystyle \frac{\eta_s}{\eta_s+\eta_p}$. In this study, we obtain all the results using $\alpha=0.1$. We also use $\alpha=0.2$ and 0.3, and find our results to be qualitatively similar suggesting the results are not sensitive to value of $\alpha$. The component-wise equations of Giesekus model in non-dimensional form for simple shear flow can be expressed as follows:

\begin{equation}\label{gxy}
    \frac{\partial \sigma_{xy}}{\partial t}=-\frac{\alpha}{(1-\bar{\eta_s})}(\sigma_{xx}+\sigma_{yy})\sigma_{xy}+[(1-\bar{\eta_s})+\sigma_{yy}]Wi\dot{\gamma}-\sigma_{xy},
\end{equation}
\begin{equation}\label{gxx}
    \frac{\partial \sigma_{xx}}{\partial t}=-\frac{\alpha}{(1-\bar{\eta_s})}(\sigma_{xx}^2+\sigma_{xy}^2)+2Wi\dot{\gamma}\sigma_{xy}-\sigma_{xx},
\end{equation}
and
\begin{equation}\label{gyy}
    \frac{\partial \sigma_{yy}}{\partial t}=-\frac{\alpha}{(1-\bar{\eta_s})}(\sigma_{yy}^2+\sigma_{xy}^2)-\sigma_{yy}.
\end{equation}

%
%\textcolor{blue}{We also analyse a model to study rheological hysteresis in thixotropic materials. We use an inelastic thixotropic model which is discussed in section. We study the thixotropic model only for comparison purposes. }

\textit{Flow protocol:} We follow the flow protocol suggested by Divoux et al.\, [\onlinecite{divoux2013rheological}] to study hysteresis. According to this flow protocol, we pre-shear, wherein we subject the model to high shear rate which is same as the maximum shear rate. For shear rate sweep experiments, shear rate is first decreased in steps with the system allowed to be in a given shear rate for a time $\delta t$ for each step. After reaching the minimum value of shear rate, it is again increased to the maximum value in steps and $\delta t$ time on each step. The time $\delta t$ is also referred to as sweep rate [\onlinecite{divoux2013rheological}] or shearing time [\onlinecite{jamali2019multiscale}] in the literature. We measure the area of hysteresis loop  as suggested by Divoux et al [\onlinecite{divoux2013rheological}] which is expressed as follows:

\begin{equation}
    A_{\sigma}= \int_{Wi_{\text{min}}}^{Wi_{\text{max}}} |\Delta {\Sigma}_{xy} (Wi)|\,d\text{log}Wi 
\end{equation}
where, $|\Delta {\Sigma}_{xy} (Wi)|=|{\Sigma}_{xy,down} (Wi)-{\Sigma}_{xy,up} (Wi)|$. We also measure area under the down-sweep $\Sigma_{xy}-Wi$ curve as follows:

\begin{equation}
    A_{d}= \int_{Wi_{\text{min}}}^{Wi_{\text{max}}} {\Sigma}^d_{xy} (Wi)\,d\text{log}Wi 
\end{equation}
where, $\Sigma^d_{xy}(Wi)$ is the stress as a function of $Wi$ during down-sweep shear flow. The $Wi$ values are equally spaced on the logarithmic scale to avoid dominance of higher shear rate values.  The number of steps per decade are fixed and is represented by $n$. As we consider inertialess flow and a planar geometry, the shear stress is assumed to be constant in the $y$ direction. We solve the above equations (Eqs. \ref{gxy}-\ref{gyy}) for down-sweep and up-sweep shear flow using the ode45 subroutine by MATLAB \textsuperscript{\textregistered}. 

\section{Results and discussion} \label{section_results}

We present results for the down-sweep followed by up-sweep paths by plotting shear stress as a function of $Wi$, wherein the shear stress value is obtained at the end of $\delta t$ time (shearing time or sweep rate) spent at each shear rate step. Furthermore, we study down-sweep and up-sweep flow of two categories of viscoelastic materials: (i) the materials with relaxation time of the order of $10^{-1}$ s, and (ii) the materials with relaxation time of the order of $10^6$ s or higher. We also investigate the possibility of hysteresis loop in these viscoelastic materials and resemblance with the hysteresis loops observed for thixotropic fluids in the literature.

\subsection{Low relaxation time viscoelastic materials} \label{subsection_lve}
 
\begin{figure}[htbp]
\centering
\subfigure[]{
\includegraphics[scale=0.19]{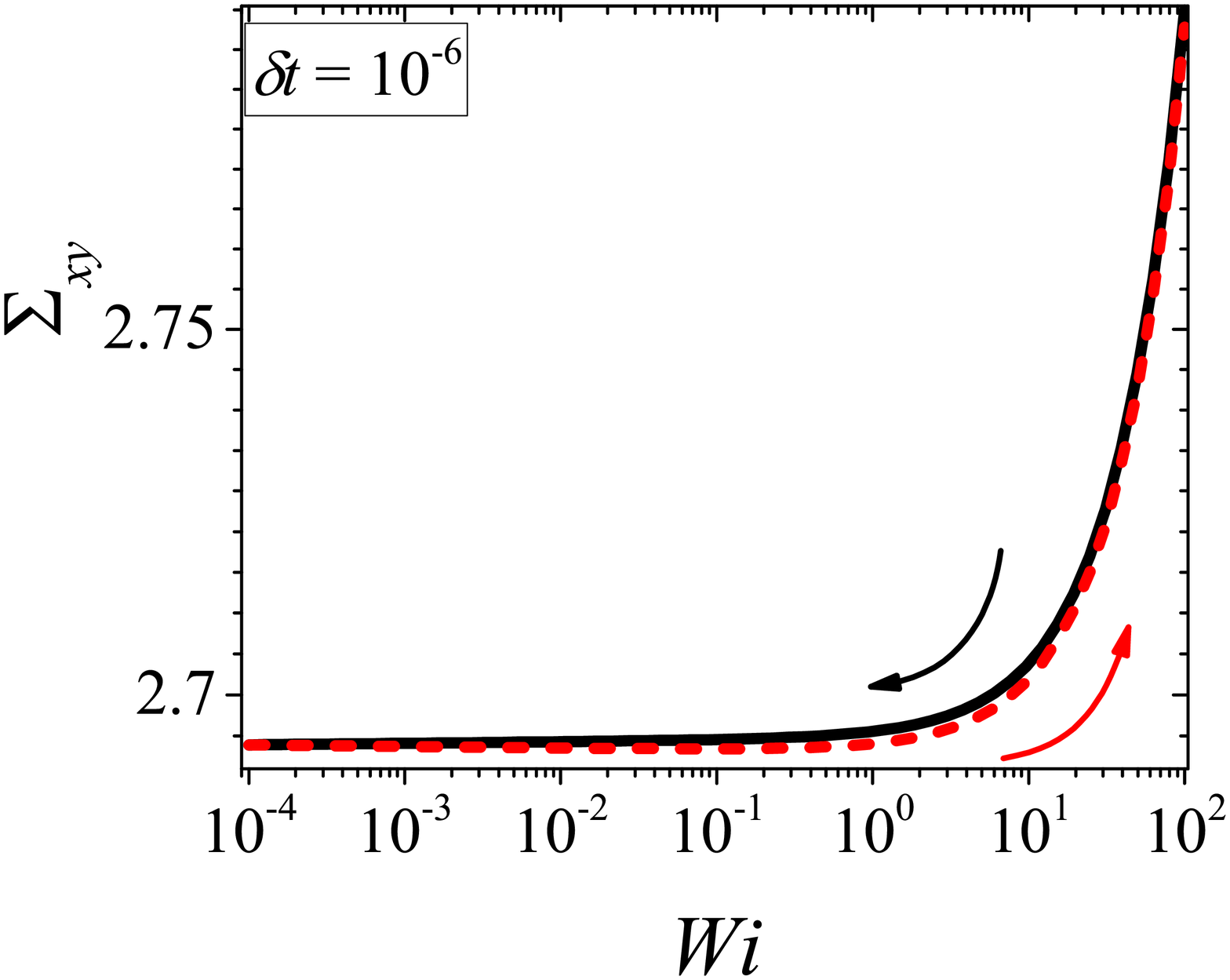}
    \label{1e_6}
}
\subfigure[]{
    \includegraphics[scale=0.19]{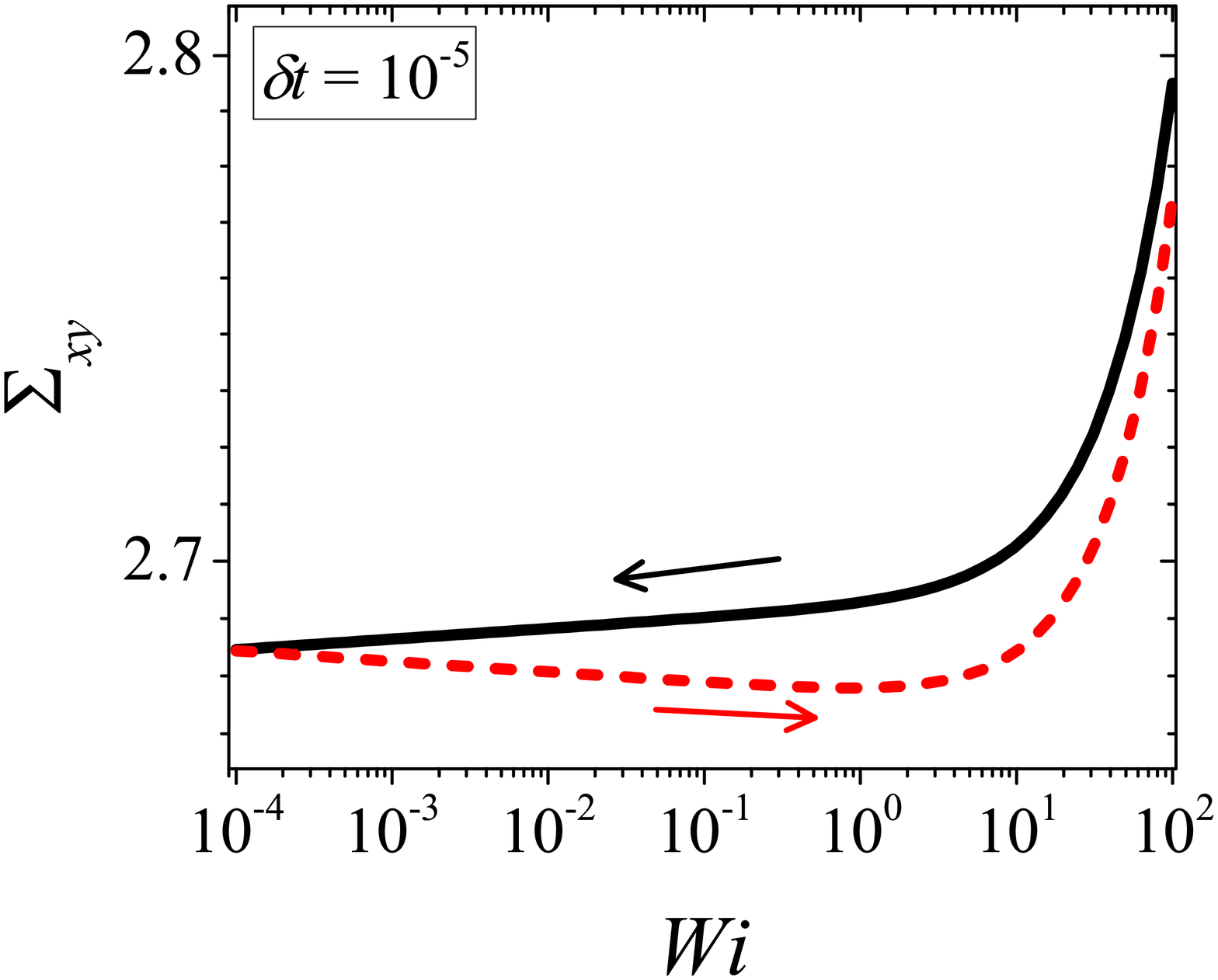}
    \label{1e_5}
}
\subfigure[]{
    \includegraphics[scale=0.19]{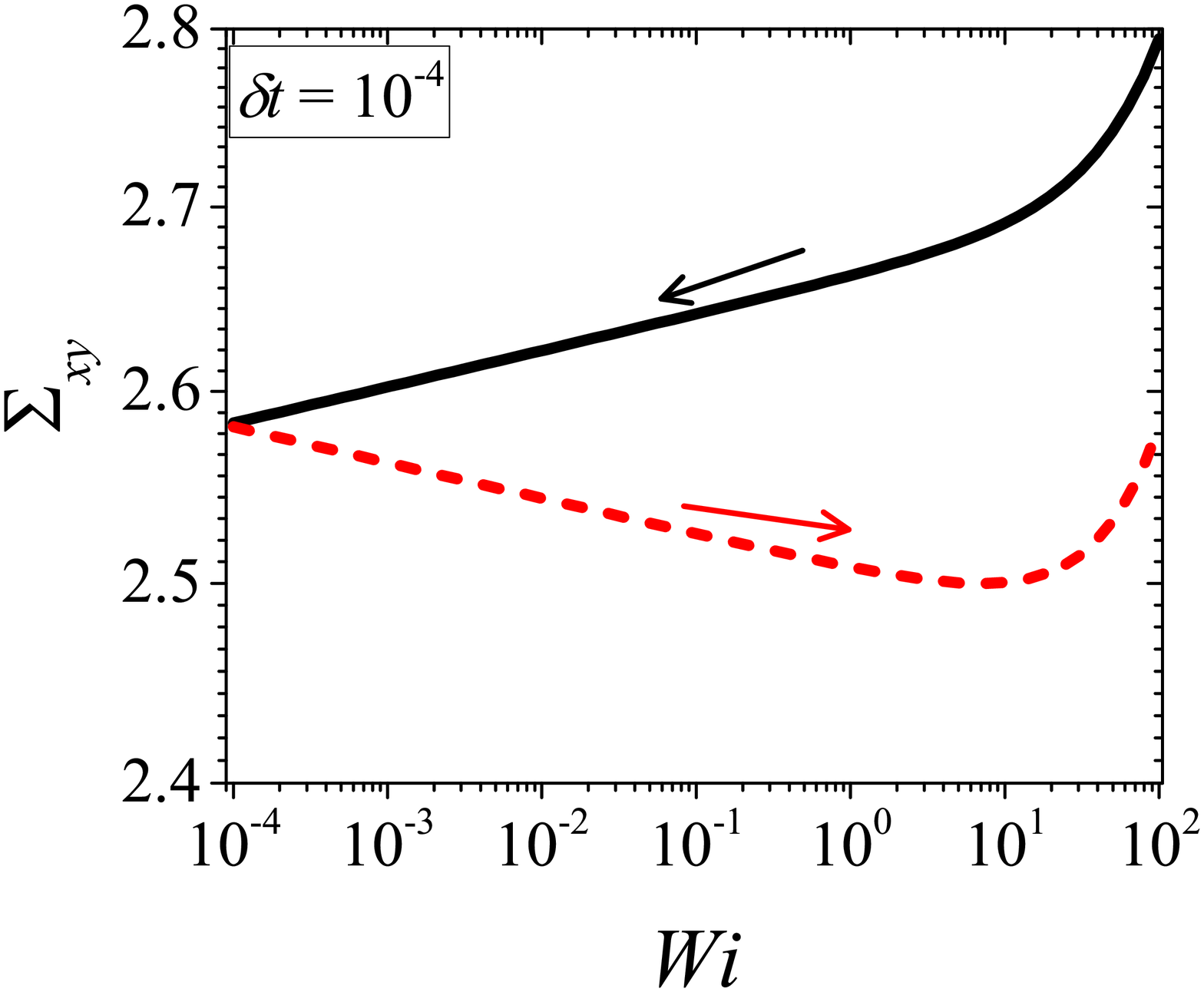}
    \label{1e_4}
}
\subfigure[]{
\includegraphics[scale=0.19]{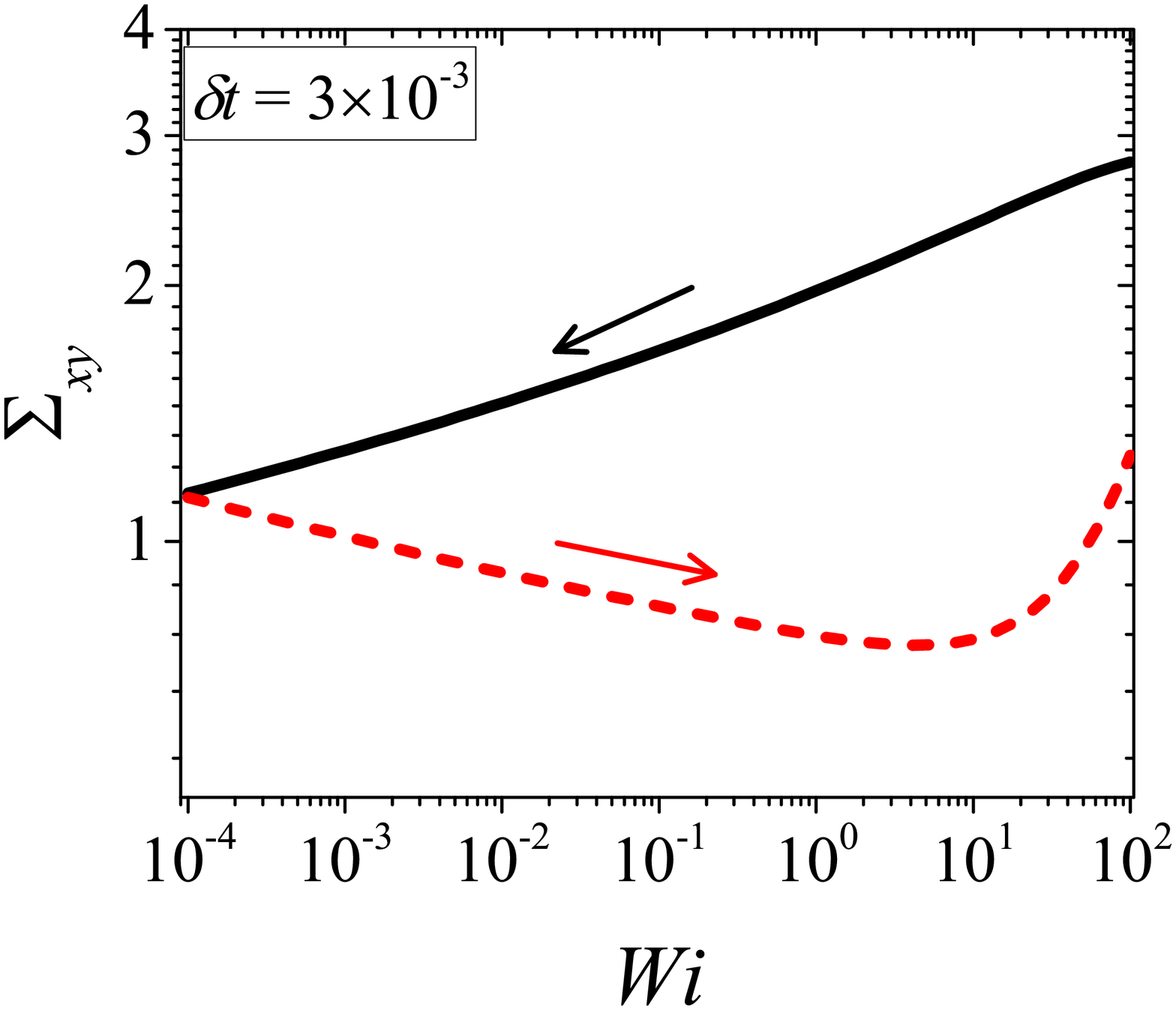}
    \label{3e_3}
}
\subfigure[]{
    \includegraphics[scale=0.19]{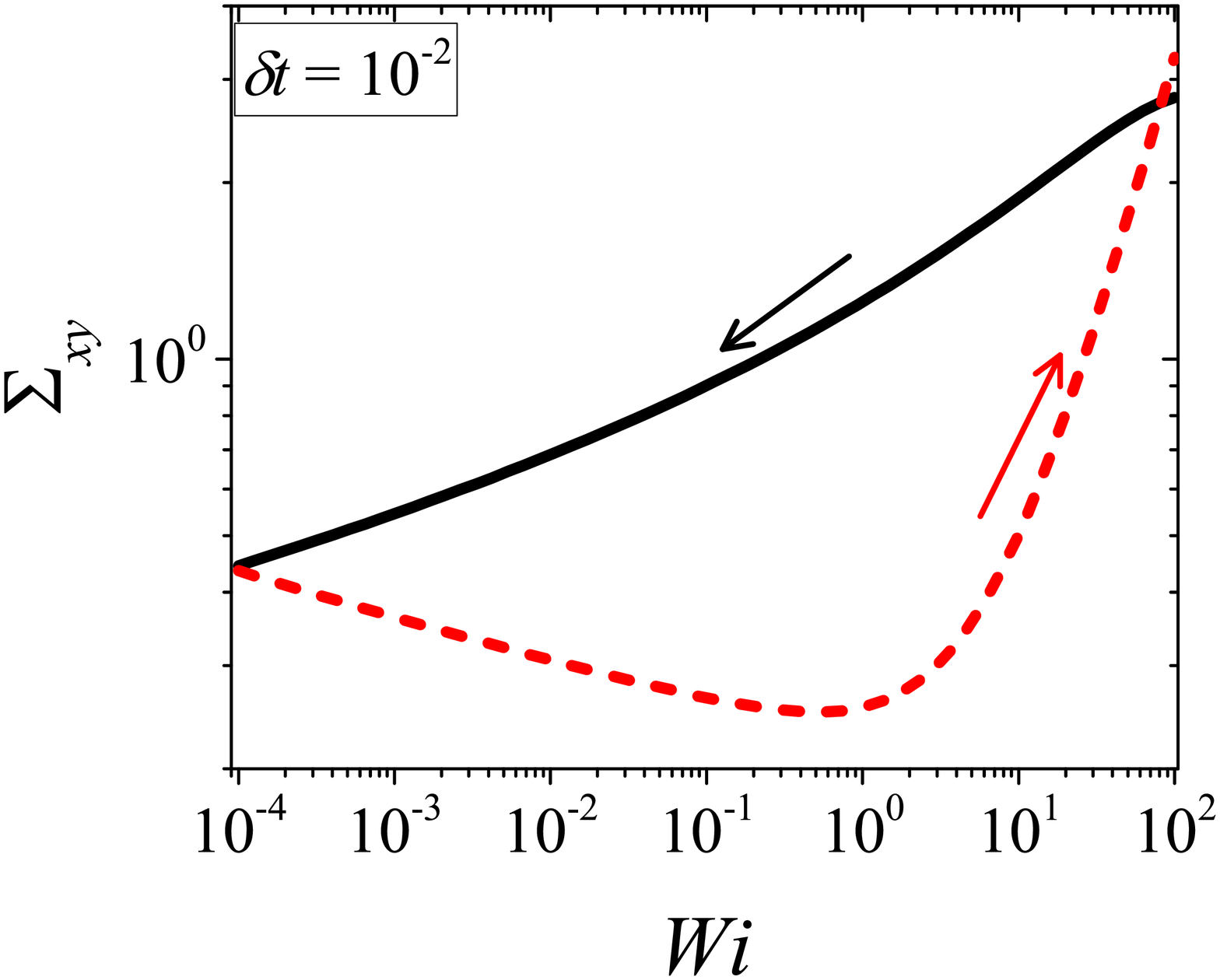}
    \label{1e_2}
}
\subfigure[]{
\includegraphics[scale=0.19]{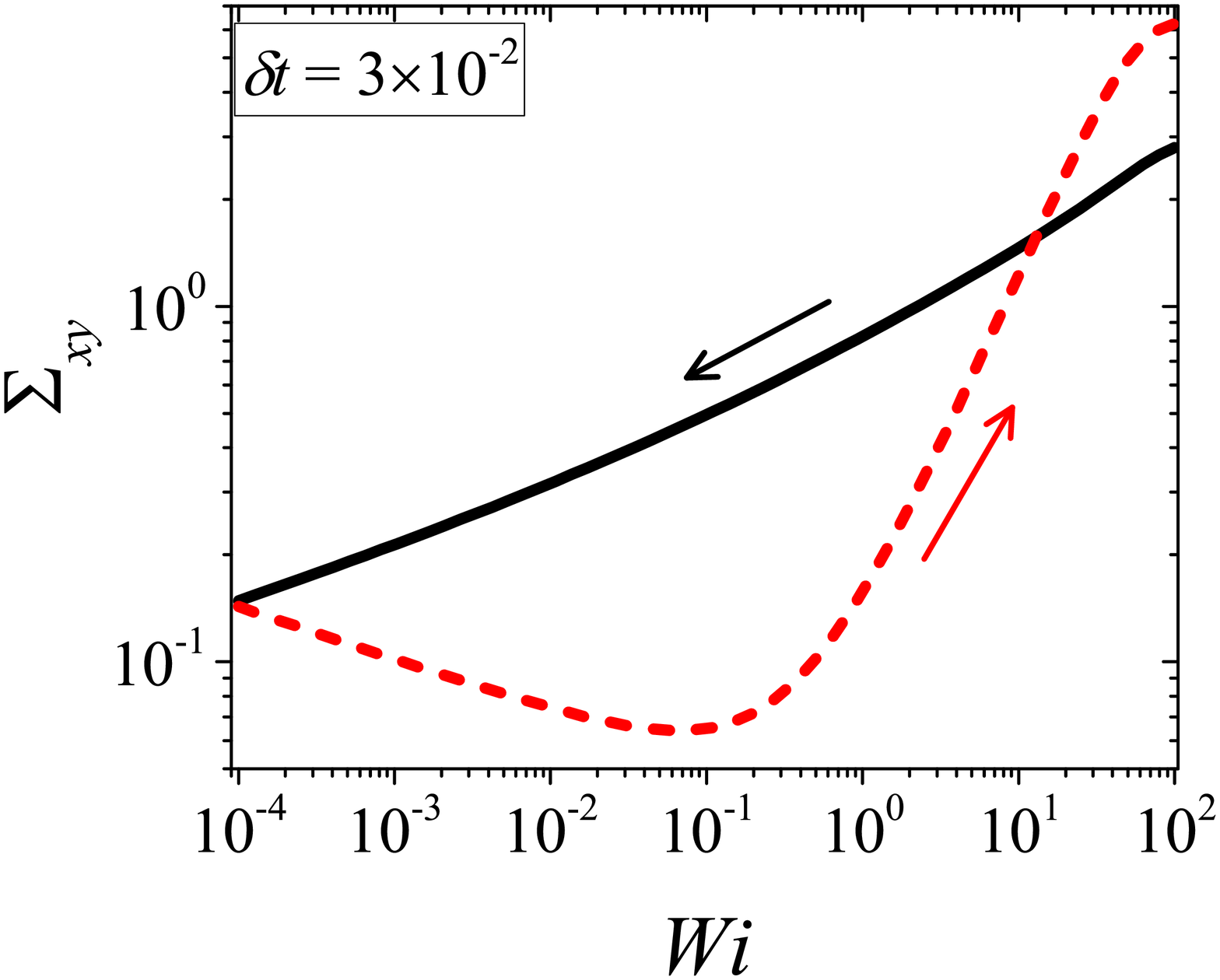}
    \label{3e_2}
}
\subfigure[]{
\includegraphics[scale=0.19]{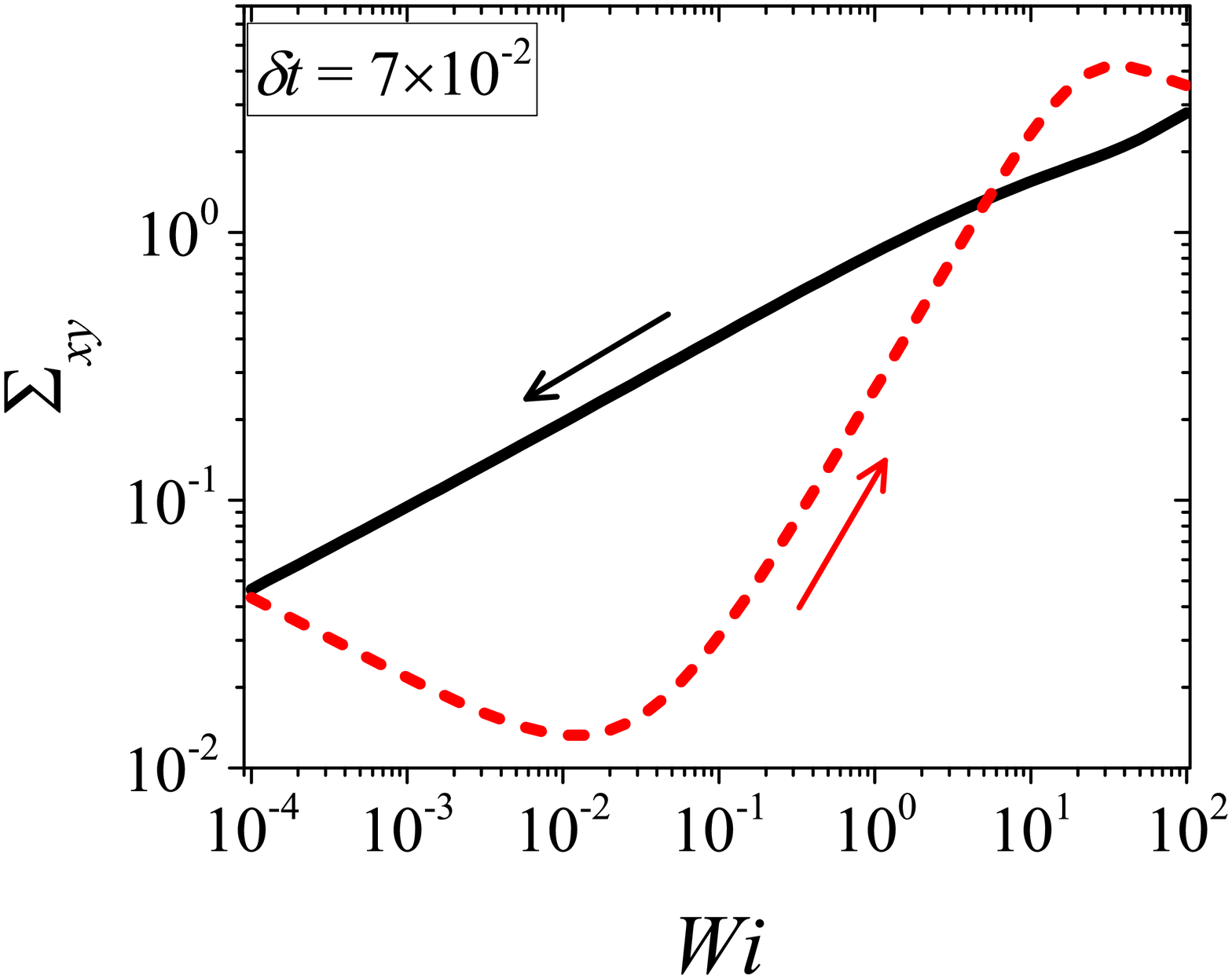}
    \label{7e_2}
}
\subfigure[]{
\includegraphics[scale=0.19]{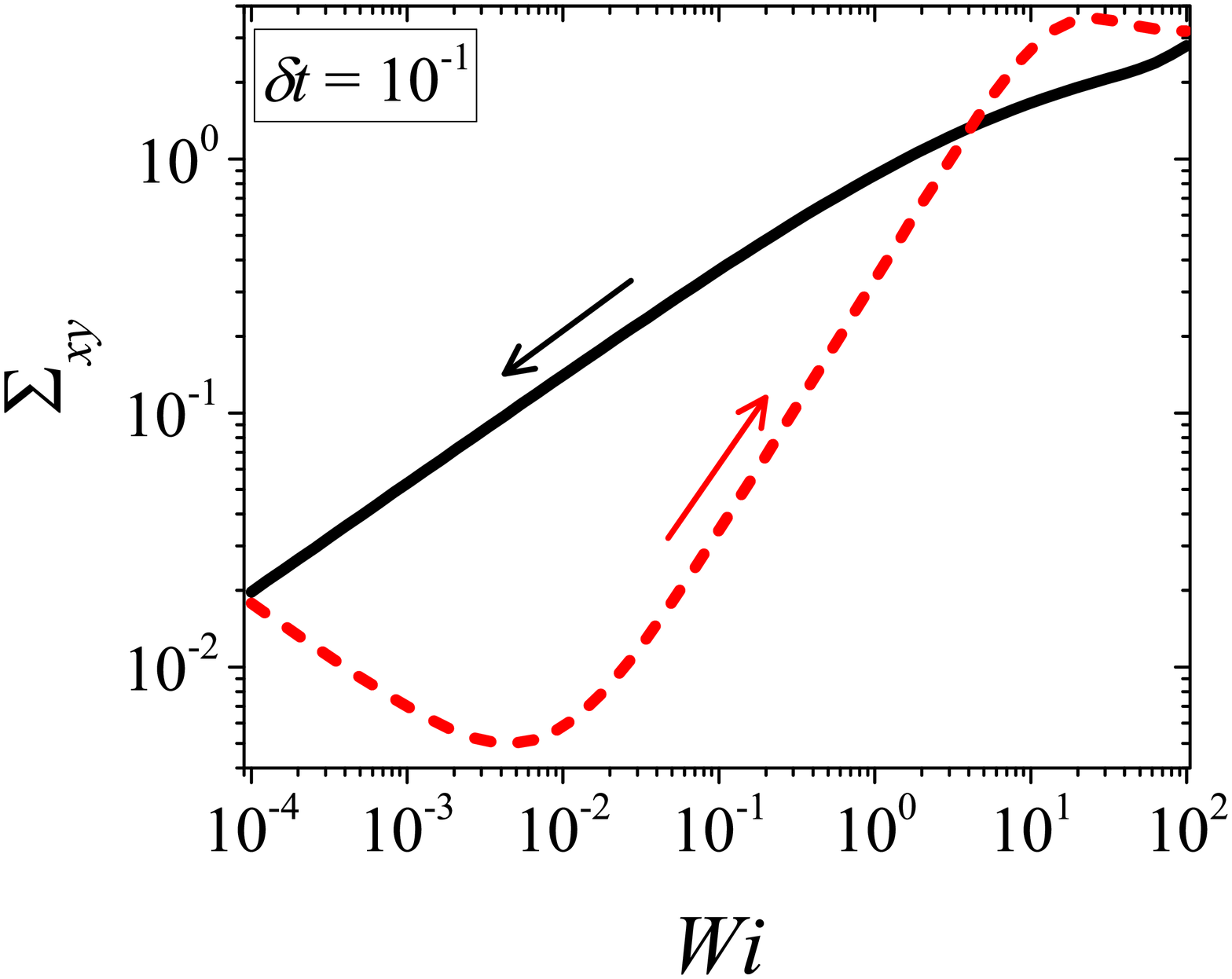}
    \label{1e_1}
}
\subfigure[]{
\includegraphics[scale=0.19]{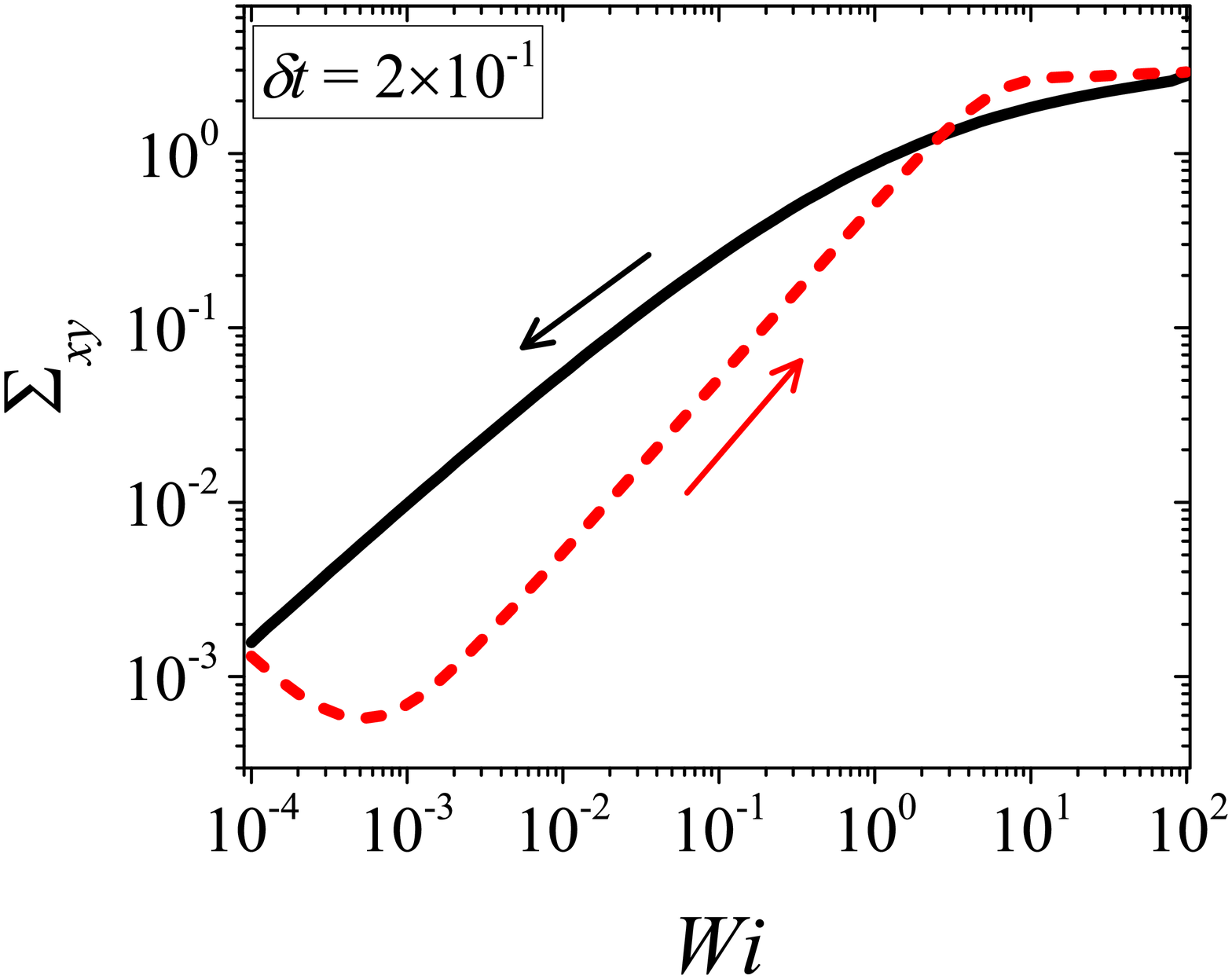}
    \label{2e_1}
}
\subfigure[]{
\includegraphics[scale=0.19]{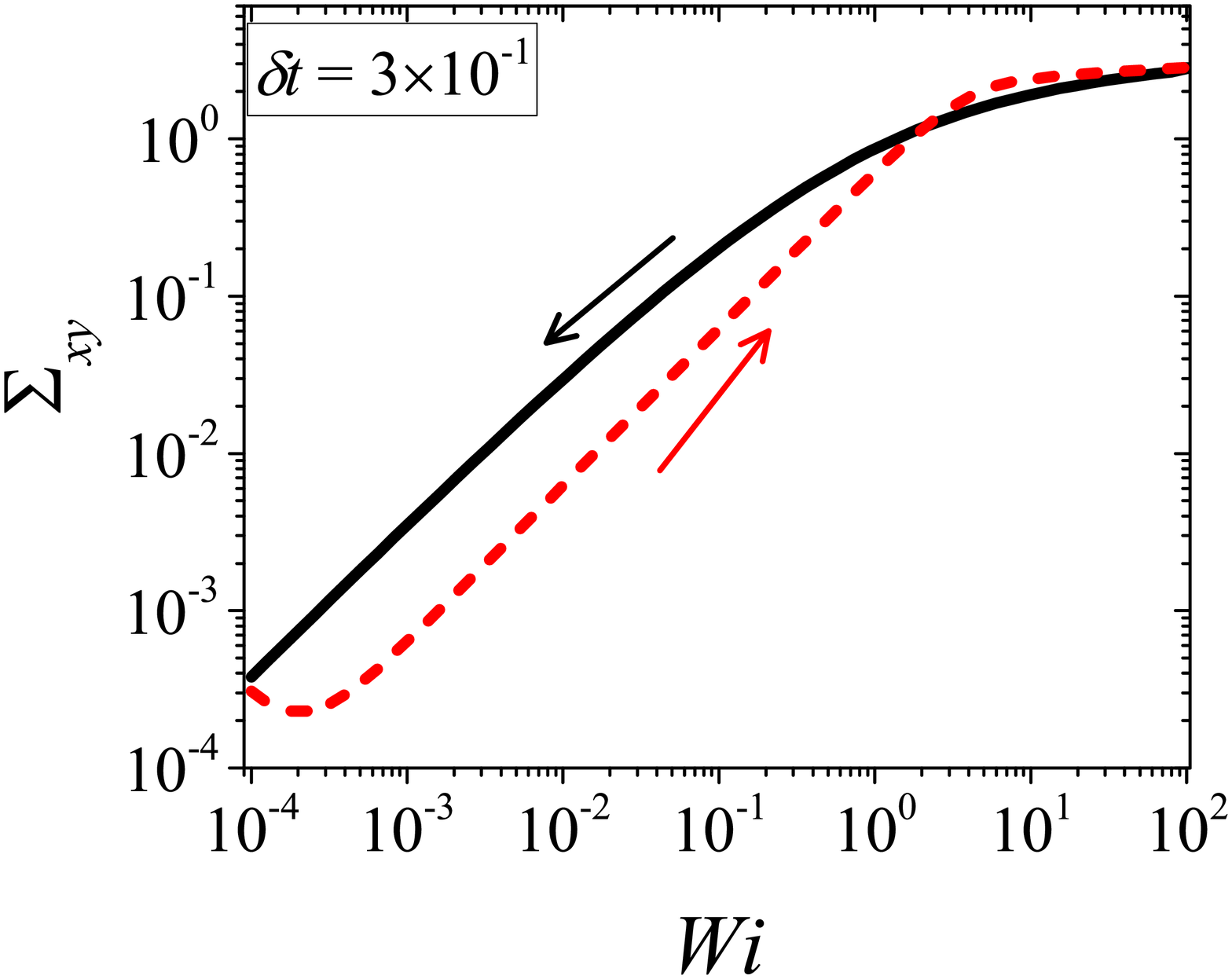}
    \label{3e_1}
}
\subfigure[]{
\includegraphics[scale=0.19]{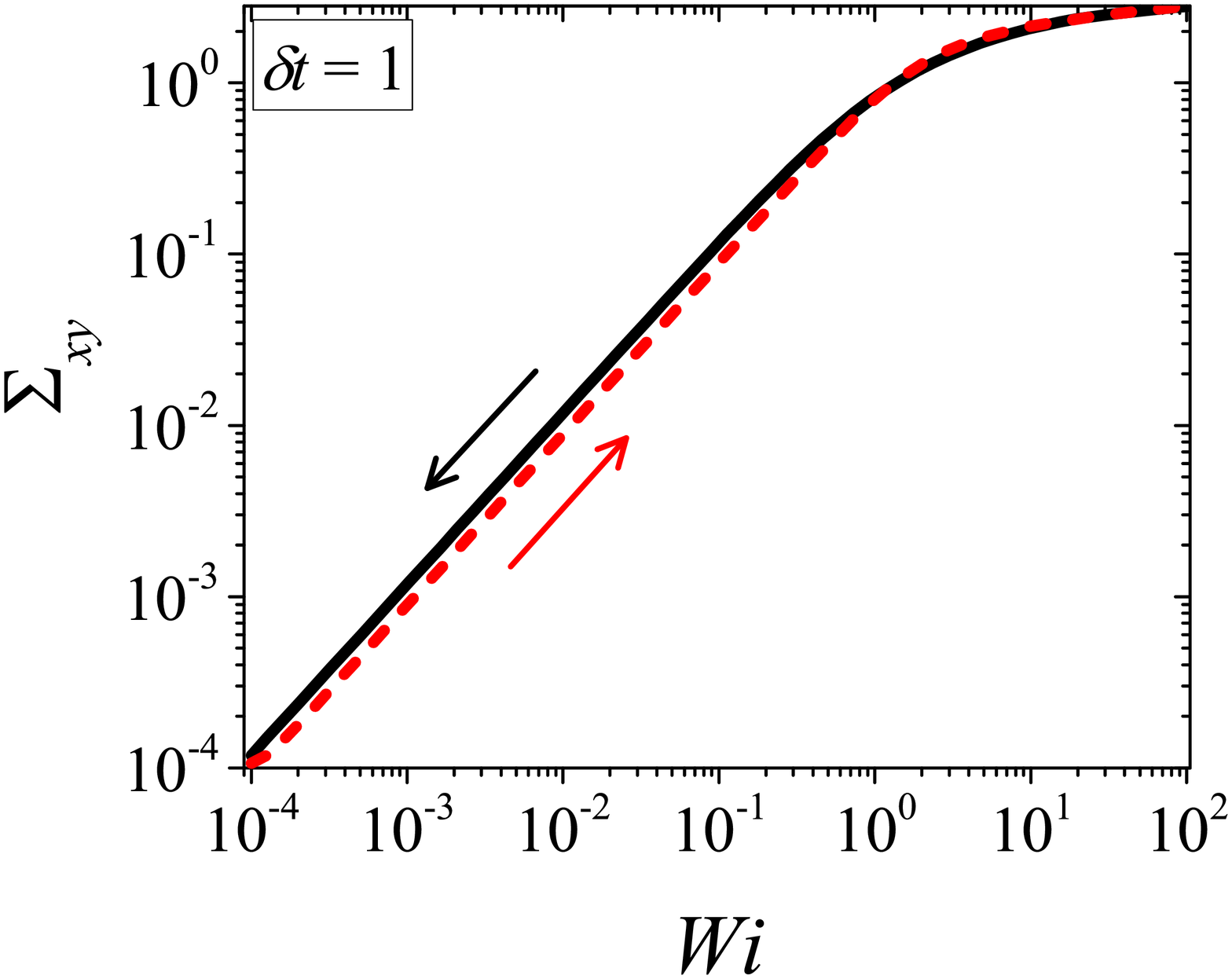}
    \label{1}
}
\subfigure[]{
\includegraphics[scale=0.19]{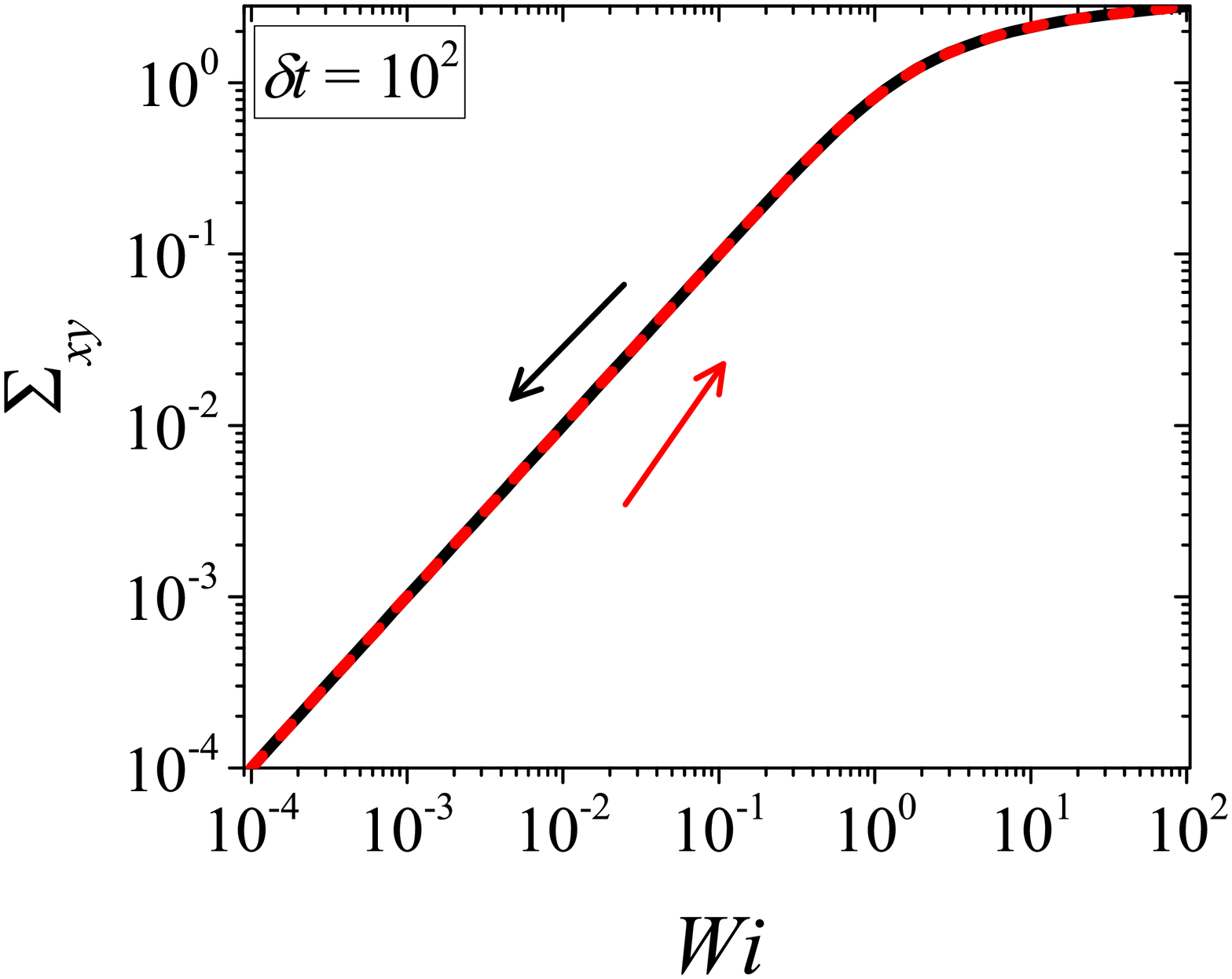}
    \label{100}
}

\caption{\scriptsize Types of hysteresis loops (in a down-up shear rate cycle) obtained for the low relaxation time $\tau=0.1$ s viscoelastic material using different values of $\delta t$ and $\bar{\eta_s}=10^{-3}$ with $n=10$. Shear stress is plotted as a function of $Wi$ for $\delta t$ equals to (a) $10^{-6}$, (b) $10^{-5}$, (c) $10^{-4}$, (d) $3\times10^{-3}$, (e) $10^{-2}$, (f) $3\times10^{-2}$, (g) $7\times10^{-2}$, (h) $10^{-1}$, (i) $2\times10^{-1}$, (j) $3\times10^{-1}$, (k) $1$, and (l) $10^2$. Solid lines shows down-sweep and dashed line shows the up-sweep shear flow results. (All the variables in this figure are dimensionless as mentioned in section \ref{section_model}.)}
\label{fig:overall}
\end{figure}

\begin{figure}[htbp]
\centering
     \subfigure[]{
\includegraphics[scale=0.189]{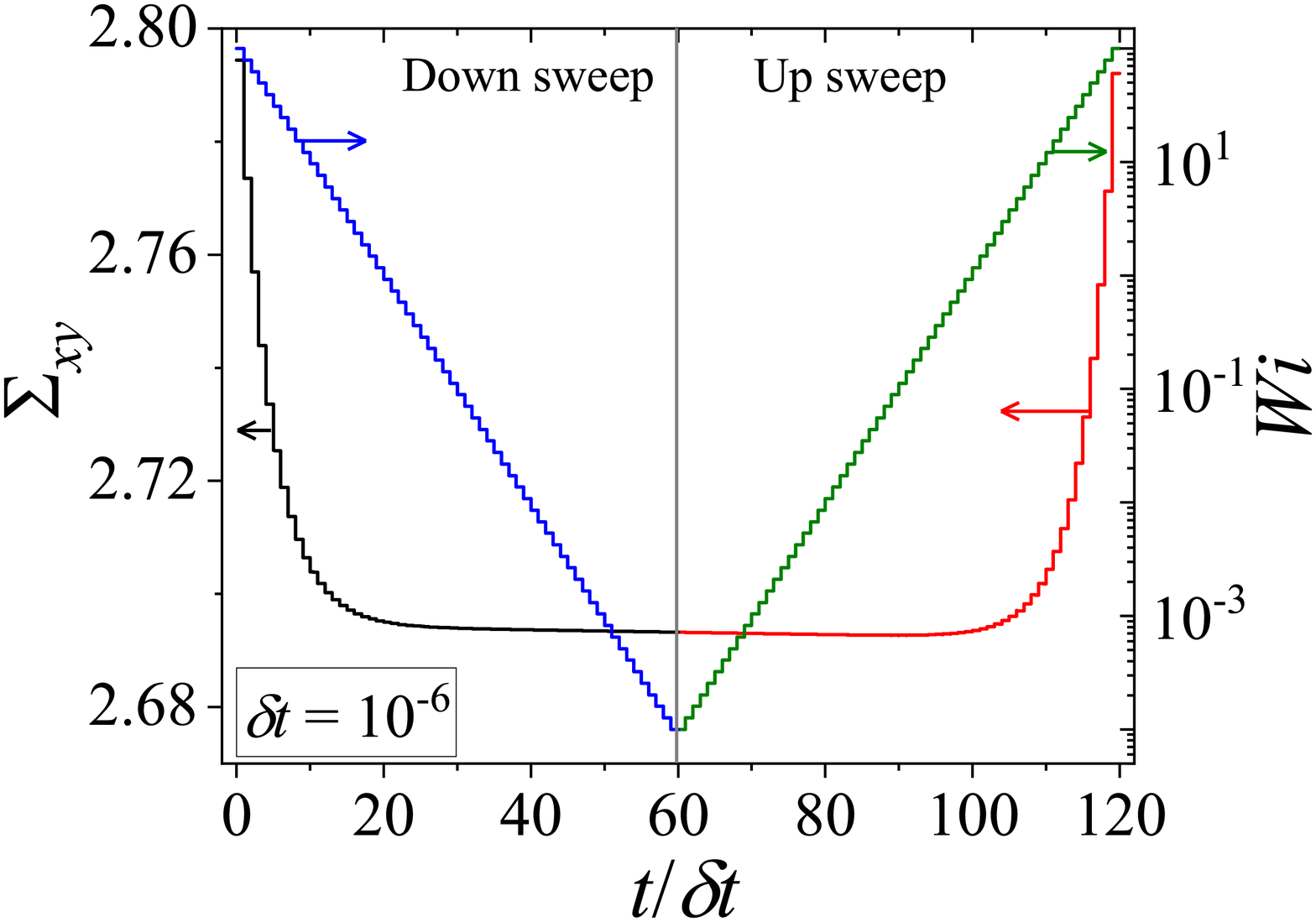}
    \label{ss1e_6}
  }
  \subfigure[]{
    \includegraphics[scale=0.19]{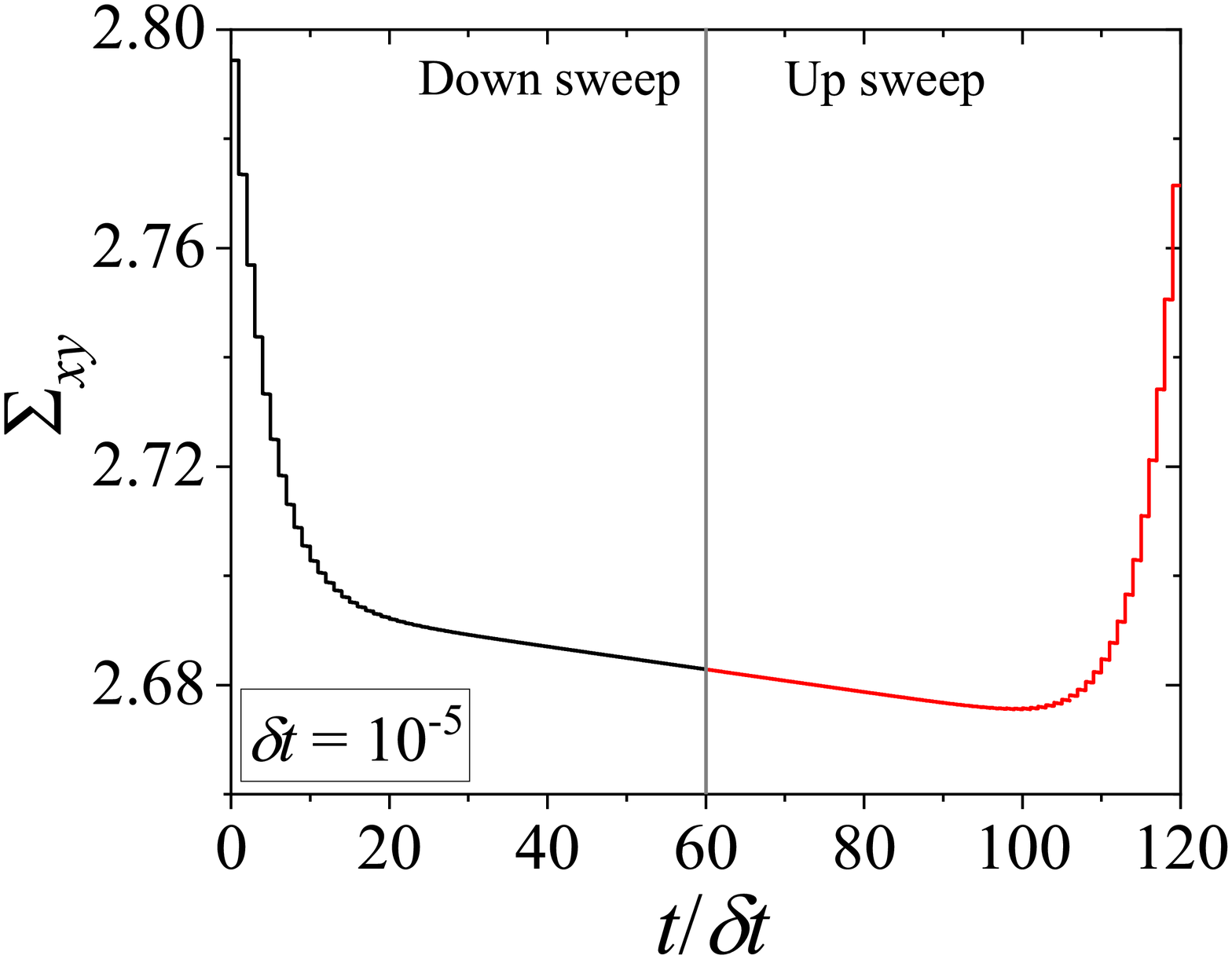}
    \label{ss1e_5}
  }
  \subfigure[]{
    \includegraphics[scale=0.19]{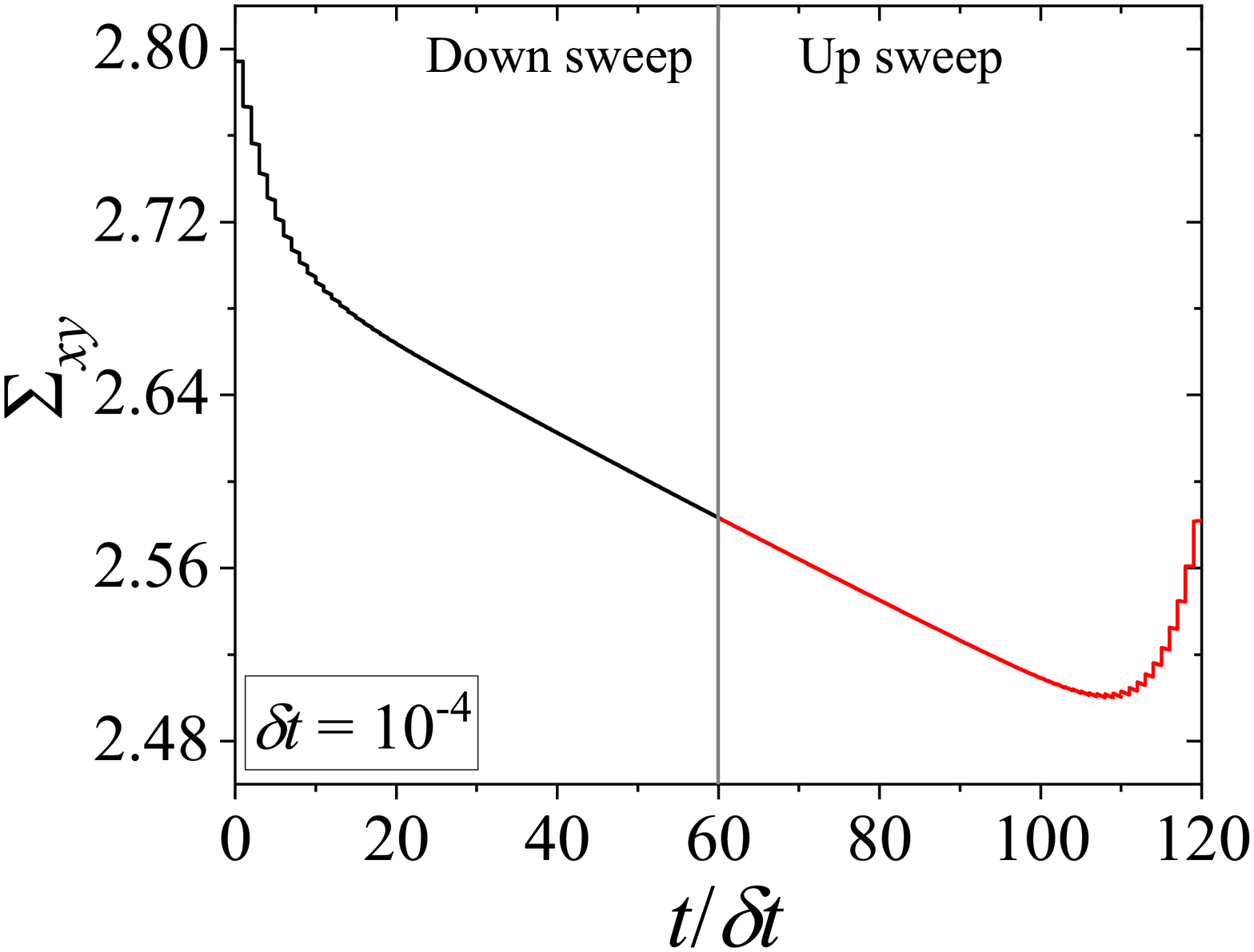}
    \label{ss1e_4}
  }
   \subfigure[]{
\includegraphics[scale=0.19]{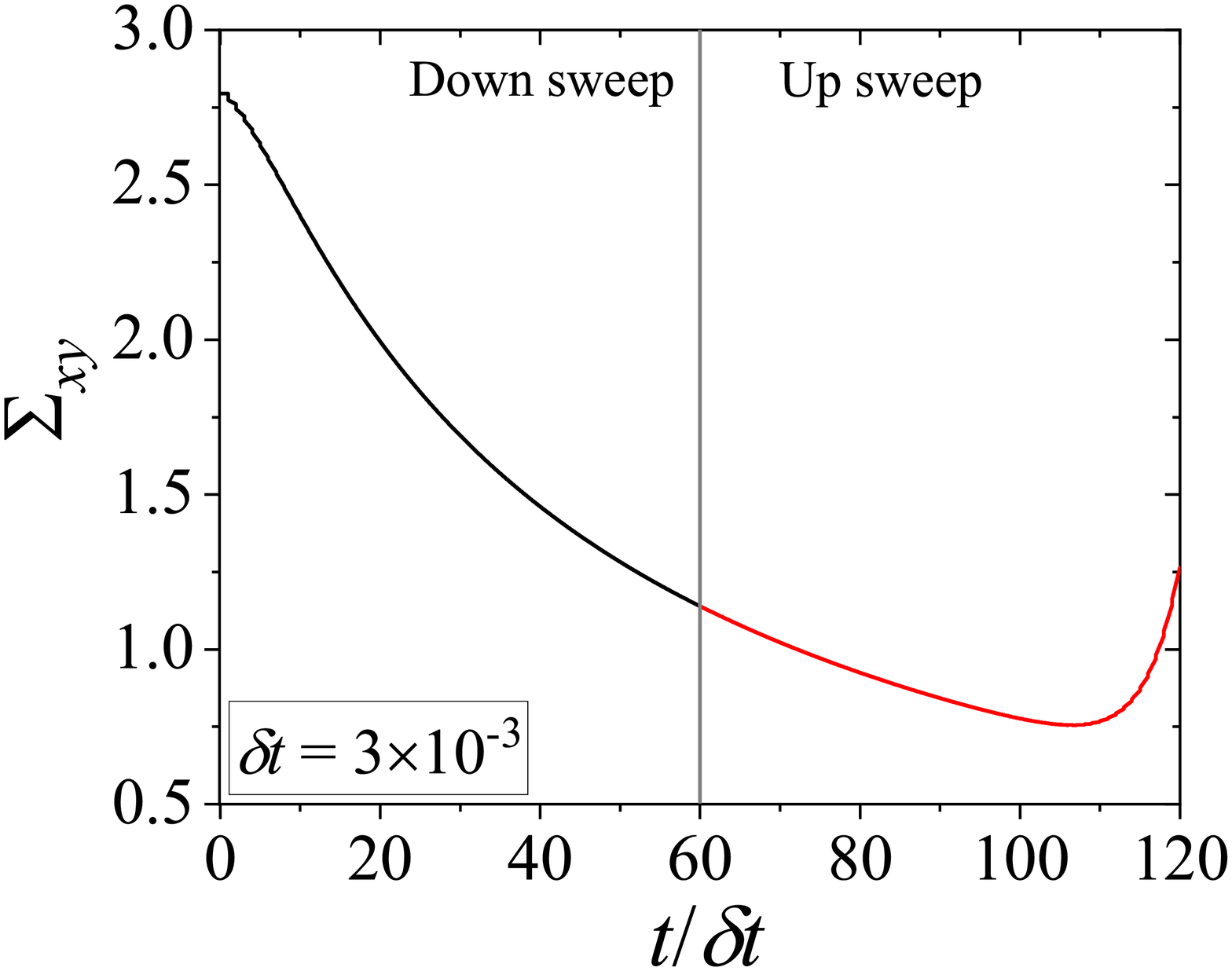}
    \label{ss3e_3}
  }
    \subfigure[]{
    \includegraphics[scale=0.19]{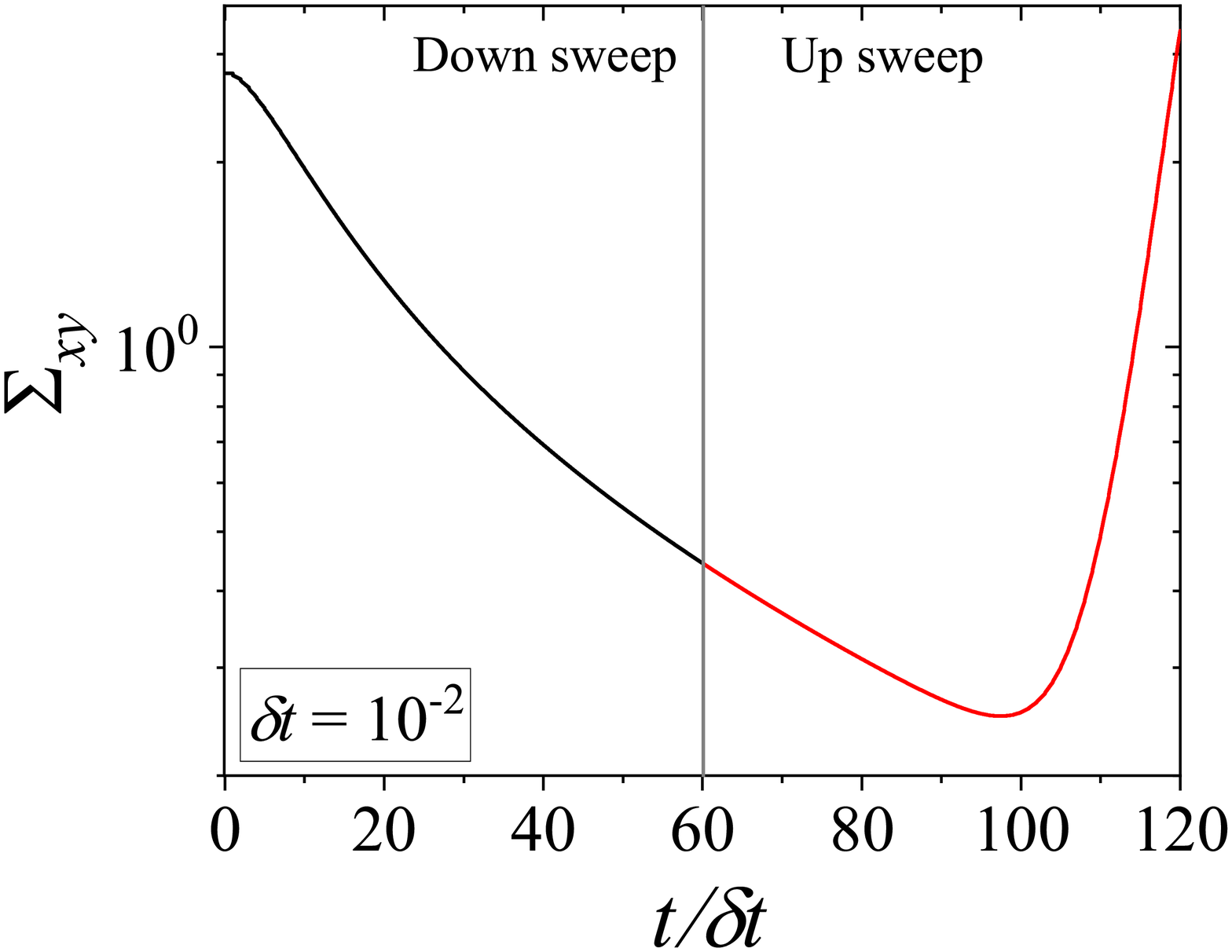}
    \label{ss1e_2}
  }
   \subfigure[]{
\includegraphics[scale=0.19]{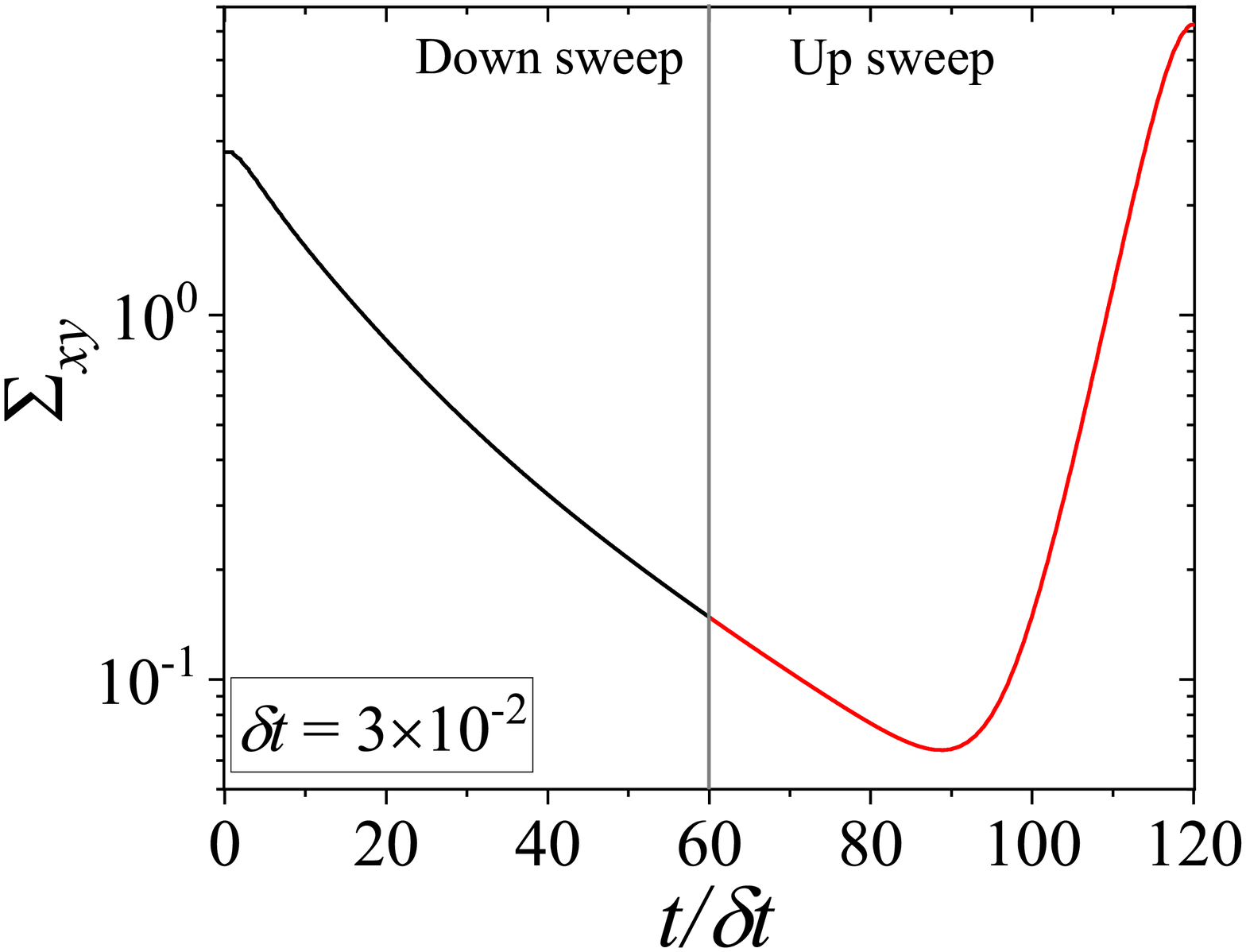}
    \label{ss3e_2}
  }
       \subfigure[]{
\includegraphics[scale=0.19]{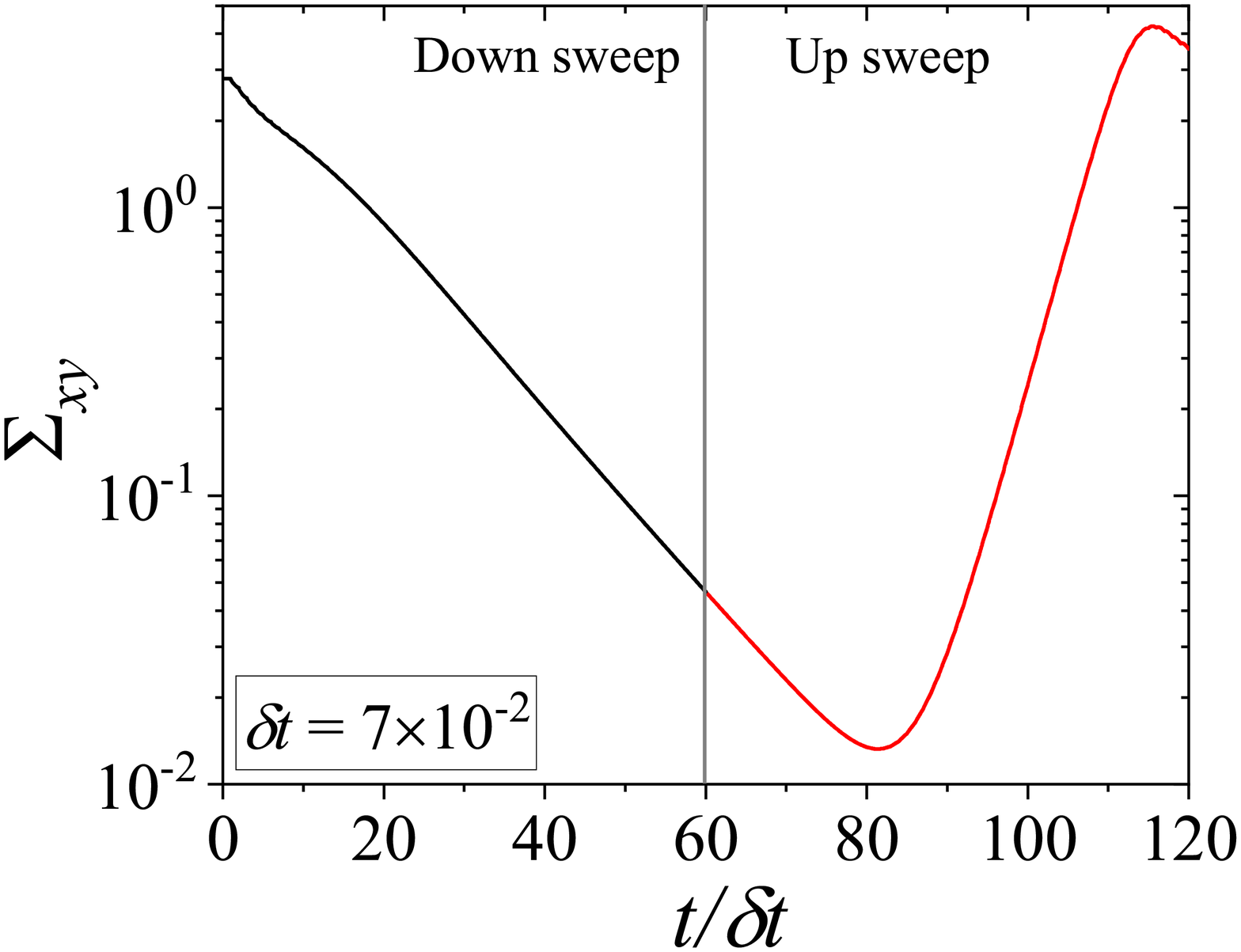}
    \label{ss7e_2}
  }
     \subfigure[]{
\includegraphics[scale=0.19]{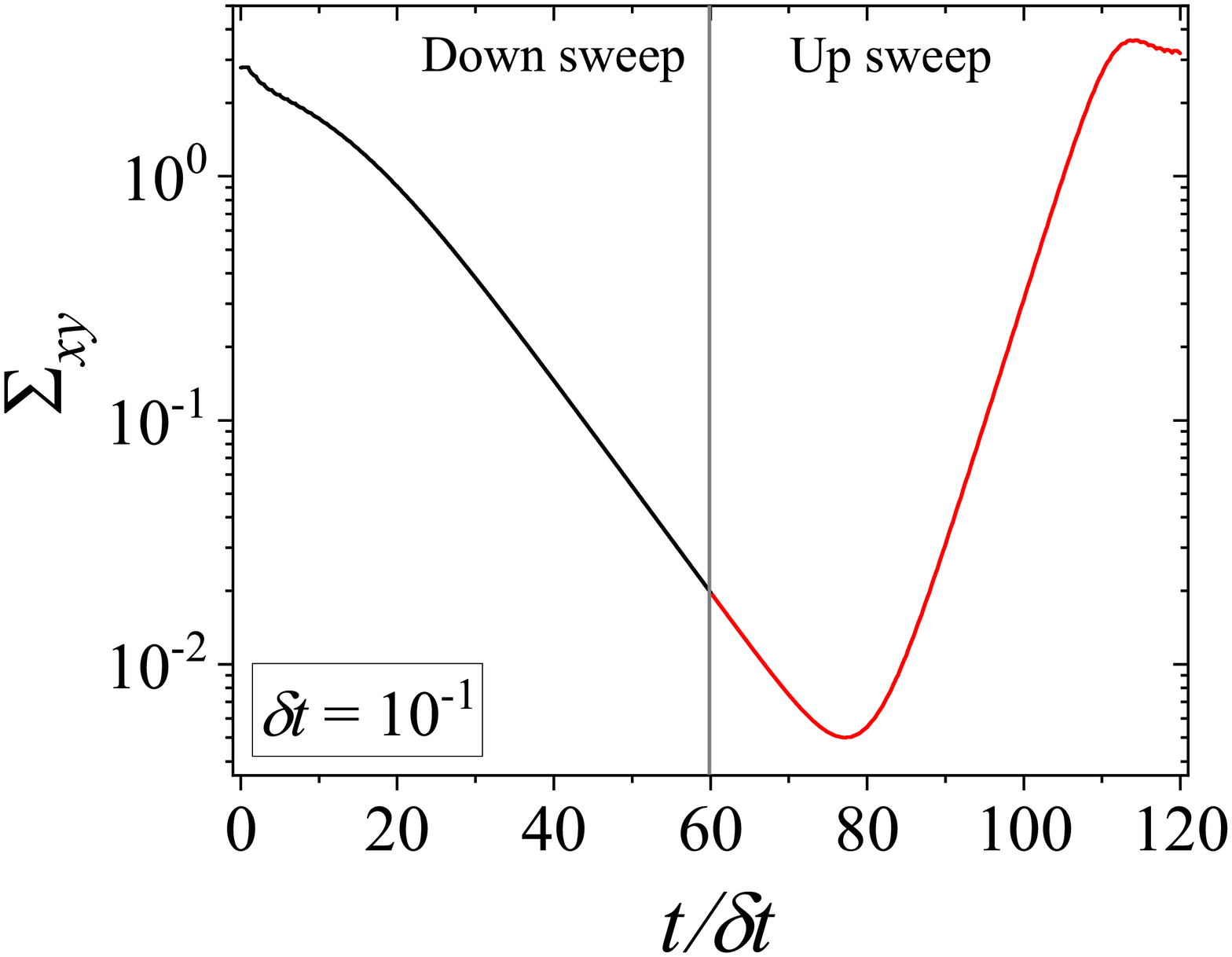}
    \label{ss1e_1}
  }
     \subfigure[]{
\includegraphics[scale=0.19]{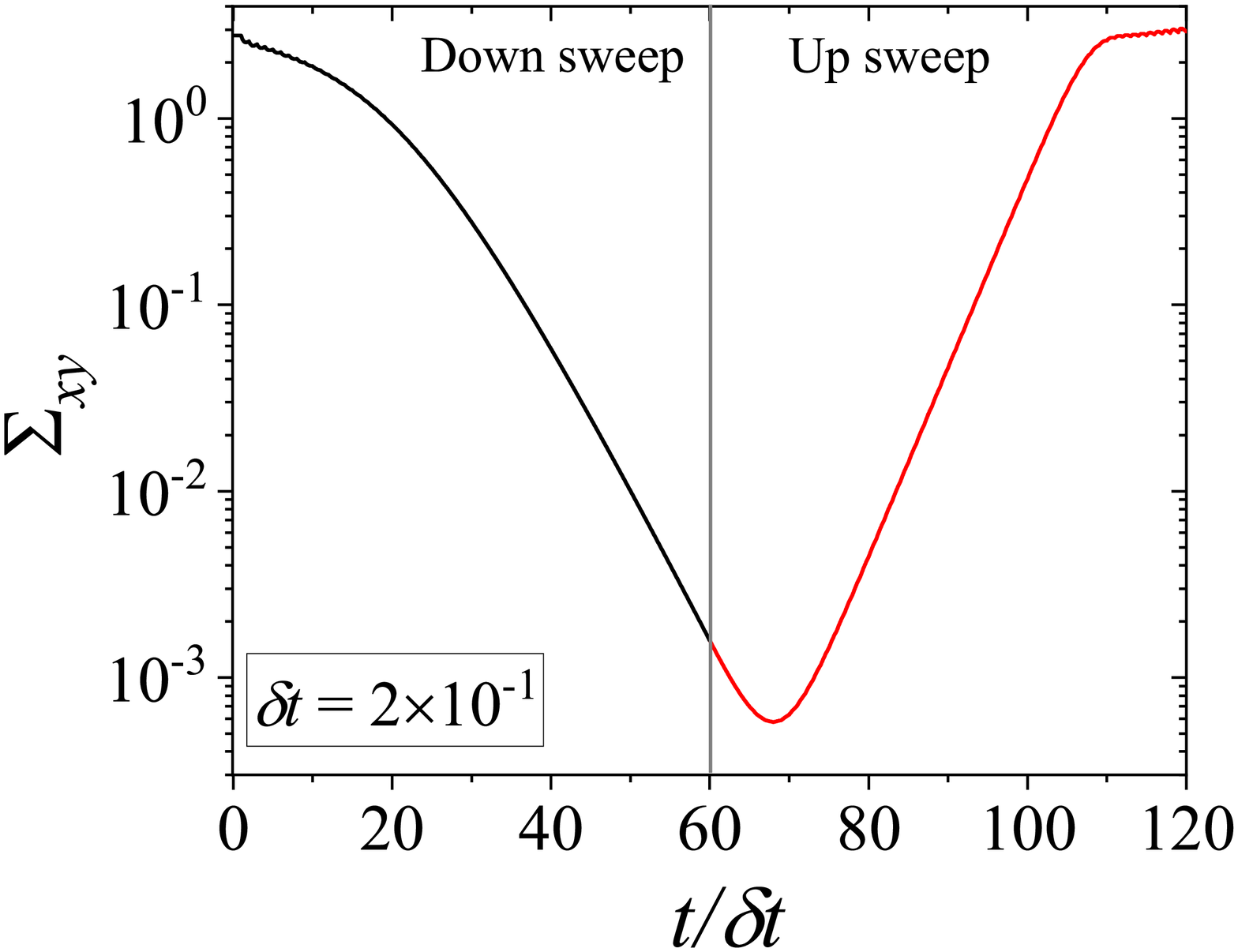}
    \label{ss2e_1}
  }
     \subfigure[]{
\includegraphics[scale=0.19]{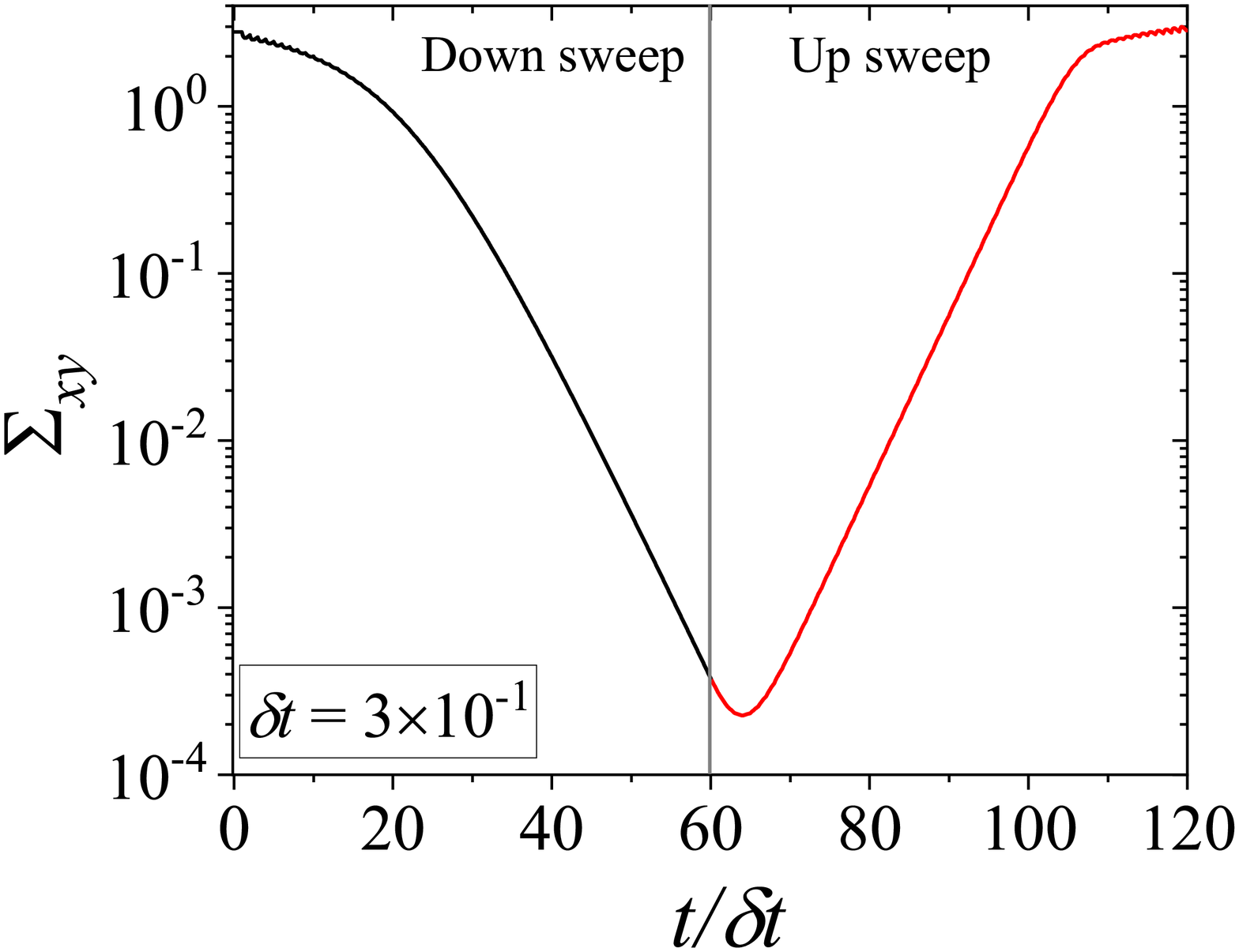}
    \label{ss3e_1}
  }
     \subfigure[]{
\includegraphics[scale=0.19]{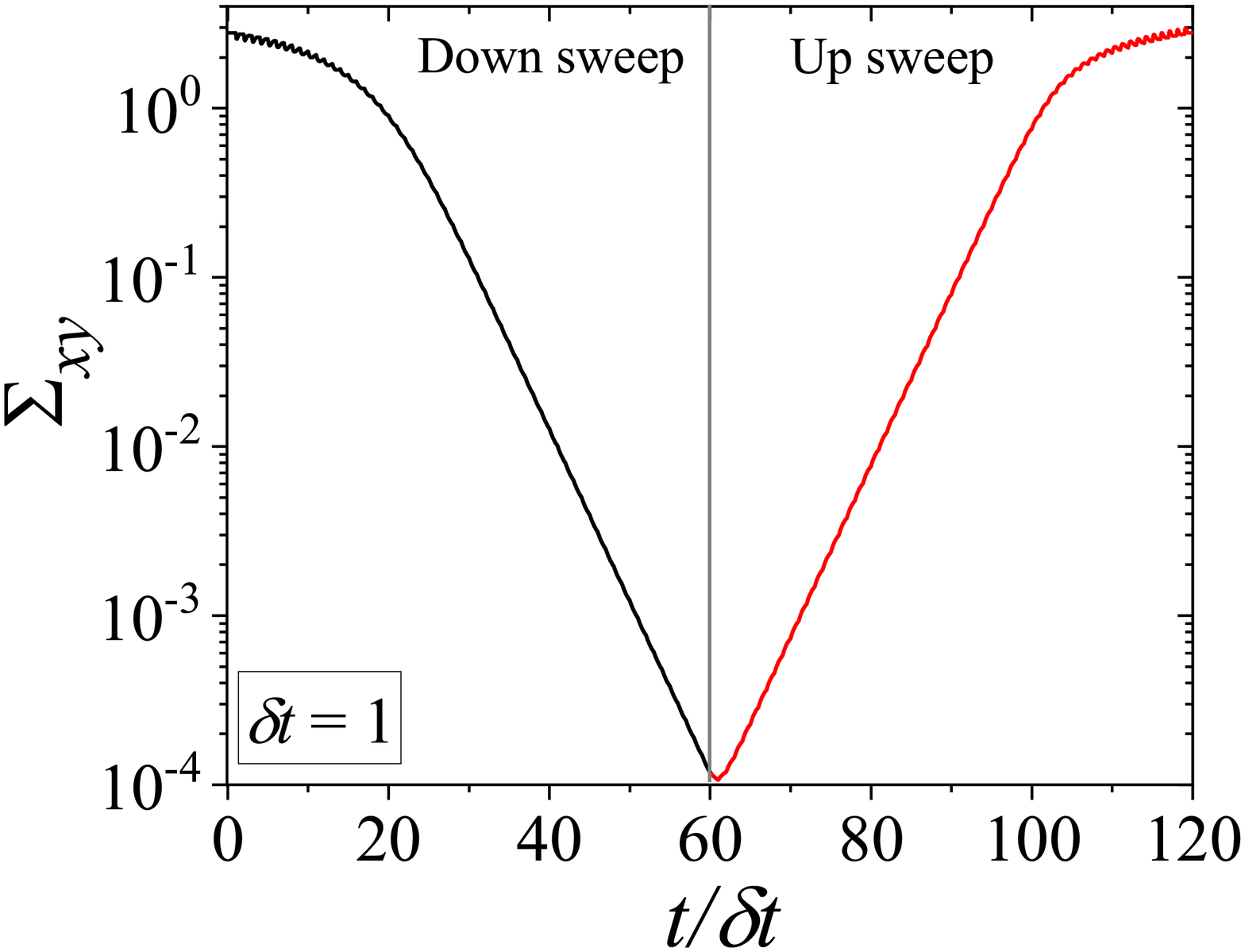}
    \label{ss1}
  }
      \subfigure[]{
\includegraphics[scale=0.19]{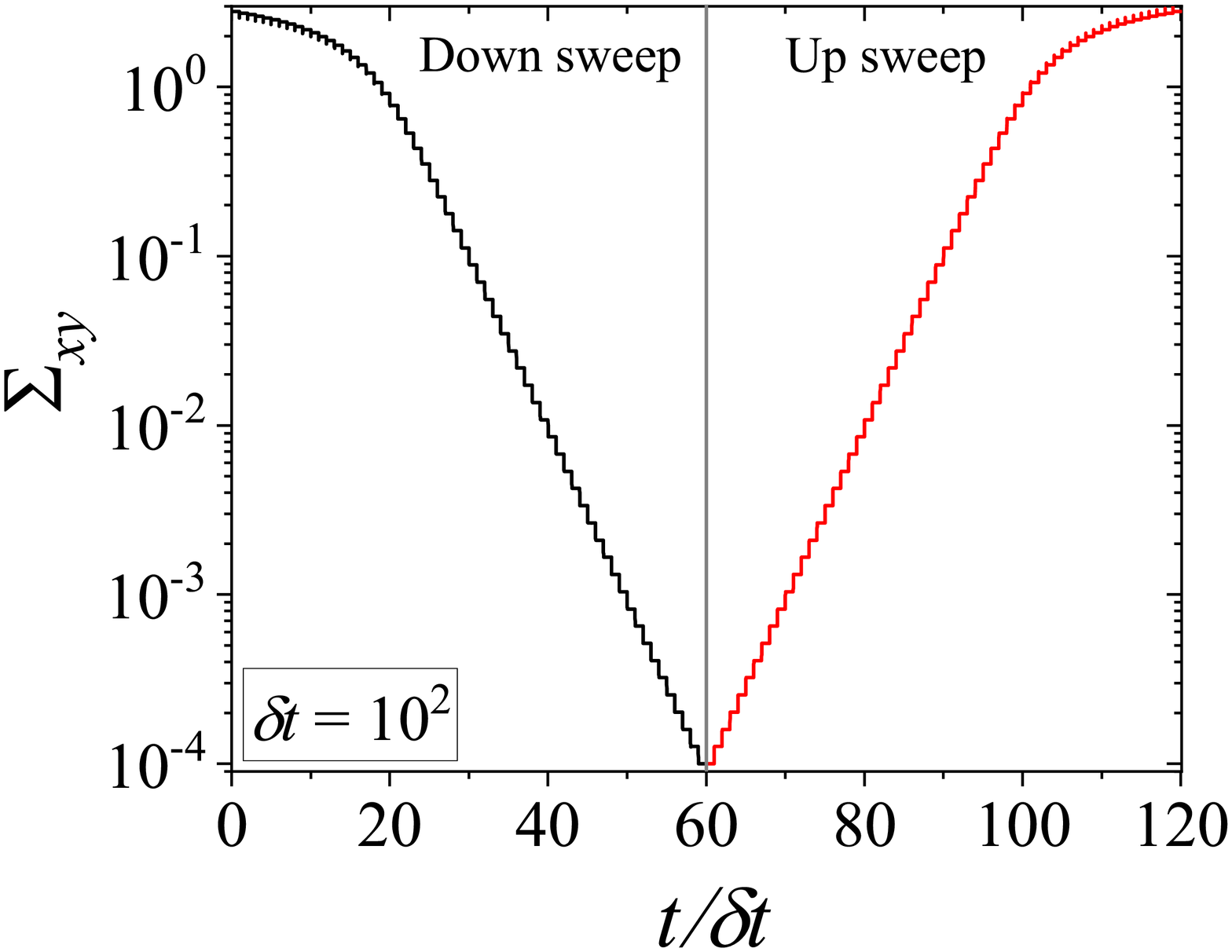}
    \label{ss100}
  }
\caption{\scriptsize Shear stress (left y-axis) is plotted as a function of time normalised by $\delta t$ for a high relaxation time material ($\tau=10^6$ s) with $n=10$. The different values of $\delta t$ in (a)  $10^{-6}$, (b) $10^{-5}$, (c) $10^{-4}$, (d) $3\times10^{-3}$, (e) $10^{-2}$, (f) $3\times10^{-2}$, (g) $7\times10^{-2}$, (h) $10^{-1}$, (i) $2\times10^{-1}$, (j) $3\times10^{-1}$, (k) $1$, and (l) $10^2$. This result is corresponding to hysteresis loops presented in Fig.\,\ref{fig:overall}. (All the variables in this figure are dimensionless as mentioned in section \ref{section_model}.) The gray line at $t/\delta t=60$ is to demarcate the down sweep and up sweep shear stress evolution during shear rate sweep flow. For clarity, variation of $Wi$ is shown only in Fig.\,(a). }
\label{fig:overall_stress_time}
\end{figure}

Consider a non-thixotropic viscoelastic material with a relaxation time $(\tau)$ of the order of $10^{-1}\,s$. The lowest value of shear rate ($\dot{\gamma^{*}}$) generally attainable in a standard rheometer is $10^{-3}$ s$^{-1}$ while the higher value of shear rate that is attainable is $10^{3}$ s$^{-1}$. Therefore, we study the range of the value of $Wi$ $=\tau\dot{\gamma^{*}}$ for down-sweep and up-sweep experiments from $10^{-4}$ to $10^{2}$. We study down-sweep and up-sweep flow for different values of $\delta t$ varying from $10^{-6}$ to $10^{2}$. In our simulation protocol, we first decrease the shear rate (down-sweep) in a step-wise manner to attain a minimum value (from $Wi = 10^{2}$ to $Wi = 10^{-4}$). After reaching the minimum shear rate, we increase the shear rate (up-sweep) to the initial maximum shear rate value (i.e., from $Wi = 10^{-4}$ to $Wi = 10^{2}$) in a step-wise manner. Figure \ref{fig:overall} shows shear stress as a function of $Wi$ for different values of $\delta t$ and $Wi$ is varying from $10^{-4}$ to $10^{2}$. As mentioned above, shear stress value is obtained at the end of $\delta t$ time at each step. We also show the corresponding shear stress and $Wi$ evolution as a function of time that is normalised with $\delta t$ in Fig.\,\ref{fig:overall_stress_time}.

Cyclic down-up sweep shear flow using Giesekus model at different values of $\delta t$ results in hysteresis loops as down-up sweep stress does not coincide for some values of $\delta t$ as shown in Fig.\,\ref{fig:overall}. There are four types of hysteresis loops that can be obtained using a viscoelastic material of relaxation time of the order of $10^{-1}$ s. All four types of hysteresis loops show decrease in stress during down-sweep (decrease in shear rate). During up-sweep (increase in shear rate), stress first decreases and then increases. This feature is clearly visible in Figs.\,\ref{fig:overall} and \ref{fig:overall_stress_time}. In Fig.\,\ref{1e_6}, hysteresis loop comprises of a closed loop in which there is a negligible change in stress over the cyclic shear rate sweep. This minimal change in stress is also clearly depicted in stress versus time plot in Fig.\,\ref{ss1e_6}. We term these kind of loops as `type 1'. Figures \ref{1e_5}-\ref{1e_2} shows loops that have a relatively higher change in stress in cyclic shear rate sweep. These loops are termed as `type 2'. The loops can be open (Fig.\,\ref{1e_5}-\ref{3e_3}) for lower values of $\delta t$ or closed (Fig.\,\ref{1e_2}) for higher values of $\delta t$. Figures \ref{3e_2}-\ref{3e_1} depict loops that have one crossover of down-up sweep stress because of stress overshoot observed at higher values of $Wi$ in up-sweep. These loops are categorised as `type 3'. Stress overshoot is clearly visible in Figs. \ref{ss3e_2}-\ref{ss3e_1}. Type 3 loops are also can be open (Fig.\,\ref{3e_2}-\ref{7e_2}) for lower values of $\delta t$ or closed (Figs.\ref{1e_1}-\ref{3e_1}) for higher values of $\delta t$. Figure \ref{1} shows a loop in which there is no crossover of down-up sweep stress and the loop is closed. Such loops are labelled as `type 4'. This type of loop gets formed because of difference in down-sweep and up-sweep stress only for lower values of $Wi$. For higher values of $Wi$, down-up stress overlaps each other. For $\delta t=10^2$, down-up sweep stress completely overlap each other resulting in absence of hysteresis loop as shown in Fig.\,\ref{100}.

The down-sweep stress always shows decrease in the type 1-4 hysteresis loops while up-sweep stress always show an initial decrease before increasing again. The decrease in down-sweep stress in type 1-4 hysteresis loop is to attain its steady state value corresponding to $Wi$ in the given $\delta t$ time at each step. If the value of stress at the end of down-sweep flow is higher than its steady state value then up-sweep stress will first show decrease. The increasing part of up-sweep stress is due to early elastic effects in high $Wi$ region driving stress away from the steady state values at each $Wi$. This stress response is governed by the initial value of $d\underset{\approx}{\Sigma}/dt$ which depends on the value of $Wi$ of previous step as also stated in literature [\onlinecite{menezes1982nonlinear,dealy1999transient,tanner2000engineering}]. Type 1, 2 and 4 hysteresis loops can be obtained by linear viscoelasticity, while type 3 hysteresis loop can only be obtained by a non-linear viscoelastic model. For dimensional values of shearing time ($\delta t^*$) less than $10^{-3}$ s, the corresponding hysteresis loops shown in Fig.\,\ref{fig:overall} may get contaminated by inertia in experimental studies such as hysteresis loops shown in Figs. \ref{1e_6}-\ref{1e_2} if $\tau=0.1$ s.

%For example, if $\tau=0.1$ s, then hysteresis loops shown in Figs. \ref{1e_6}-\ref{1e_2} may get affected by inertia.

The observation that hysteresis loops originate because of competition between finite time required for stress to reach respective steady state values during down-sweep and up-sweep shear flow and the time scale of experiment ($\delta t$) is in agreement with the results of Bird and Marsh [\onlinecite{bird1968nonlinear}] and Marsh [\onlinecite{marsh1968viscoelastic}]. The open hysteresis loops predicted at low values of $\delta t$ (Figs. \ref{1e_6}-\ref{3e_3}) without any crossover are also similar to hysteresis loops observed for simple yield stress fluids by Puisto et. al.\, [\onlinecite{puisto2015dynamic}].

\subsection{High relaxation time viscoelastic materials} \label{subsection_hve}

As discussed in Introduction, the motivation of this work is to assess whether a completely unknown material is viscoelastic or thixotropic purely based on its hysteresis behavior. For those viscoelastic materials that take a very long time to relax, it is possible to misconstrue the time dependence associated with viscoelastic relaxation with inherent time dependence associated with thixotropy. Consequently, in this subsection, we consider the case of a viscoelastic material that has relaxation time of the order of $10^6$ s ($\approx11.5$ days) or higher.  We study cyclic shear rate down-up sweep of such materials using the Giesekus model by first varying $Wi$ from $10^{9}$ to $10^{3}$ (down-sweep) in a step-wise manner. Subsequently, we increase $Wi$ from $10^{3}$ to $10^{9}$ (up-sweep) in a step-wise manner. As mentioned above, this range of $Wi$ is also in accordance with the range of shear rate values that are generally attainable in standard rheometer to obtain the flow curve of a material. In this case, we also consider $\bar{\eta_s}$ is $10^{-7}$ to magnify viscoelastic effects that can be observed at high value of $Wi$. The low value of $\bar{\eta_s}$ i.e., a very high value of zero shear viscosity($\eta_s+\eta_p$) is indeed realistic as also shown by Kulicke et al.\, [\onlinecite{kulicke1982preparation}]. Further discussion on the choice of parameters is deferred to the end of this subsection. %

\begin{figure}[htbp]
\centering
     \subfigure[]{
    \includegraphics[scale=0.19]{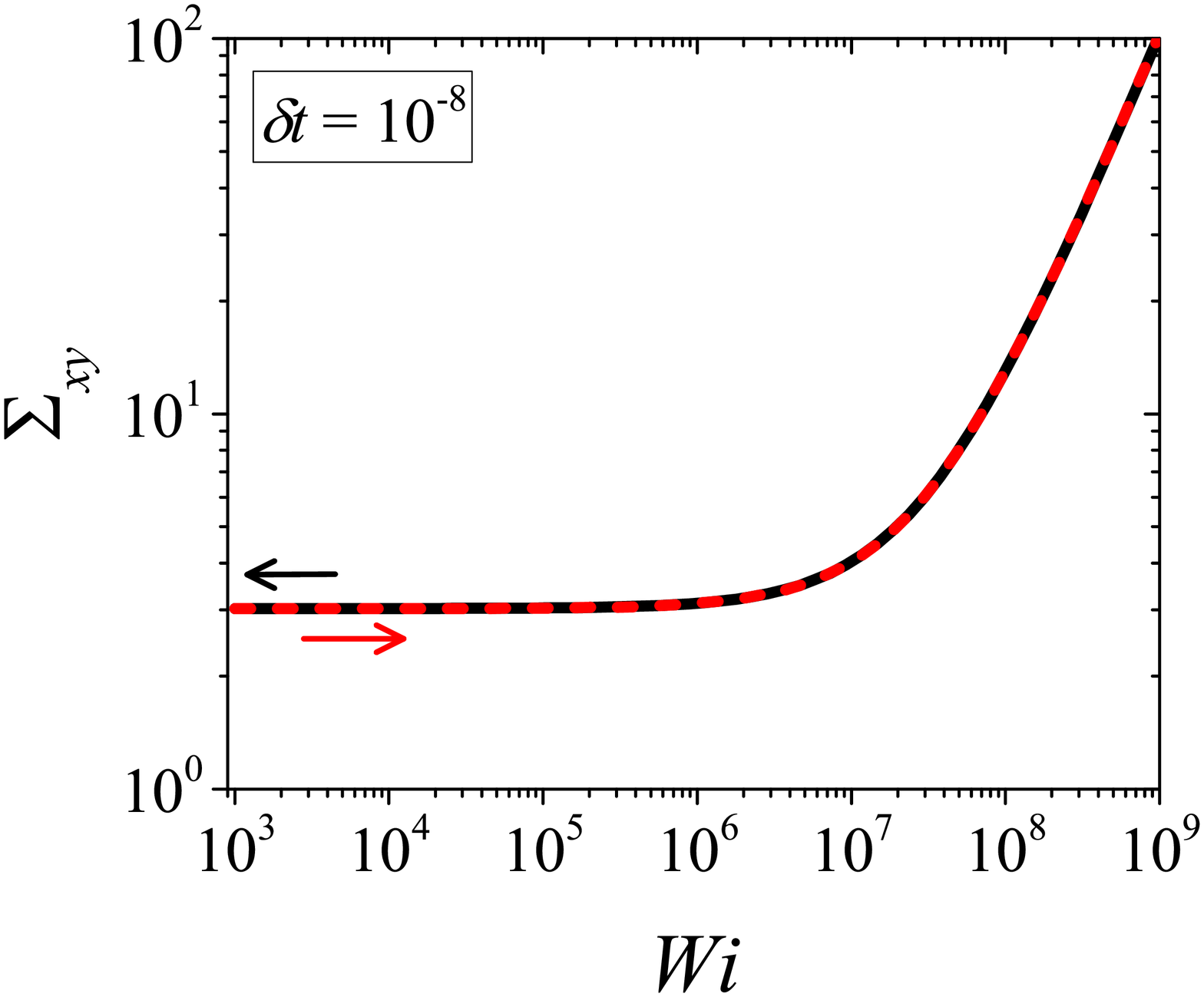}
    \label{hw1e_8}
  }
  \subfigure[]{
    \includegraphics[scale=0.19]{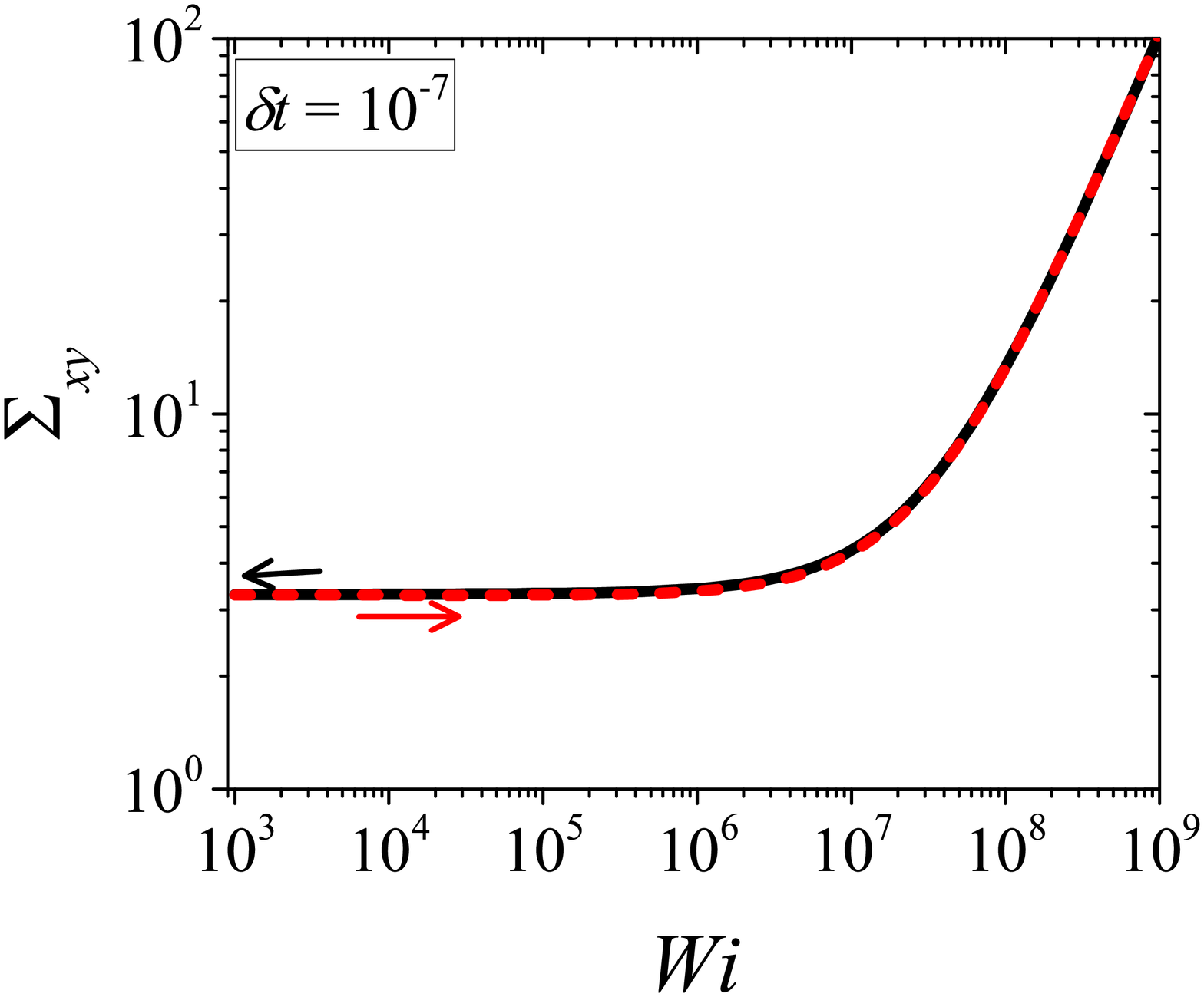}
    \label{hw1e_7}
  }
  \subfigure[]{
    \includegraphics[scale=0.19]{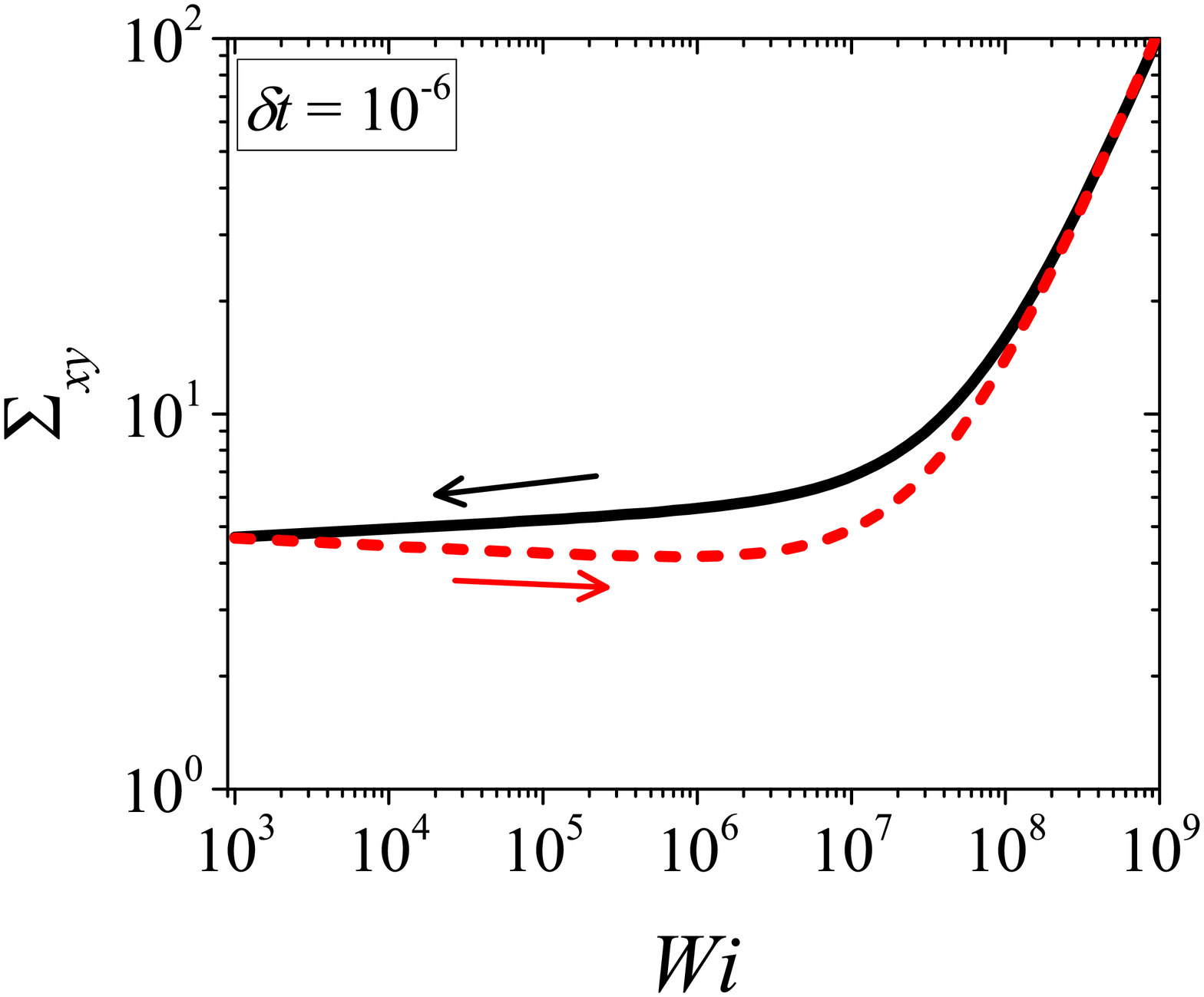}
    \label{hw1e_6}
  }
   \subfigure[]{
\includegraphics[scale=0.19]{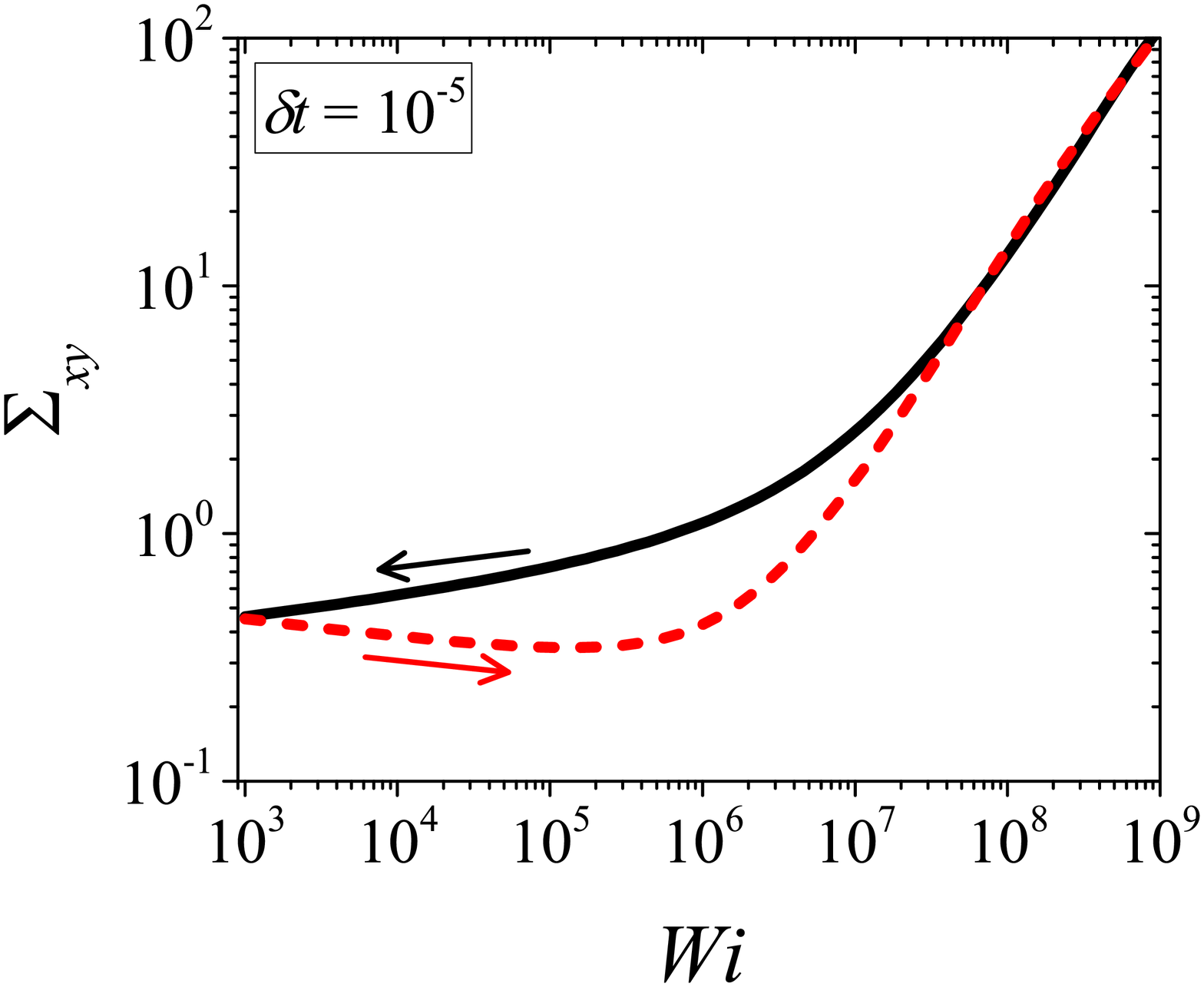}
    \label{hw1e_5}
  }
    \subfigure[]{
    \includegraphics[scale=0.19]{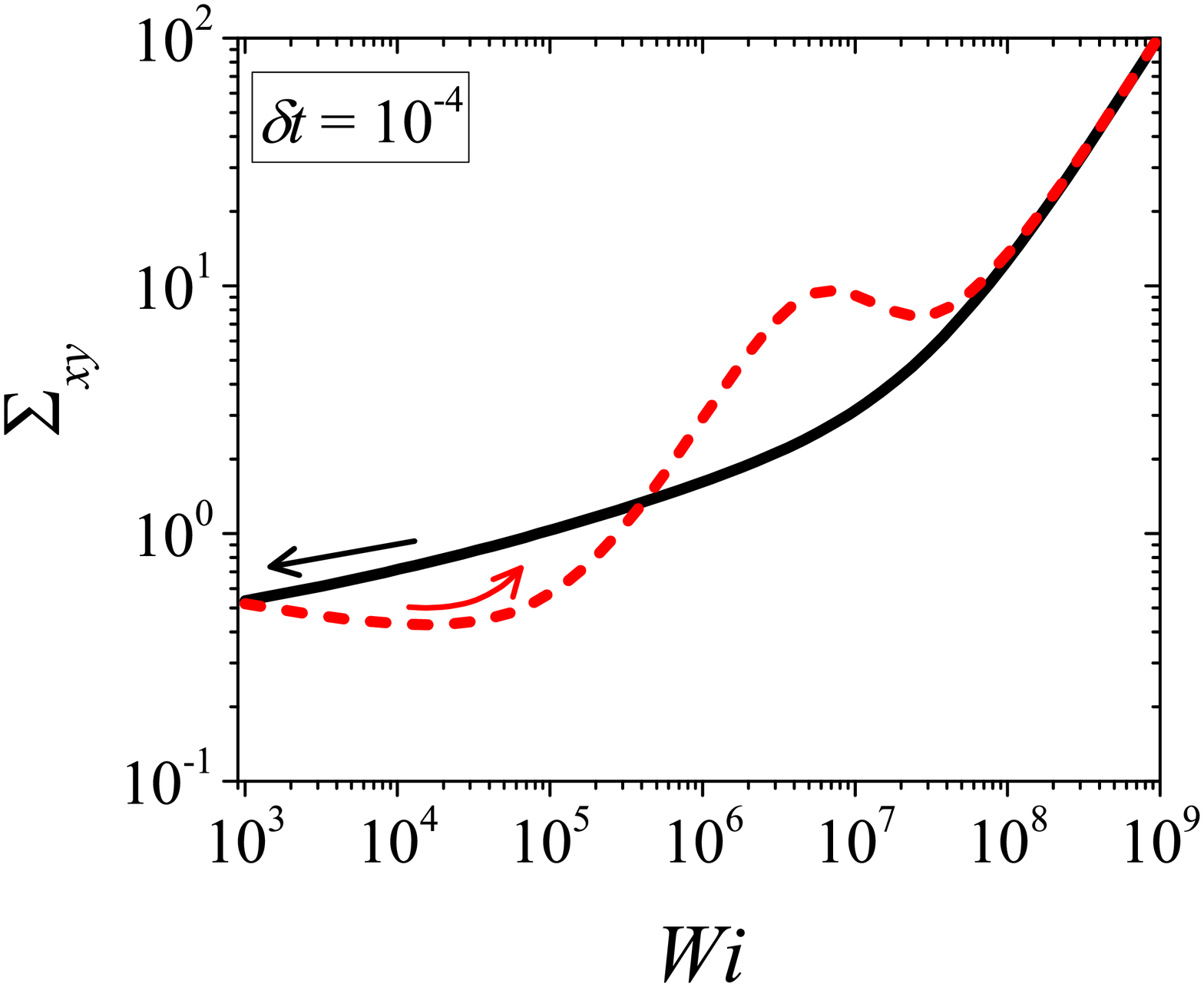}
    \label{hw1e_4}
  }
   \subfigure[]{
\includegraphics[scale=0.19]{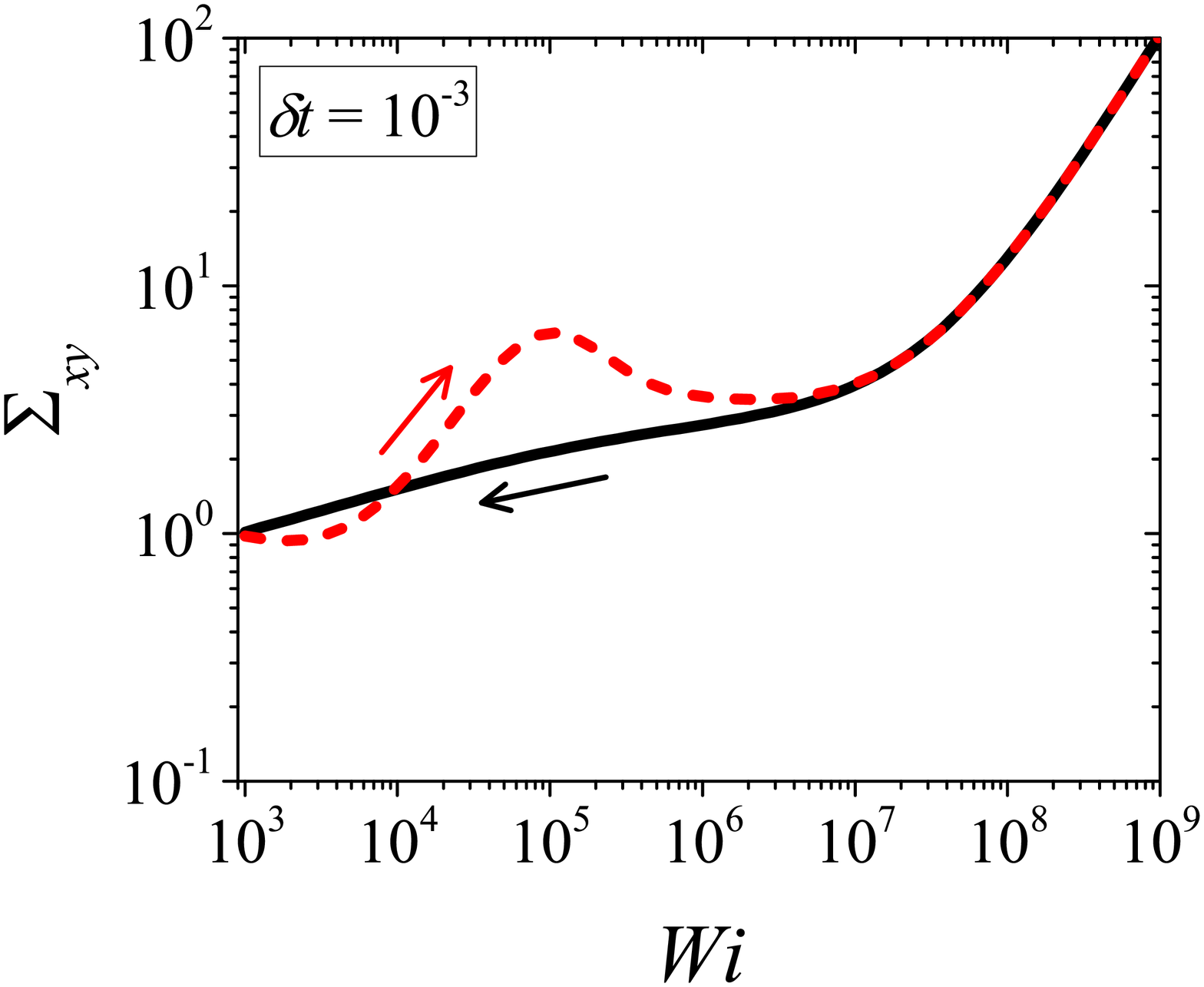}
    \label{hw1e_3}
  }
       \subfigure[]{
\includegraphics[scale=0.19]{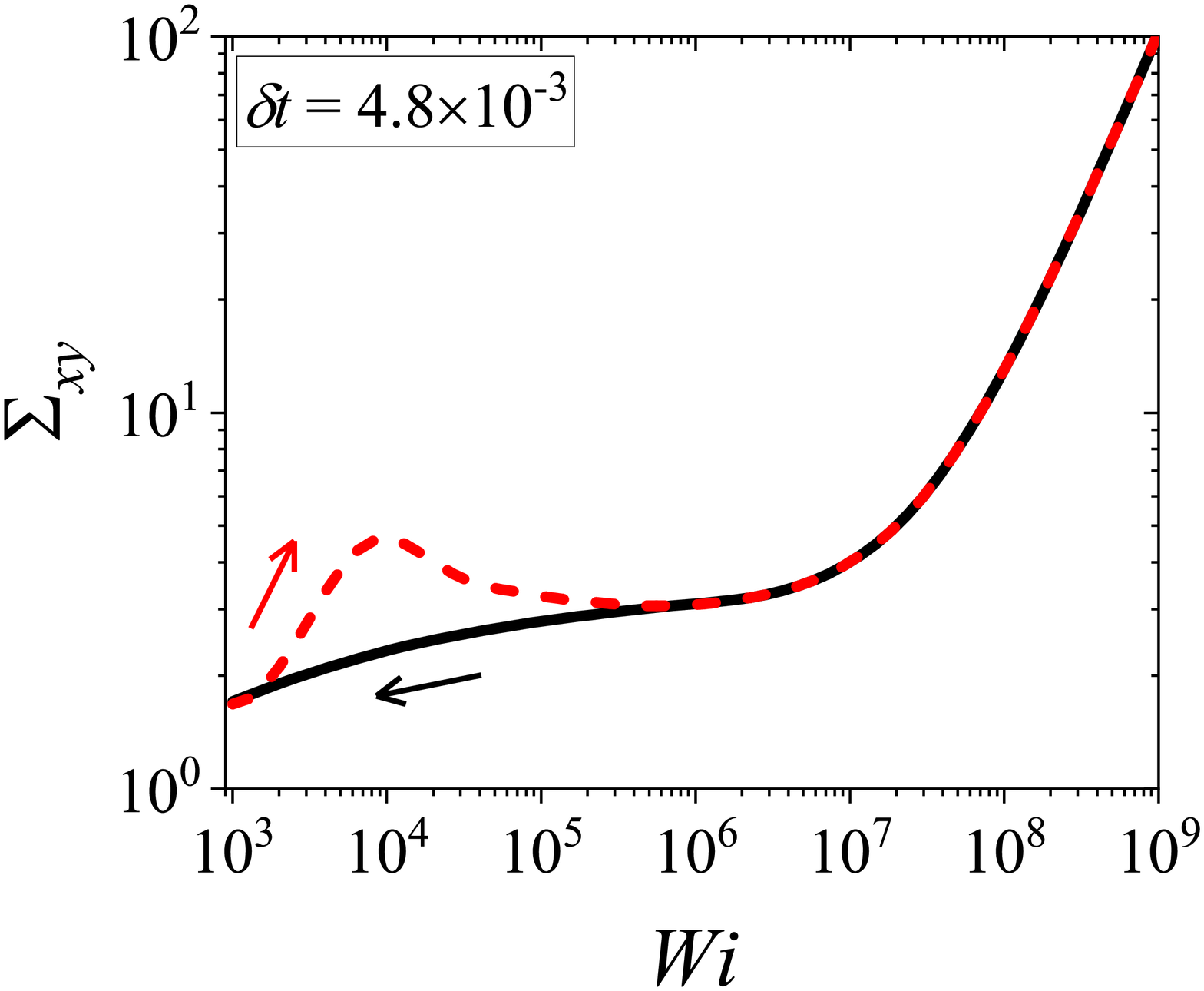}
    \label{hw45e_3}
  }
         \subfigure[]{
\includegraphics[scale=0.19]{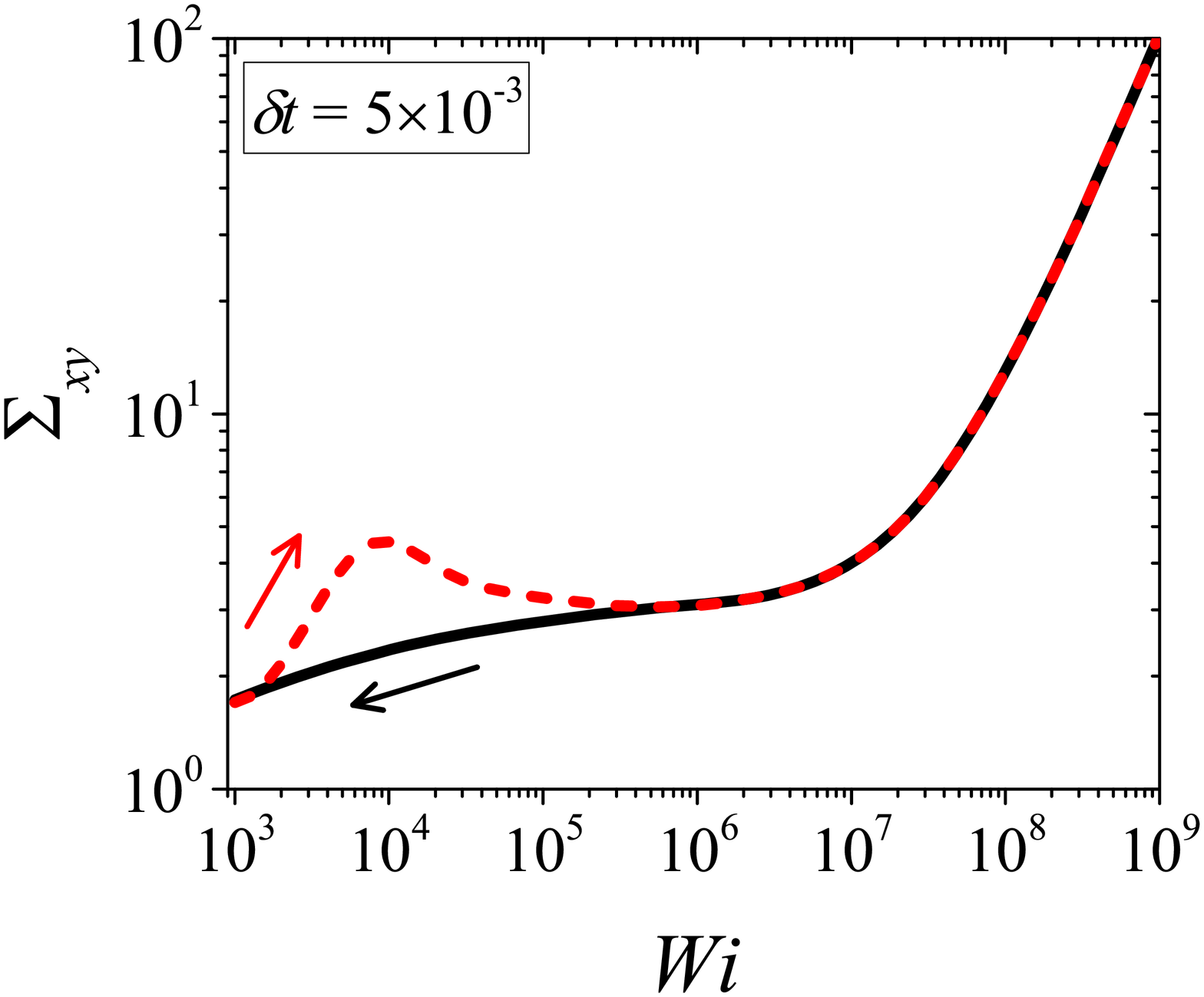}
    \label{hw5e_3}
  }
       \subfigure[]{
\includegraphics[scale=0.19]{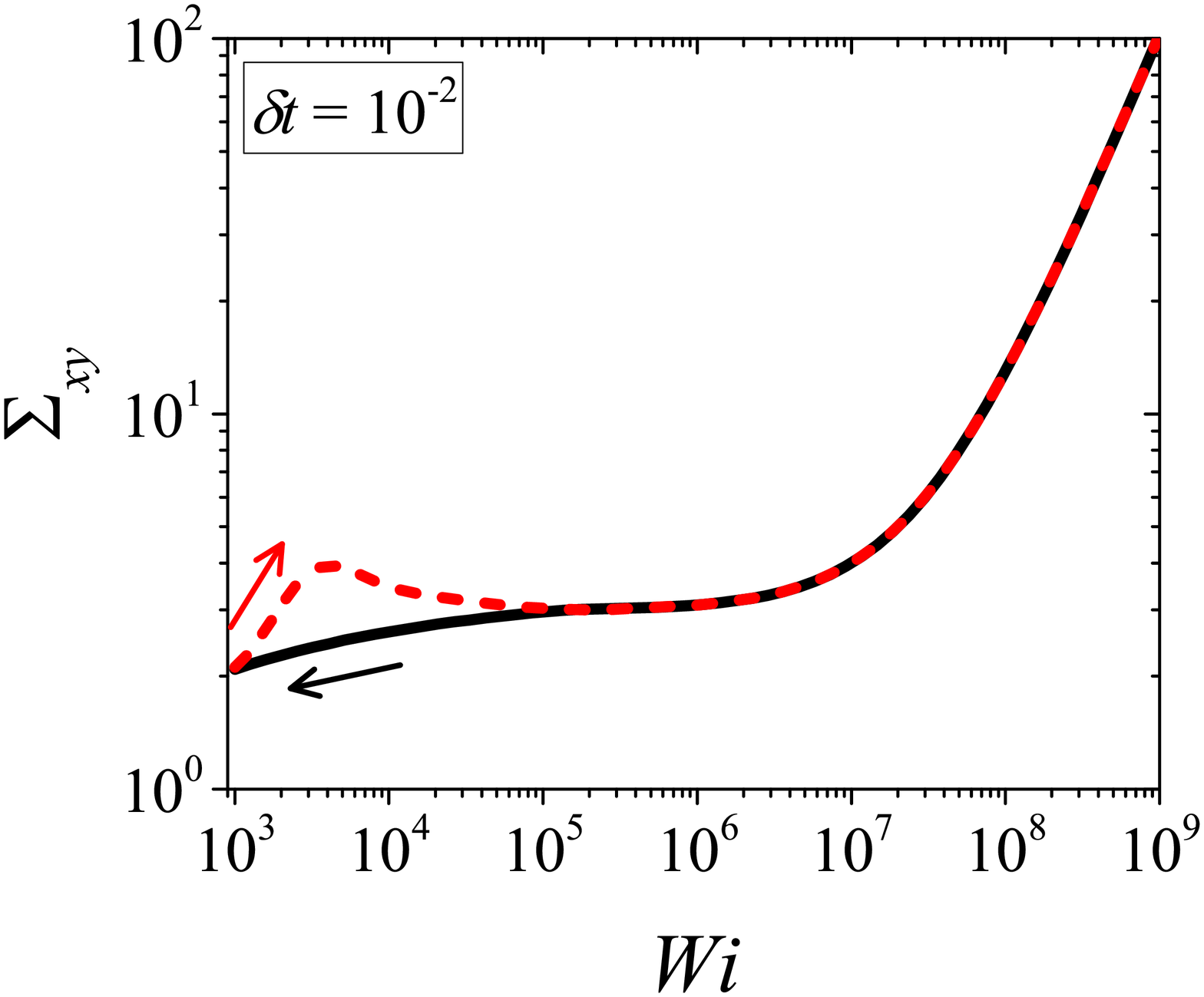}
    \label{hw1e_2}
  }
       \subfigure[]{
\includegraphics[scale=0.19]{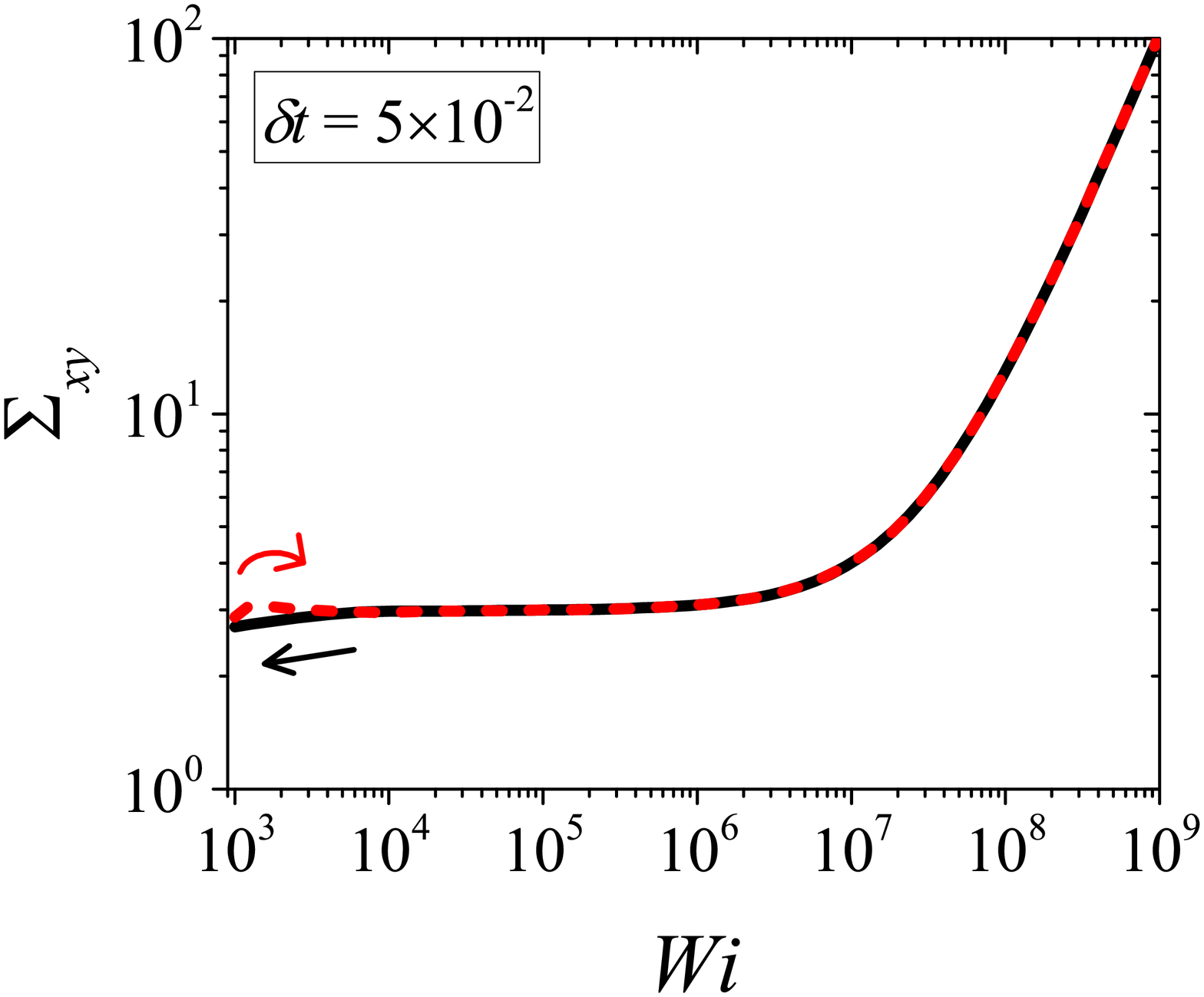}
    \label{hw5e_2}
  }
     \subfigure[]{
\includegraphics[scale=0.19]{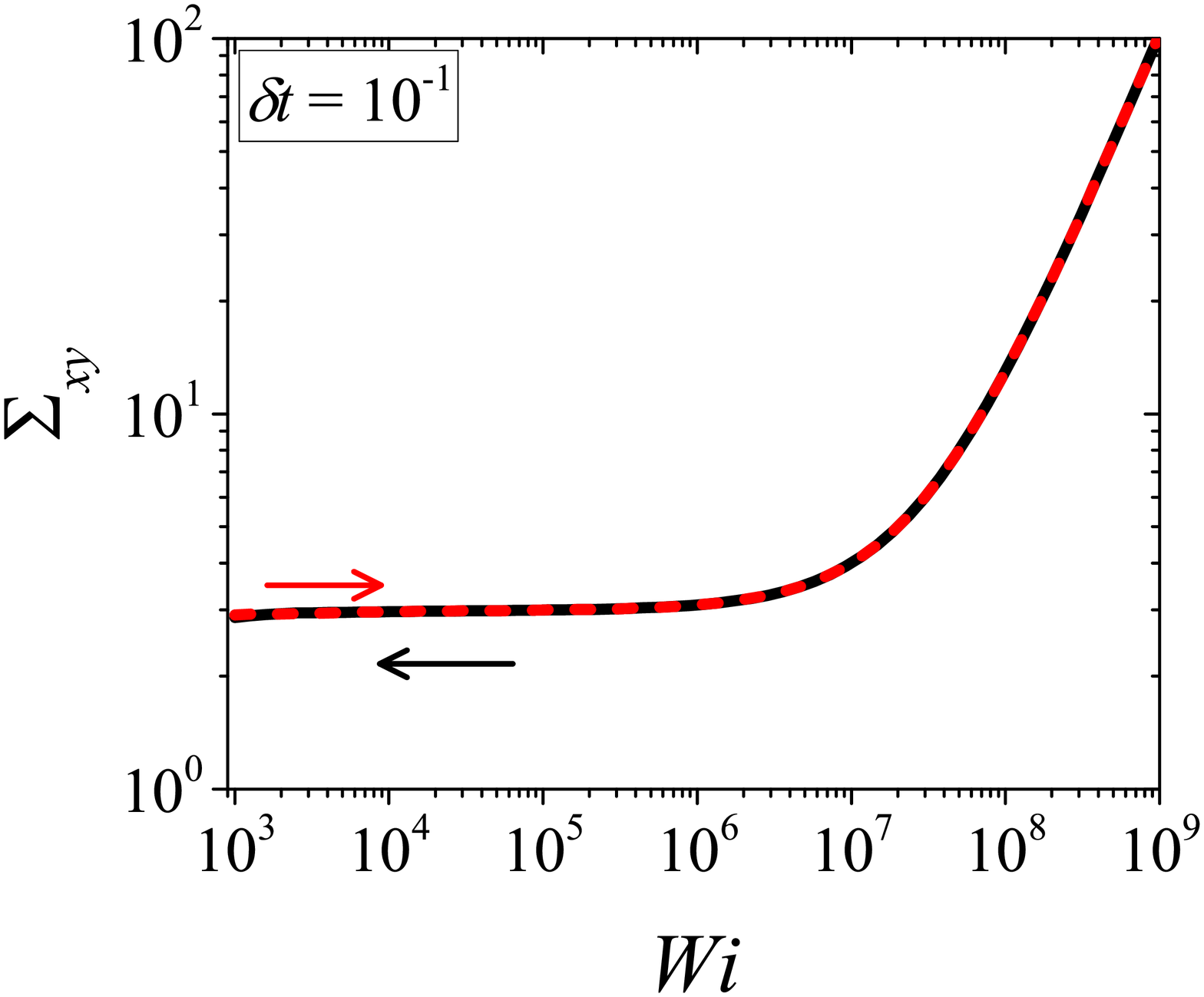}
    \label{hw1e_1}
  }
     \subfigure[]{
\includegraphics[scale=0.19]{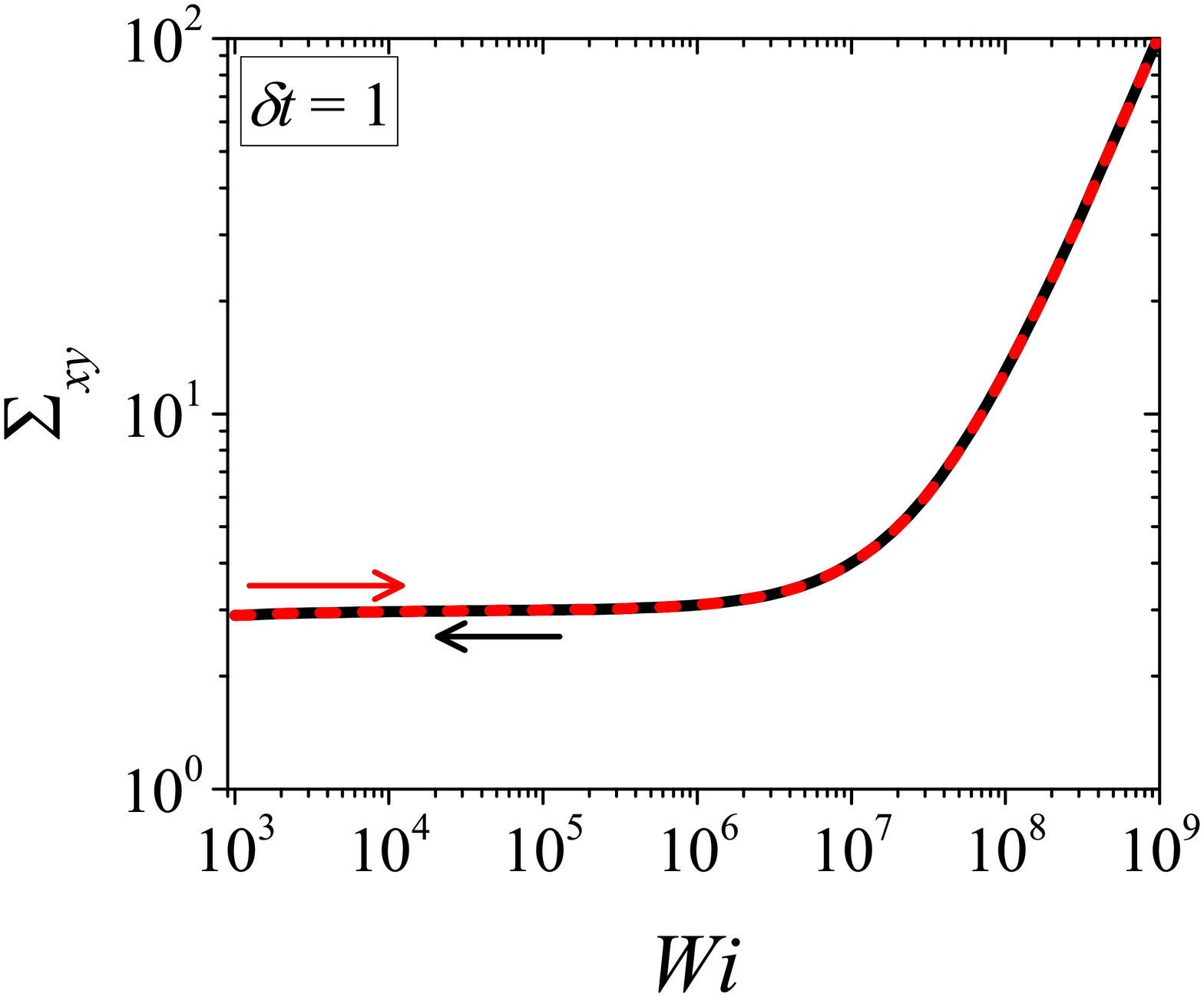}
    \label{hw1}
  }
\caption{\scriptsize Types of hysteresis loops (in a down-up shear rate cycle) obtained for the high relaxation time$\tau=10^6$ s viscoelastic material using different values of $\delta t$ and $\bar{\eta_s}=10^{-7}$ with $n=10$. Shear stress is plotted as a function of shear rate $(Wi)$ for $\delta t$ (a) $10^{-8}$, (b) $10^{-7}$, (c) $10^{-6}$, (d) $10^{-5}$, (e) $10^{-4}$, (f) $10^{-3}$, (g) $4.8\times10^{-3}$, (h) $5\times10^{-3}$, (i) $10^{-2}$, (j) $5\times10^{-2}$, (k) $10^{-1}$, and (l) $1$. Solid lines shows down-sweep and dashed line shows the up-sweep shear flow results. (All the variables in this figure are dimensionless as mentioned in section \ref{section_model}.)}
\label{fig:hw_overall}
\end{figure}

\begin{figure}[htbp]
\centering
     \subfigure[]{
\includegraphics[scale=0.19]{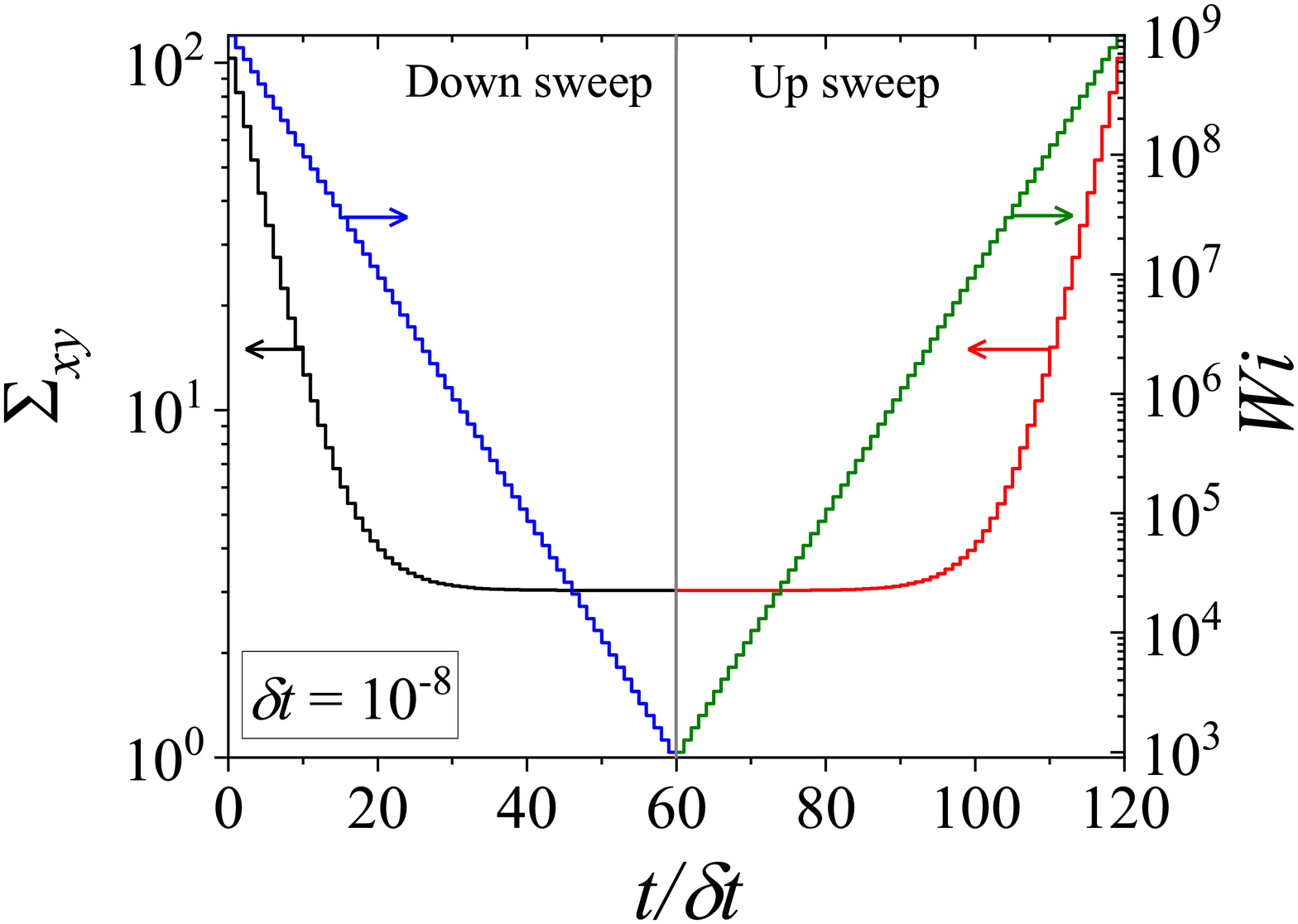}
    \label{hwss1e_8}
  }
  \subfigure[]{
    \includegraphics[scale=0.19]{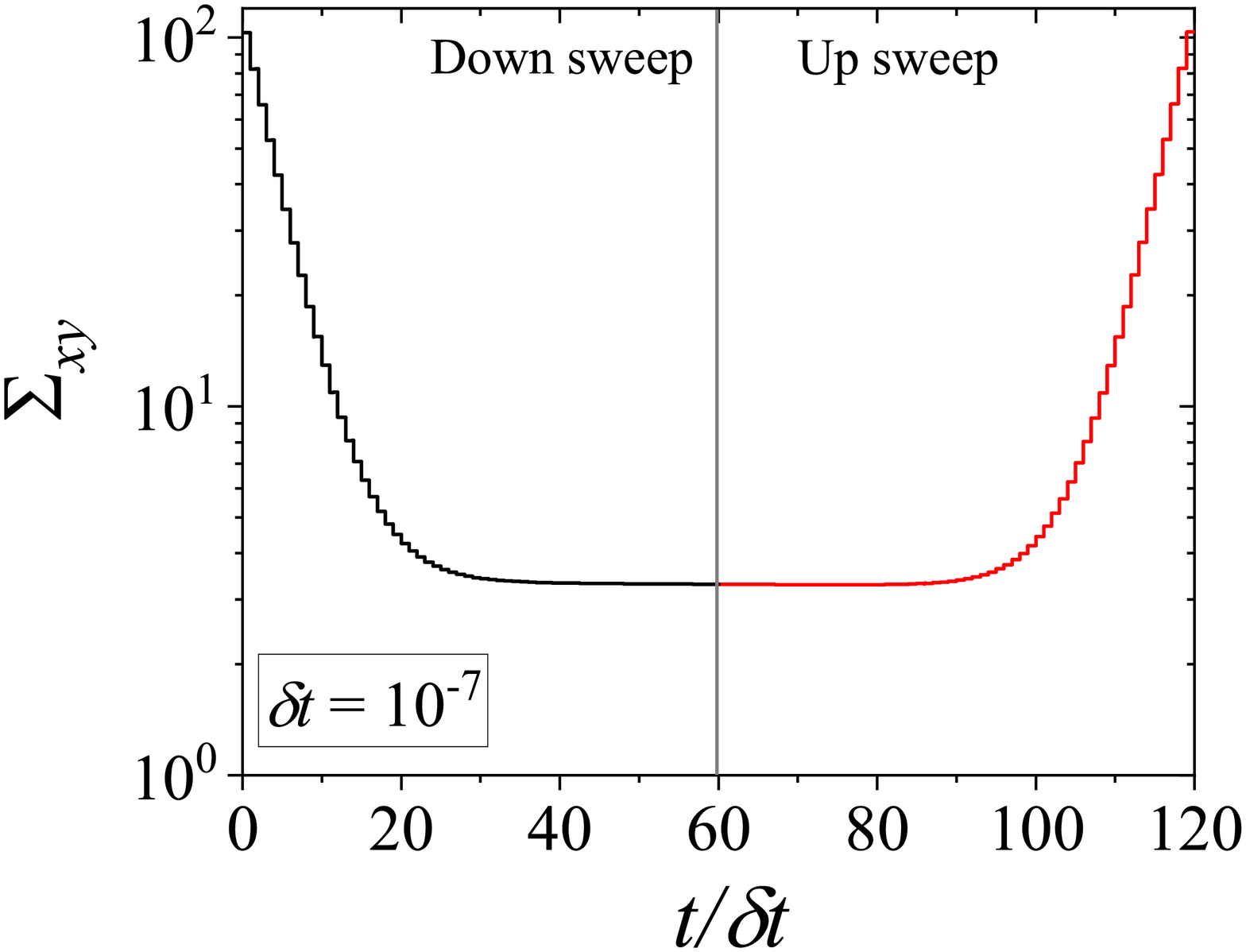}
    \label{hwss1e_7}
  }
  \subfigure[]{
    \includegraphics[scale=0.19]{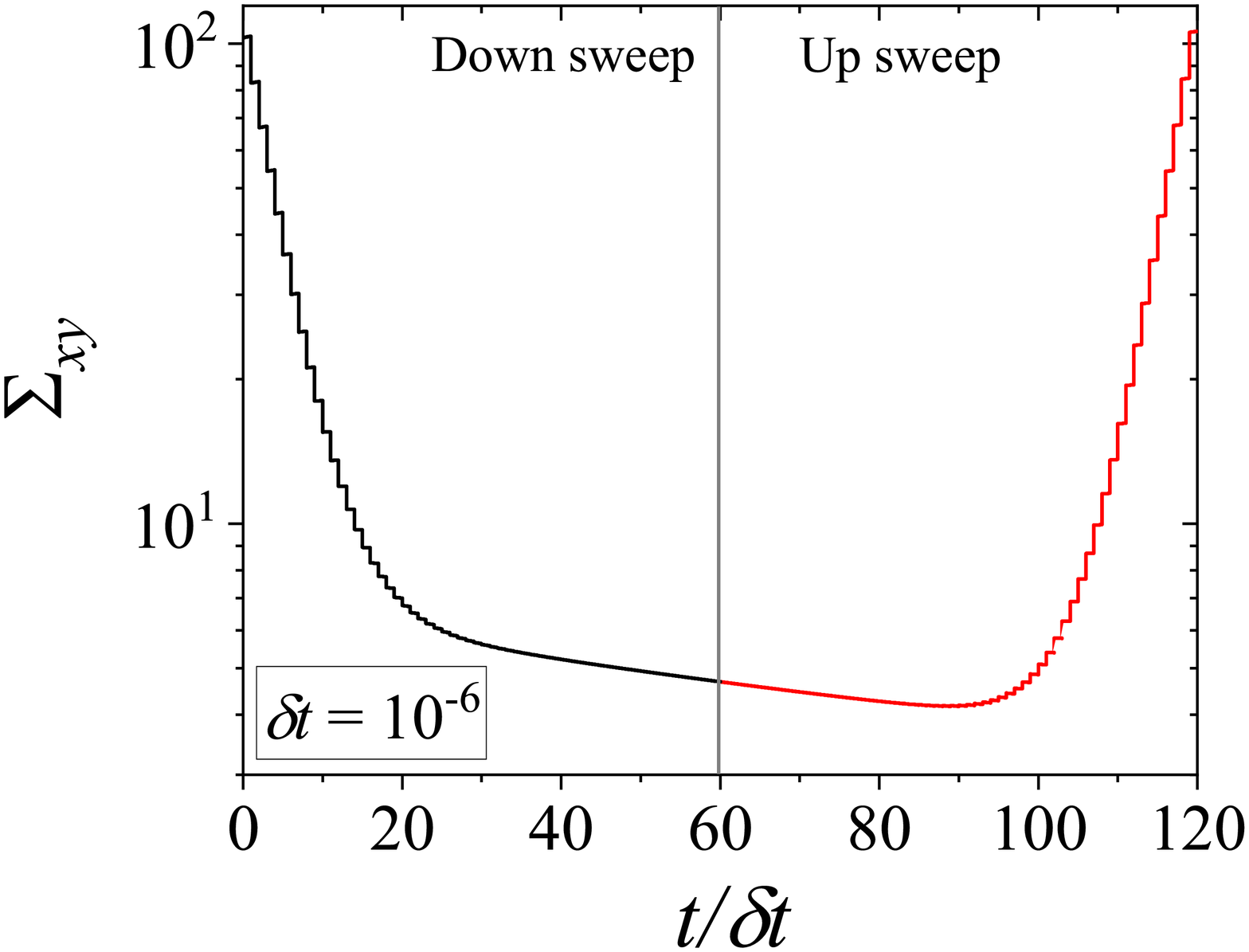}
    \label{hwss1e_6}
  }
   \subfigure[]{
\includegraphics[scale=0.19]{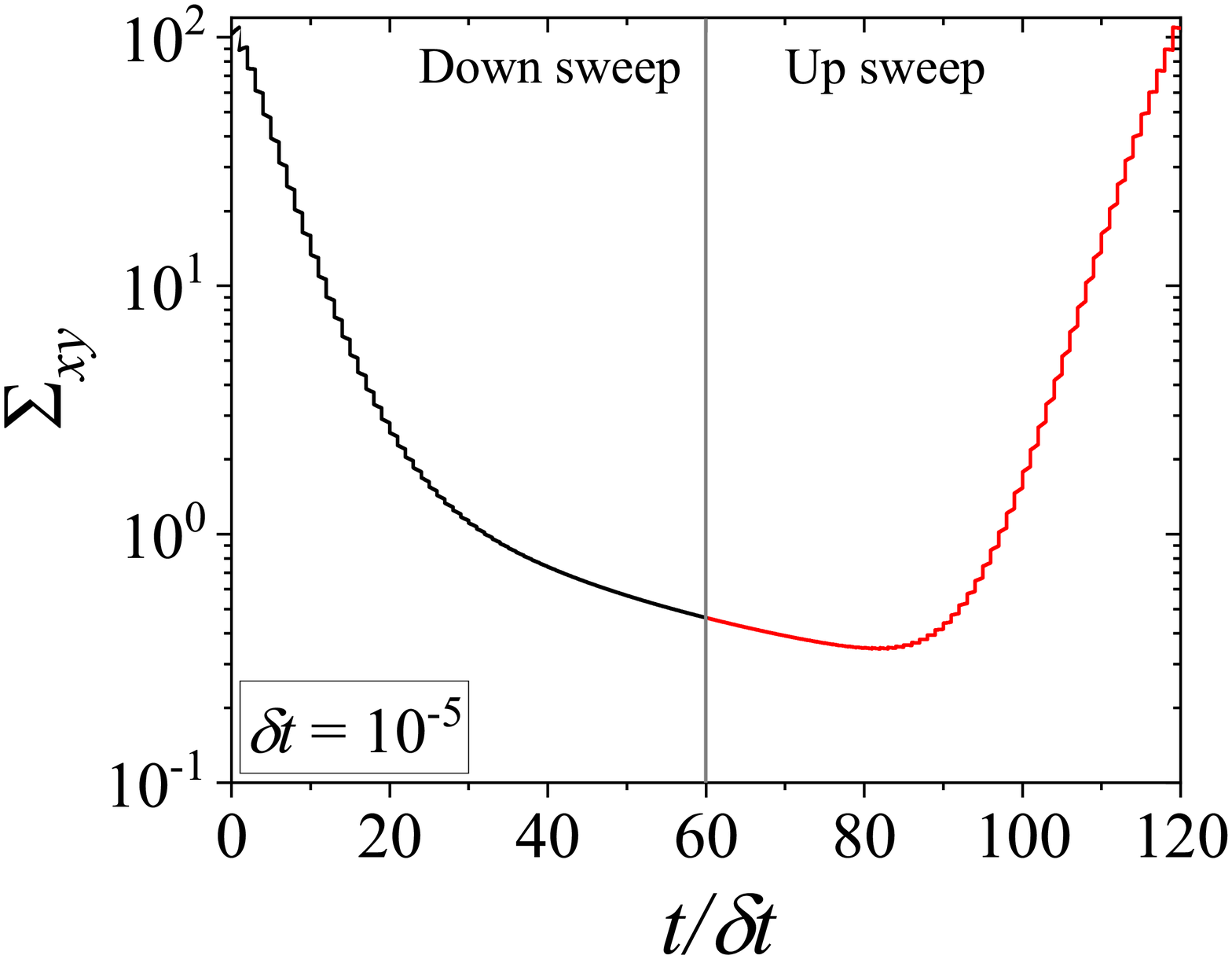}
    \label{hwss1e_5}
  }
    \subfigure[]{
    \includegraphics[scale=0.19]{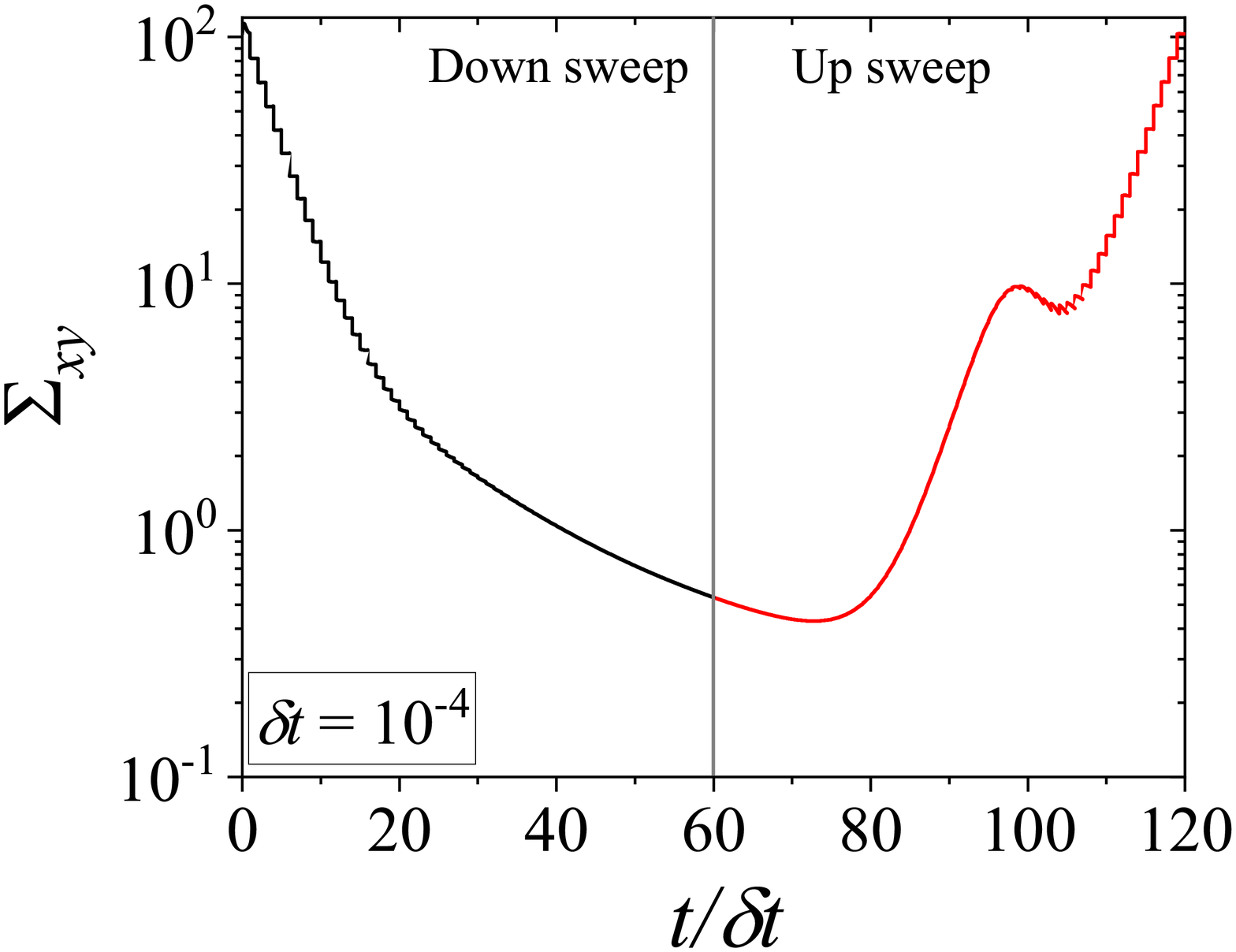}
    \label{hwss1e_4}
  }
   \subfigure[]{
\includegraphics[scale=0.19]{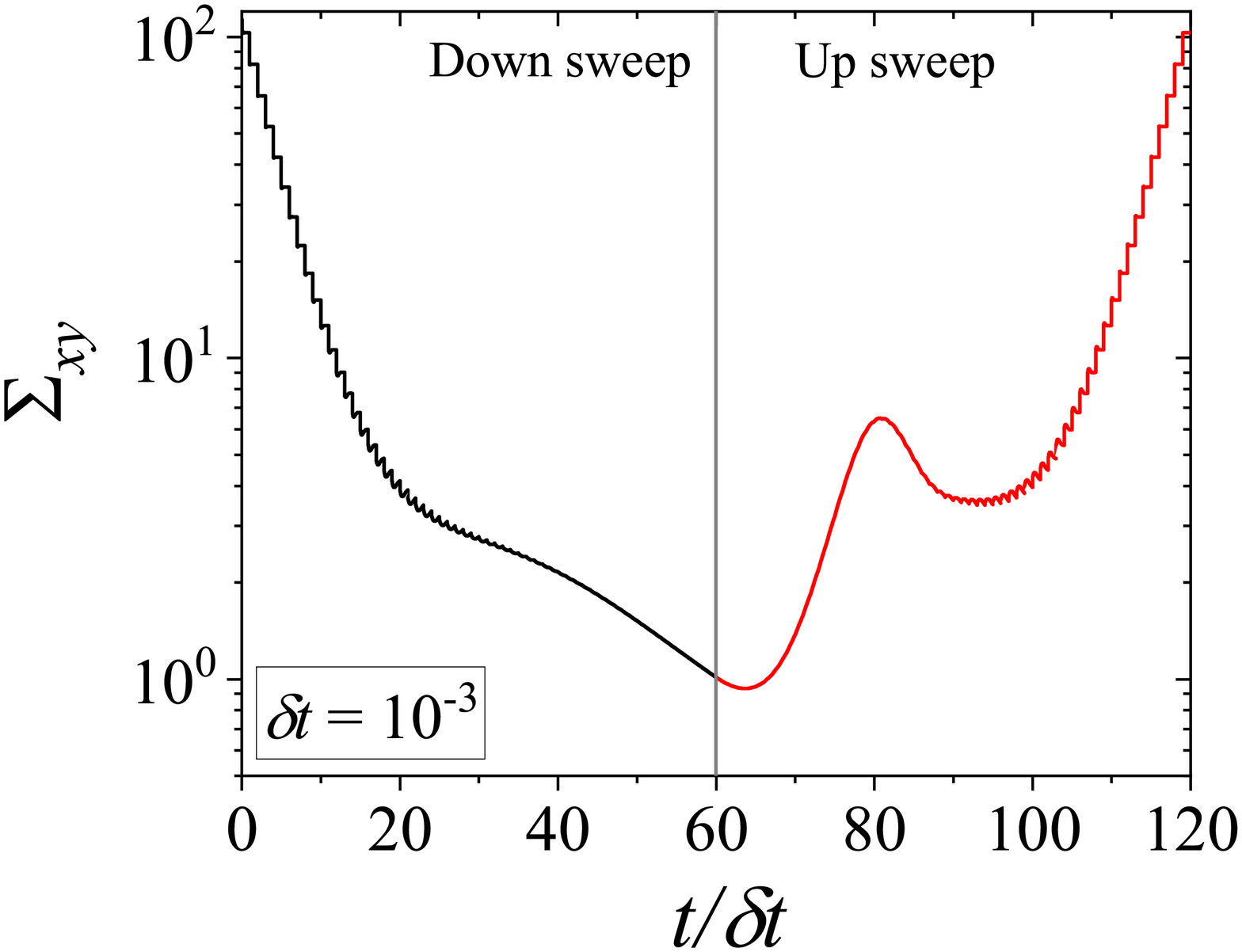}
    \label{hwss1e_3}
  }
       \subfigure[]{
\includegraphics[scale=0.19]{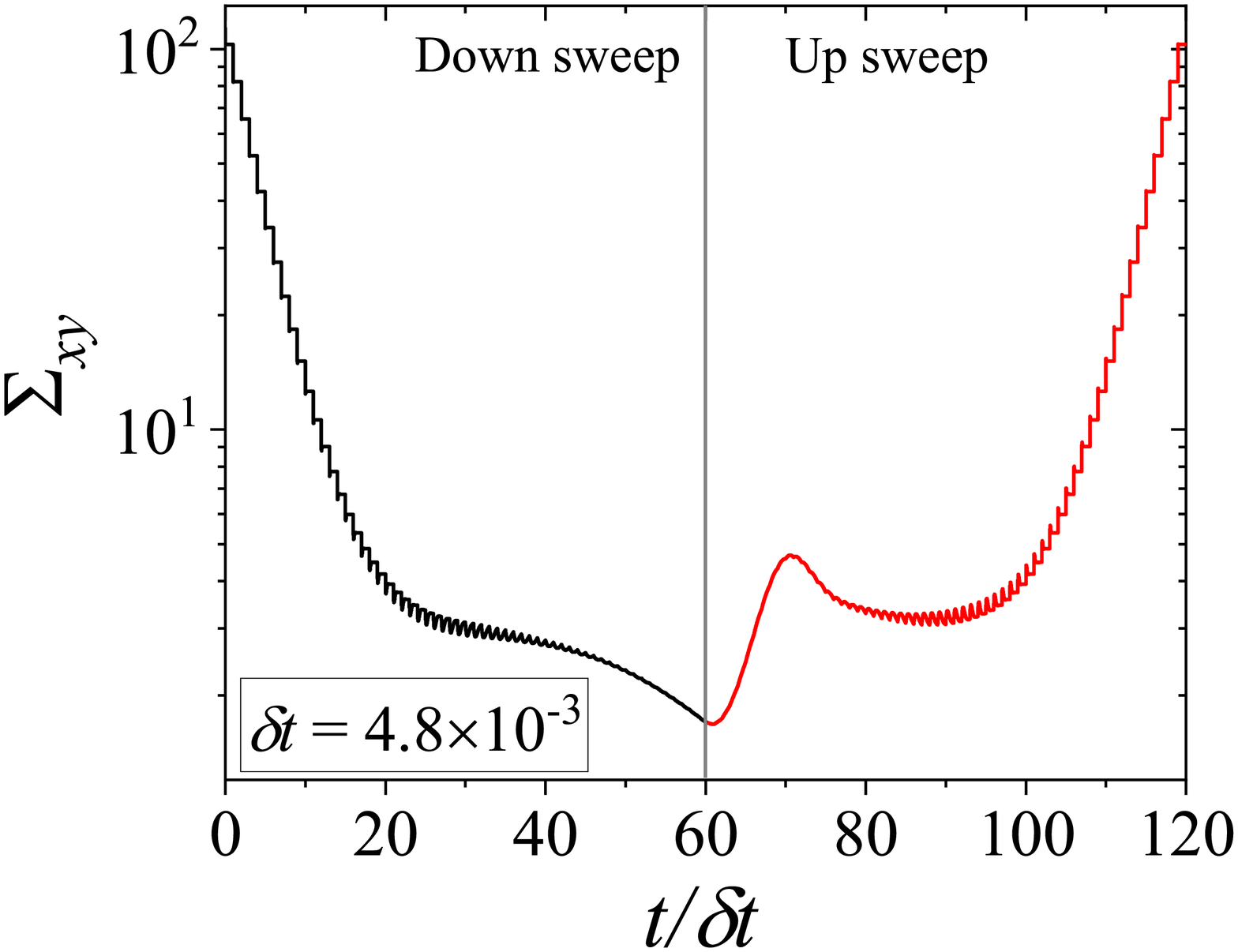}
    \label{hwss48e_3}
  }
     \subfigure[]{
\includegraphics[scale=0.19]{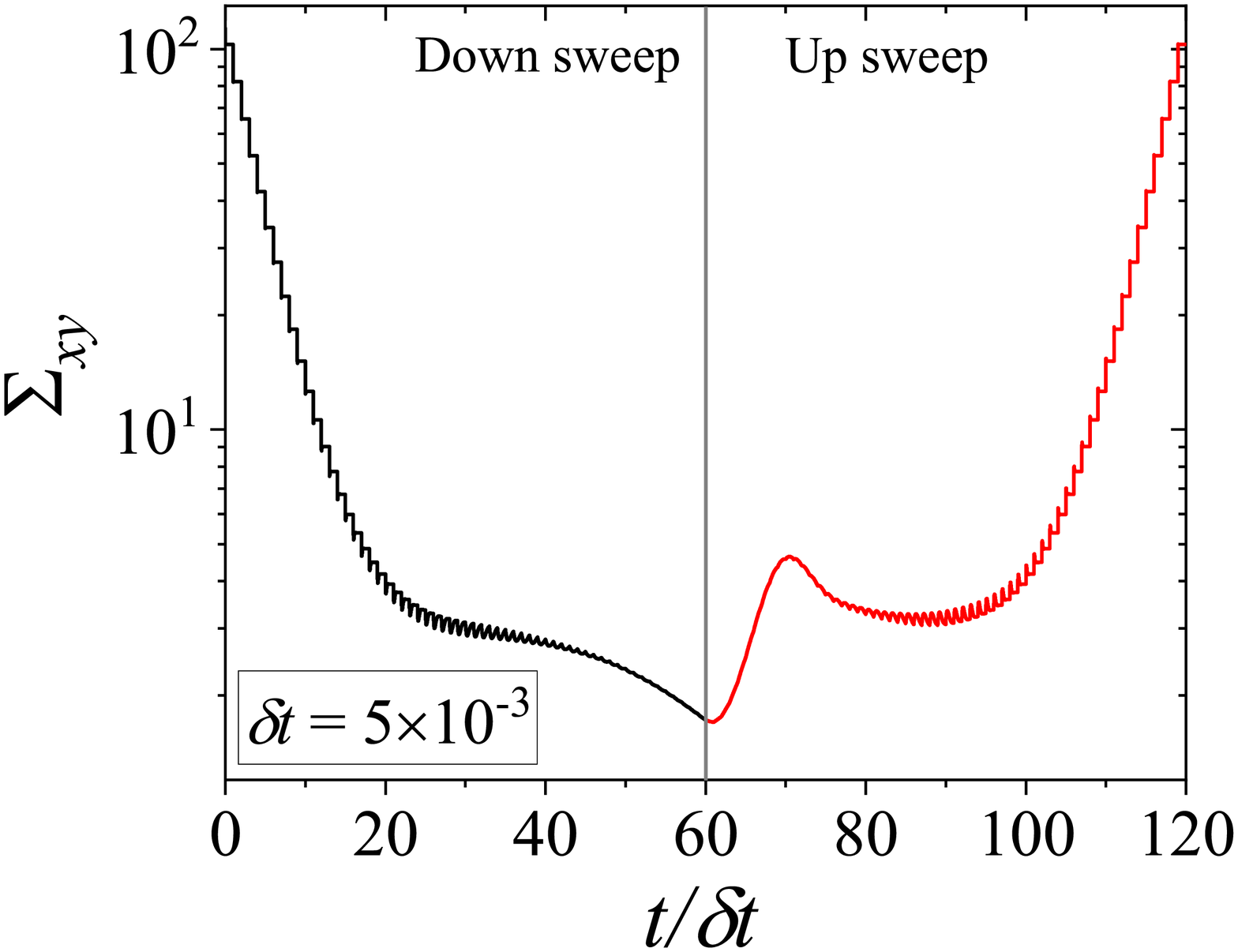}
    \label{hwss5e_3}
  }
     \subfigure[]{
\includegraphics[scale=0.19]{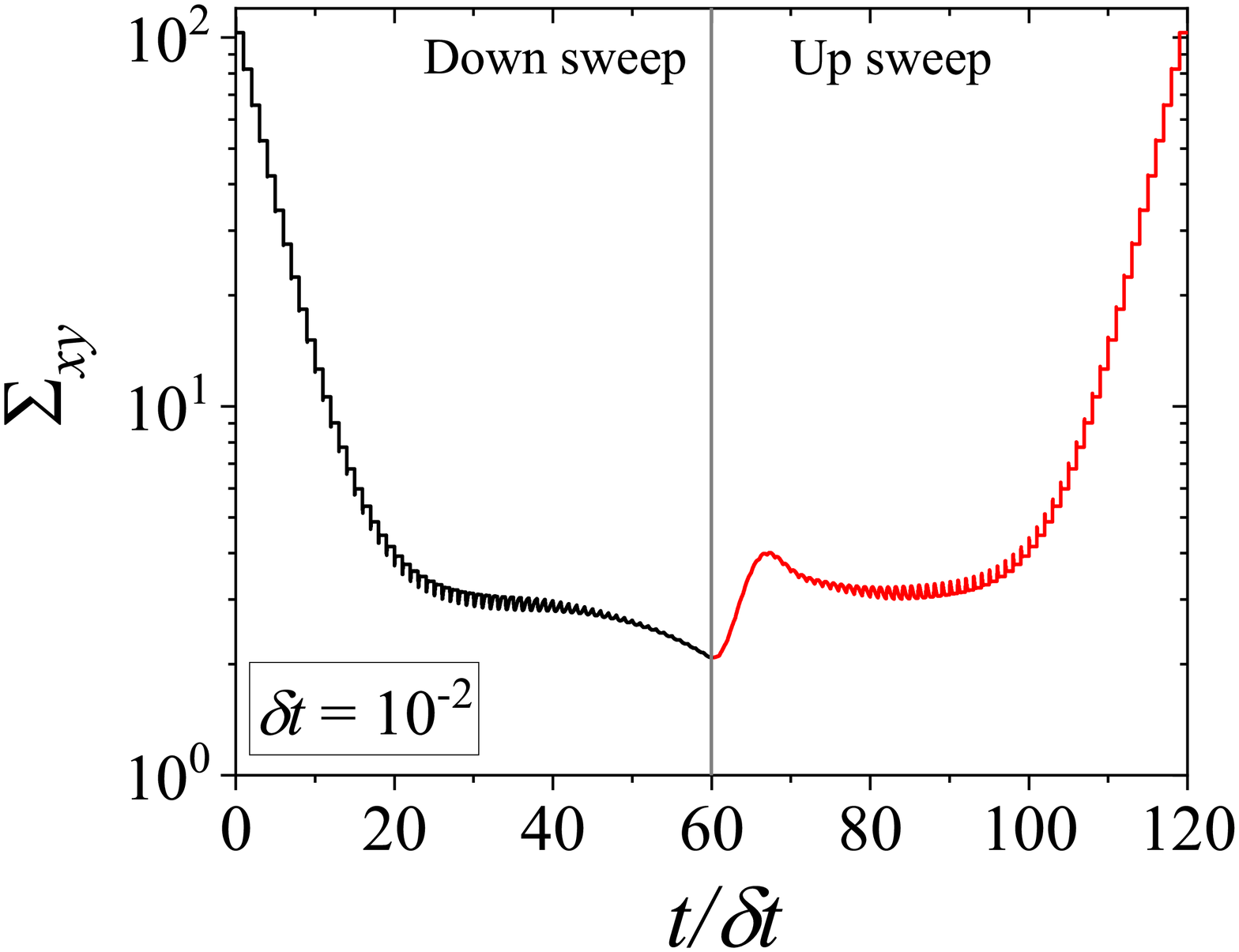}
    \label{hwss1e_2}
  }
     \subfigure[]{
\includegraphics[scale=0.19]{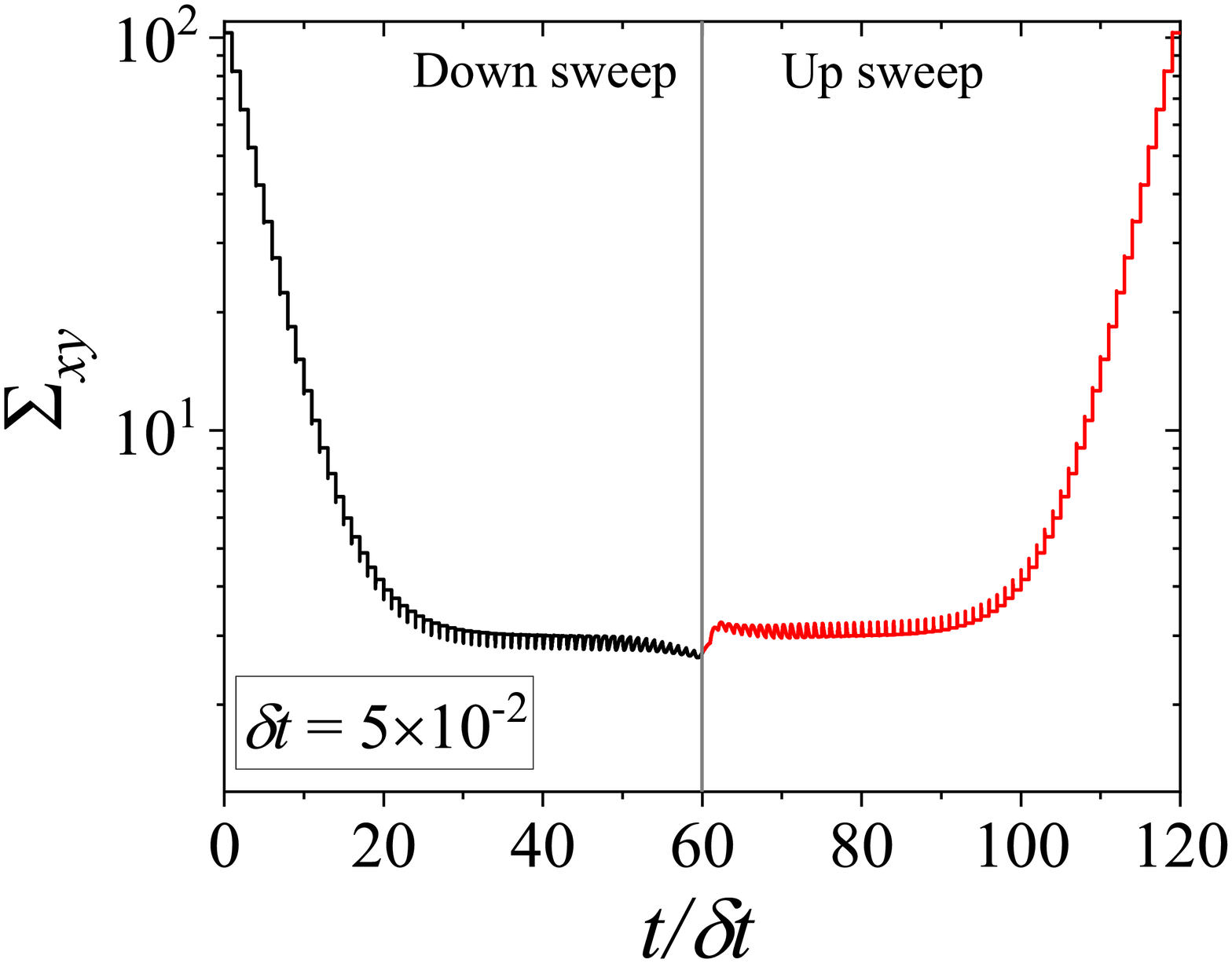}
    \label{hwss5e_2}
  }
     \subfigure[]{
\includegraphics[scale=0.19]{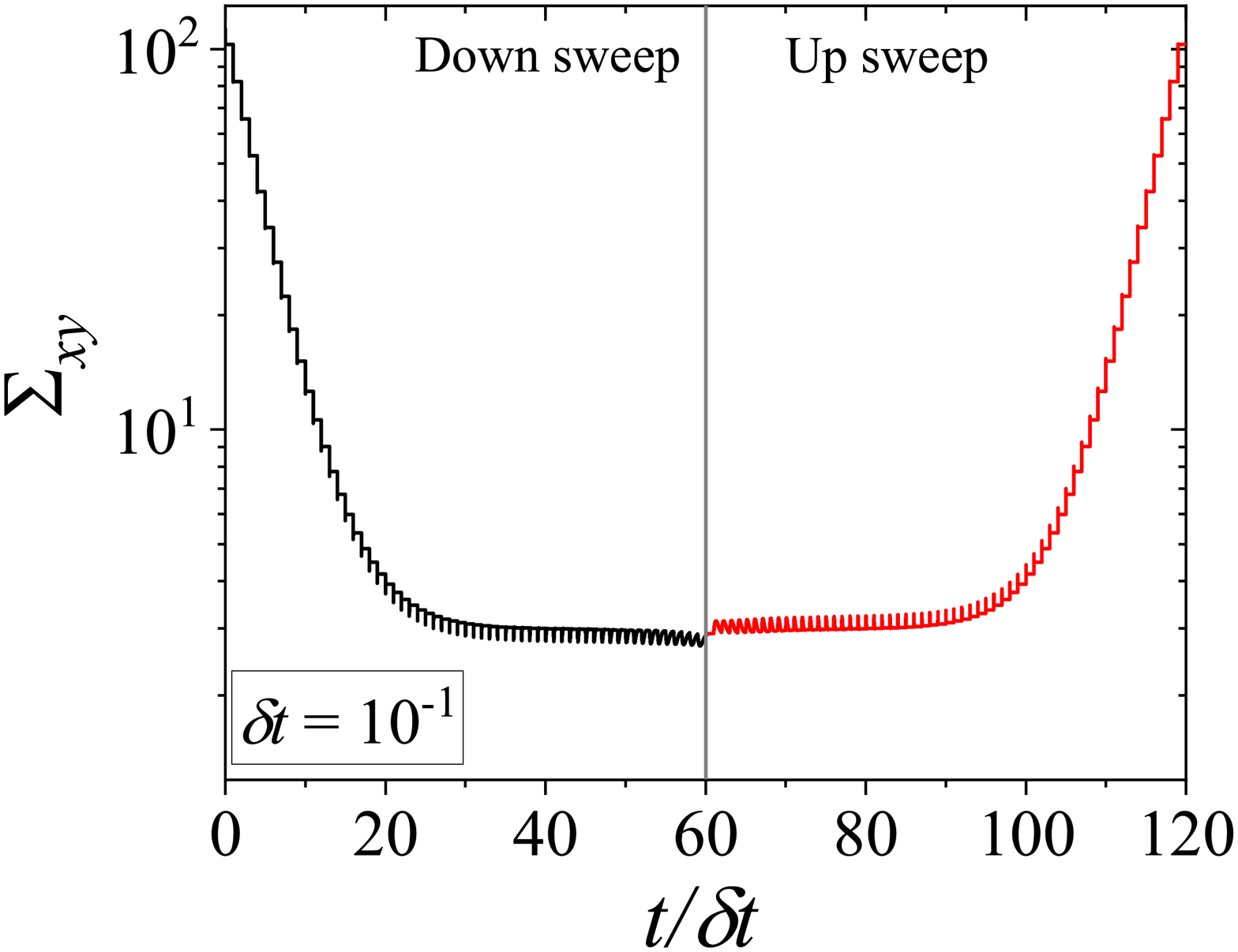}
    \label{hwss1e_1}
  }
      \subfigure[]{
\includegraphics[scale=0.19]{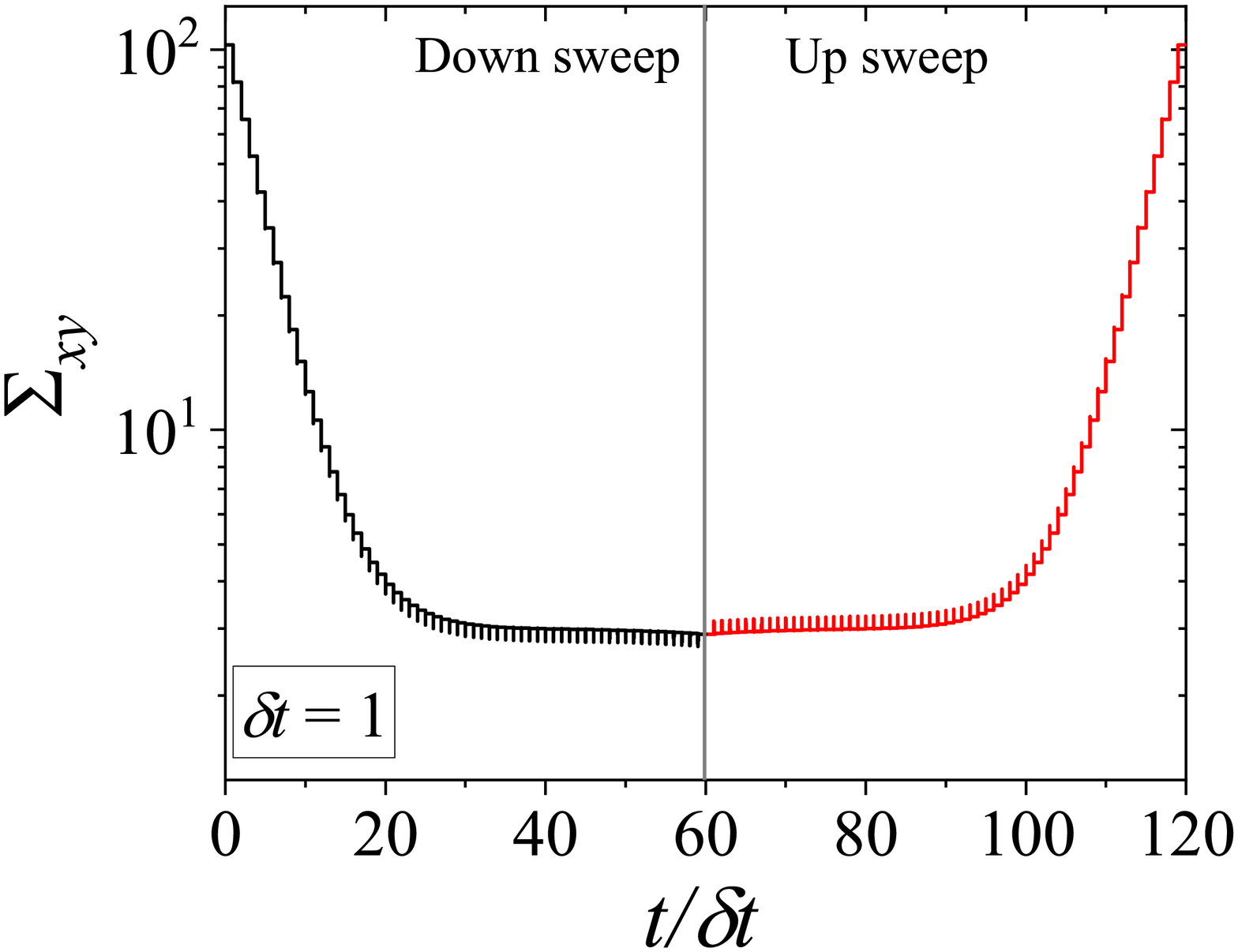}
    \label{hwss1}
  }
\caption{\scriptsize Shear stress (left y-axis) is plotted as a function of time normalised by $\delta t$ for a high relaxation time material ($\tau=10^6$ s) with $n=10$. The different values of $\delta t$ in (a) $10^{-8}$, (b) $10^{-7}$, (c) $10^{-6}$, (d) $10^{-5}$, (e) $10^{-4}$, (f) $10^{-3}$, (g) $4.8\times10^{-3}$, (h) $5\times10^{-3}$, (i) $5\times10^{-2}$, (j) $5\times10^{-2}$, (k) $10^{-1}$, and (l) $1$.This result is corresponding to hysteresis loops presented in Fig.\,\ref{fig:hw_overall}.  (All the variables in this figure are dimensionless as mentioned in section \ref{section_model}.) The gray line at $t/\delta t=60$ is to demarcate the down sweep and up sweep shear stress evolution during shear rate sweep flow. For clarity, variation of $Wi$ is shown only in Fig.\,(a).}  
\label{fig:hw_overall_stress_time}
\end{figure}

Figure \ref{fig:hw_overall} shows results for down-sweep and up-sweep shear flow using Giesekus model for different values of $\delta t$. We also show the corresponding shear stress and $Wi$ evolution as a function of  $t/\delta t$ in Fig.\,\ref{fig:hw_overall_stress_time}. In this case, three types of hysteresis loops can be obtained using a high relaxation time viscoelastic material. In all these hysteresis loops, down-sweep (decrease in $Wi$) stress always shows decrease while the behaviour of up-sweep (increase in $Wi$) stress is different for each type of loop. Figures \ref{hw1e_8}-\ref{hw1e_6} show hysteresis loops in which down-sweep stress is always above the up-sweep stress. These hysteresis loops can be termed as `type 5', which is similar to type 2 closed hysteresis loop of low relaxation time viscoelastic materials. The up-sweep stress in type 5 hysteresis loop shows a decrease and then increase for the same reason as noted in the case of low relaxation time viscoelastic material. Hysteresis loops shown in Figs. \ref{hw1e_5}-\ref{hw1e_3} are labelled as `type 6'. These loops show a crossover of down-sweep and up-sweep stress. The up-sweep stress first shows decrease and then increase that is also accompanied with an overshoot. Hysteresis loops depicted in in Figs. \ref{hw45e_3}-\ref{hw1} are termed as `type 7'. In these loops, up-sweep stress is always above the down-sweep stress.

Interestingly, stress at the end of down-sweep first decreases (with increase in $\delta t$) and then increases to finally attain its steady state for $\delta t=1$ (for type 5-7 hysteresis loops shown in Fig.\,\ref{fig:hw_overall}). This is in contrast to a gradual decrease of stress at the end of down-sweep to attain its steady state value for low relaxation time viscoelastic materials (Fig.\,\ref{fig:overall}). This is due to the stress undershoots in the shear thinning region (least steeper region of constitutive curve). These stress undershoots are clearly visible in Fig.\,\ref{fig:hw_overall_stress_time}. Similarly, stress overshoot in the shear thinning region in the up-sweep leads to type 7 hysteresis loops, wherein only stress overshoot causes difference in down and up sweep stresses. The stress overshoot in the up-sweep shear is also visible in Fig.\,\ref{fig:hw_overall_stress_time}. We also observe that the value of stress remains the same (at steady state) for high shear rate branch in all the hysteresis loops. This is due to the fact that in this region viscosity remains constant with shear rate, and hence no undershoot and overshoot in stress is present (as also shown in Fig.\,\ref{fig:hw_overall_stress_time}). Also, at such high shear rates, the initial rate of change of stress is significantly high. Therefore, even if $\delta t=10^{-8}$, the stress immediately reaches steady state at each step during down-sweep and up-sweep. This effect always leads to a closed hysteresis loop as compared to open hysteresis loop obtained in case of low relaxation time viscoelastic materials shown in Fig.\,\ref{fig:overall}. The closed hysteresis loops show resemblance to loops generally observed experimentally for thixotropic materials. The hysteresis loops observed for high relaxation time materials also show a particular kind of loop, wherein stress during up-sweep shear flow is higher, and not lower, than stress during down-sweep shear flow in the region of loop formation (Figs. \ref{hw45e_3}-\ref{hw1e_2}). This is observed due to the fact that stress overshoot is triggered only in the shear-thinning part of the constitutive curve. In this case, type 5 loop can be obtained by linear viscoelastic effects, however, type 6 and 7 loops can only be obtained using non-linear viscoelastic effects. Also, the loops shown in Fig.\,\ref{fig:hw_overall} may not have any significant effect of inertia because of the high relaxation time of the material. For example, if $\tau=10^6$ s, then the lowest dimensional value of $\delta t^*$ shown in Fig.\,\ref{fig:hw_overall} is $10^{-2}$ s which is sufficient for usual commercial rheometers to attain the lowest value of shear rates. 

As mentioned by Larson [\onlinecite{larson2015constitutive}], there are many reports of nonthixotropic viscoelastic materials in the literature with relaxation time $O\left({10}^6\right)$ s or greater. For example, Povlov et al.\, [\onlinecite{povolo1996stress}] reports the relaxation time of polyvinyl chloride (PVC) at 293 K (closer to ambient temperature) to be much greater than $10^6$ s. Furthermore, it is well known that the relaxation time and modulus of polymeric materials such as PVC can be manipulated by changing the molecular weight and plasticizer content [\onlinecite{shaw2005introduction}]. Also, it is well known from the literature on sol-gel transition that as the point of critical state is approached from the sol side, the viscosity as well as the relaxation time tends to infinity [\onlinecite{suman2020universality}]. Consequently, the state of a material in the sol phase, just before it attains the critical state, is a classic example of viscoelastic material having extremely large viscosity as well as relaxation time. The modulus, in principle, can be manipulated by changing stoichiometry and the reaction conditions [\onlinecite{Izuka}]. Therefore, rheological behavior of the sol phase, particularly desired modulus and relaxation time, can be achieved in a non-thixotropic viscoelastic polymeric solution [\onlinecite{scanlan1991evolution,suman2021phenomenological}]. The materials that undergo thermo-reversible sol-gel transition, in principle, remain in such an extremely high relaxation time (or viscosity) "sol” phase for ever if appropriate temperature is maintained [\onlinecite{negi2014viscoelasticity,joshi2020rheological,suman2021rheological}].

We can also consider suspension of colloidal particles that share hard sphere interactions with each other. It is well known that viscosity and hence relaxation time of these materials increases with increase in volume fraction [\onlinecite{mewis2012colloidal}]. At the random loose/close packing threshold the relaxation time diverges to infinity. Depending upon the size polydispersity of the particles, volume fraction – relaxation time dependence will show large variance [\onlinecite{larson1999structure}]. This point can also be generalised by considering any unknown material that appears to be a yield stress material and has a consistency of a paste. By virtue of being a yield stress material, the same has extremely large viscosity (in principle infinite viscosity) and hence equally large relaxation time (In the classic literature, a Bingham plastic material is purely elastic below the yield stress [\onlinecite{denn1998plug}]. Beyond the yield stress, the Bingham model treats a material as inelastic. However, real or practical yield stress materials are rarely inelastic and are usually viscoelastic with very large relaxation time even beyond the yield stress. Therefore, the dimensional value of relaxation time of the order of $10^6$ s utilised in this work (Figs. \ref{fig:hw_overall}-\ref{fig:hw_overall_stress_time}) to obtain the range of $Wi$ is practically realizable. 

The seminal review by Kulicke et al.\, [\onlinecite{kulicke1982preparation}], reports rheological behavior of aqueous polyacrylamide solution with extremely large zero shear viscosity. They give concentration and molecular weight dependence of viscosity that suggests small change in either can cause significant variation in viscosity. As mentioned above, a state of material very close to the critical gel state (but is still in the sol state) may have extremely large viscosity. Polyacrylamide itself can be subjected to crosslinking leading to sol-gel transition in such a fashion [\onlinecite{yilmaz2004sol}]. The value of zero shear viscosity $(\eta_s+\eta_p)={10}^4$ Pa s considered in this work is, therefore, practically realizable. If the solvent is water having viscosity  $10^{-3}$ Pa s, as is the case with polyacrylamide solution, we shall get the considered value of $\bar{\eta_s}=10^{-7}$.

\subsection{Area of hysteresis loop} \label{subsection_area}

\begin{figure}[htbp]
\centering
 \subfigure[]{
    \includegraphics[scale=0.28]{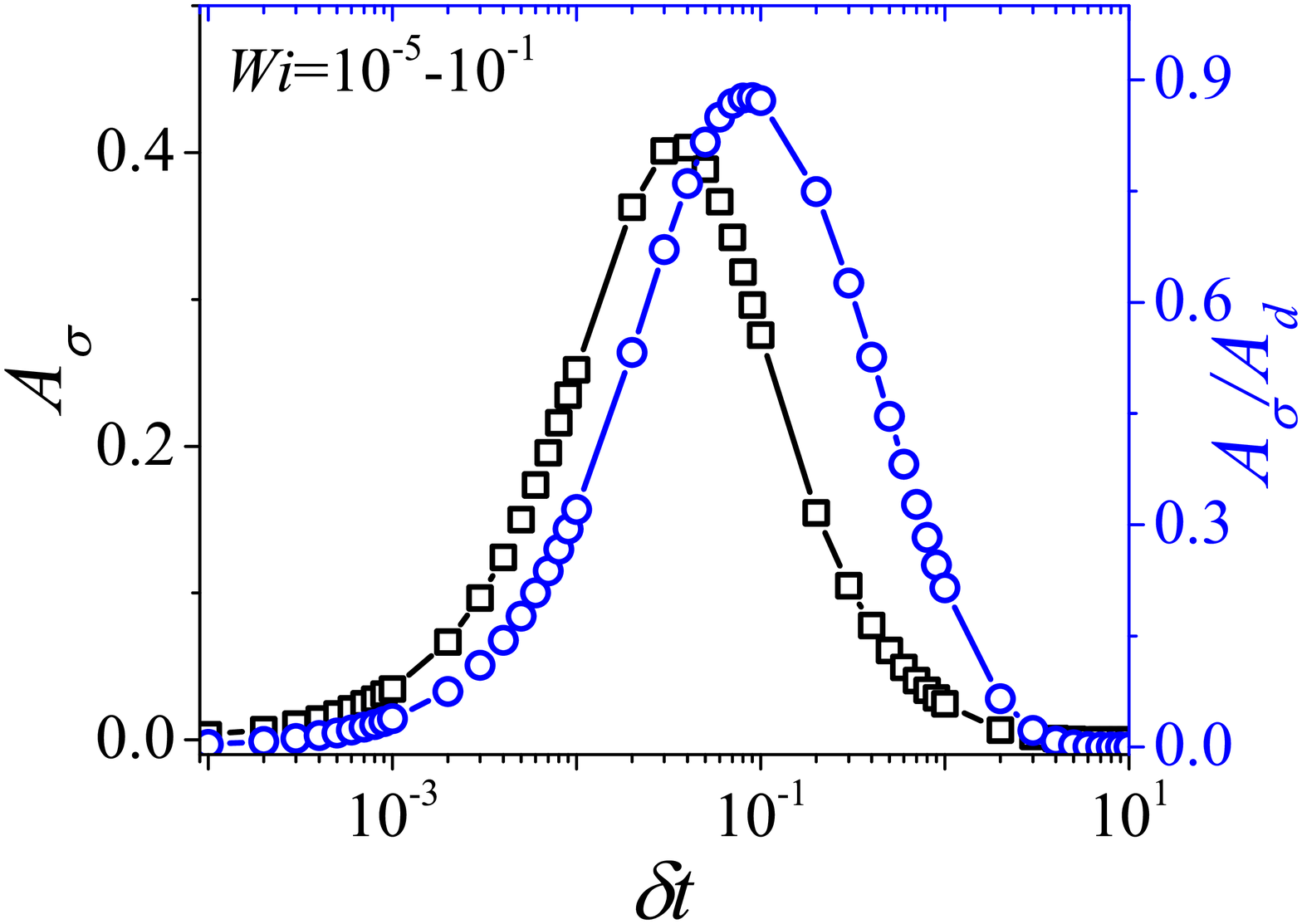}\label{area_1e_5}}
  \subfigure[]{
    \includegraphics[scale=0.28]{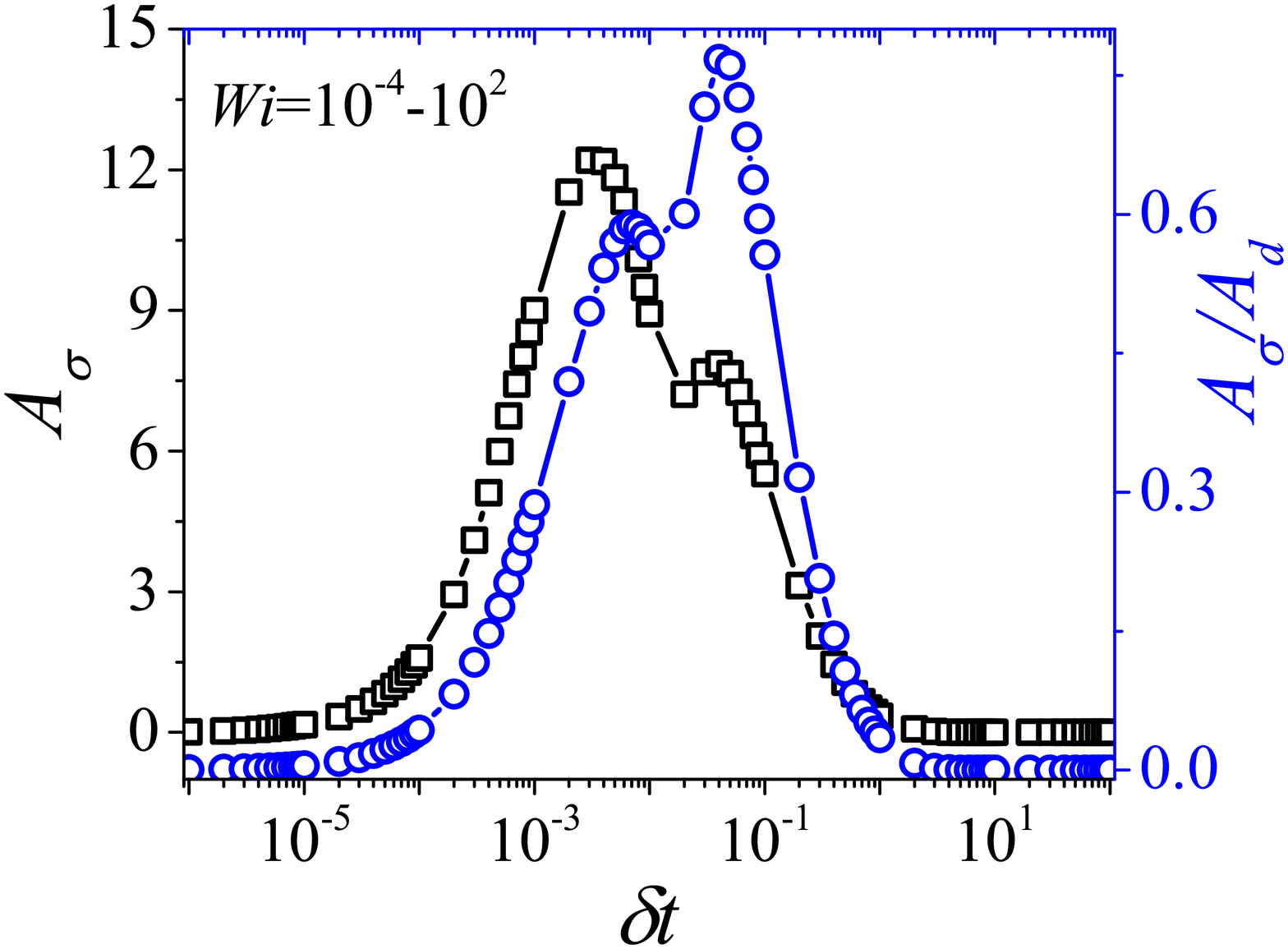}\label{area_1e_4}}  
    \subfigure[]{
        \includegraphics[scale=0.28]{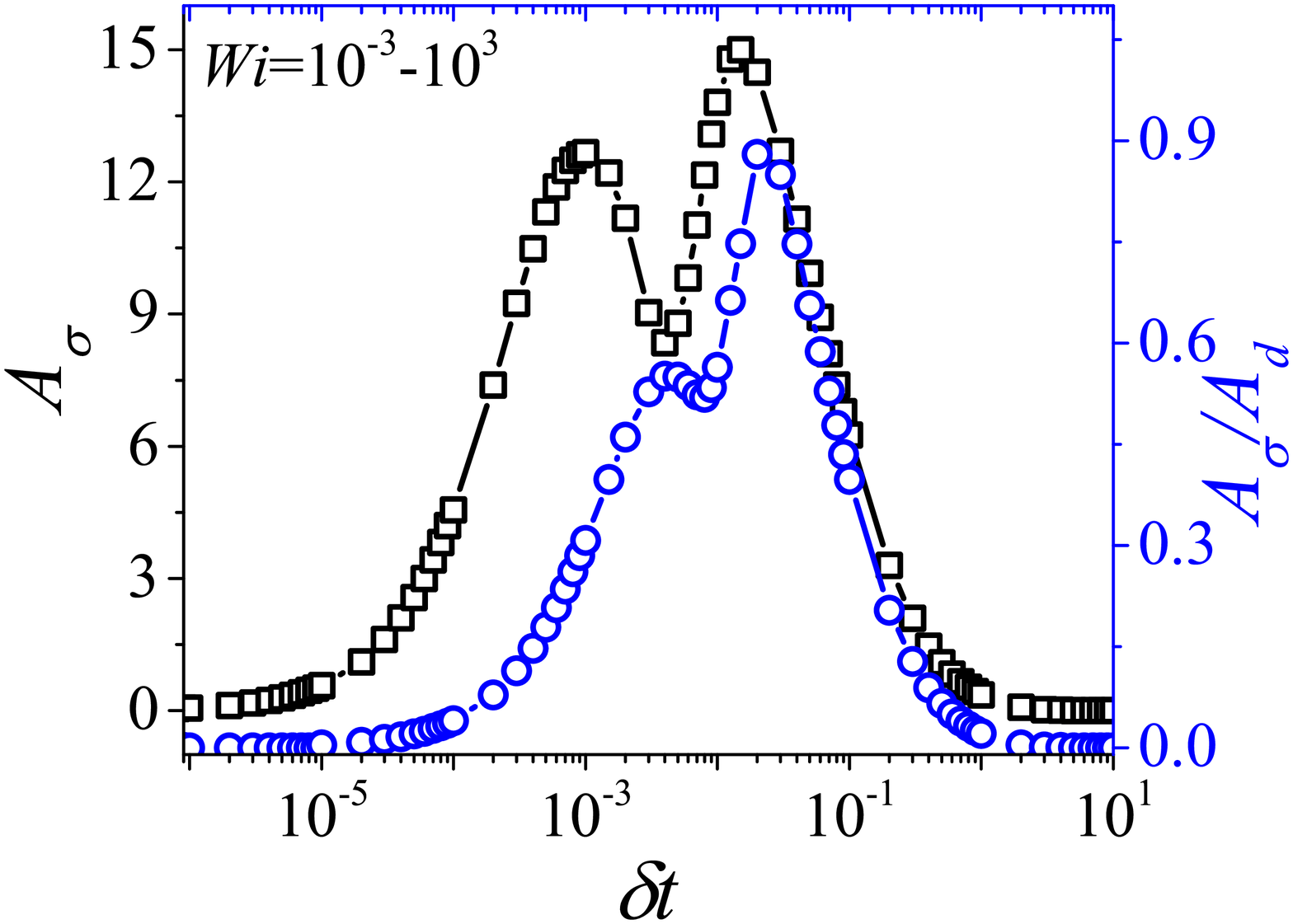}\label{area_1e_3}}
        \subfigure[]{
        \includegraphics[scale=0.28]{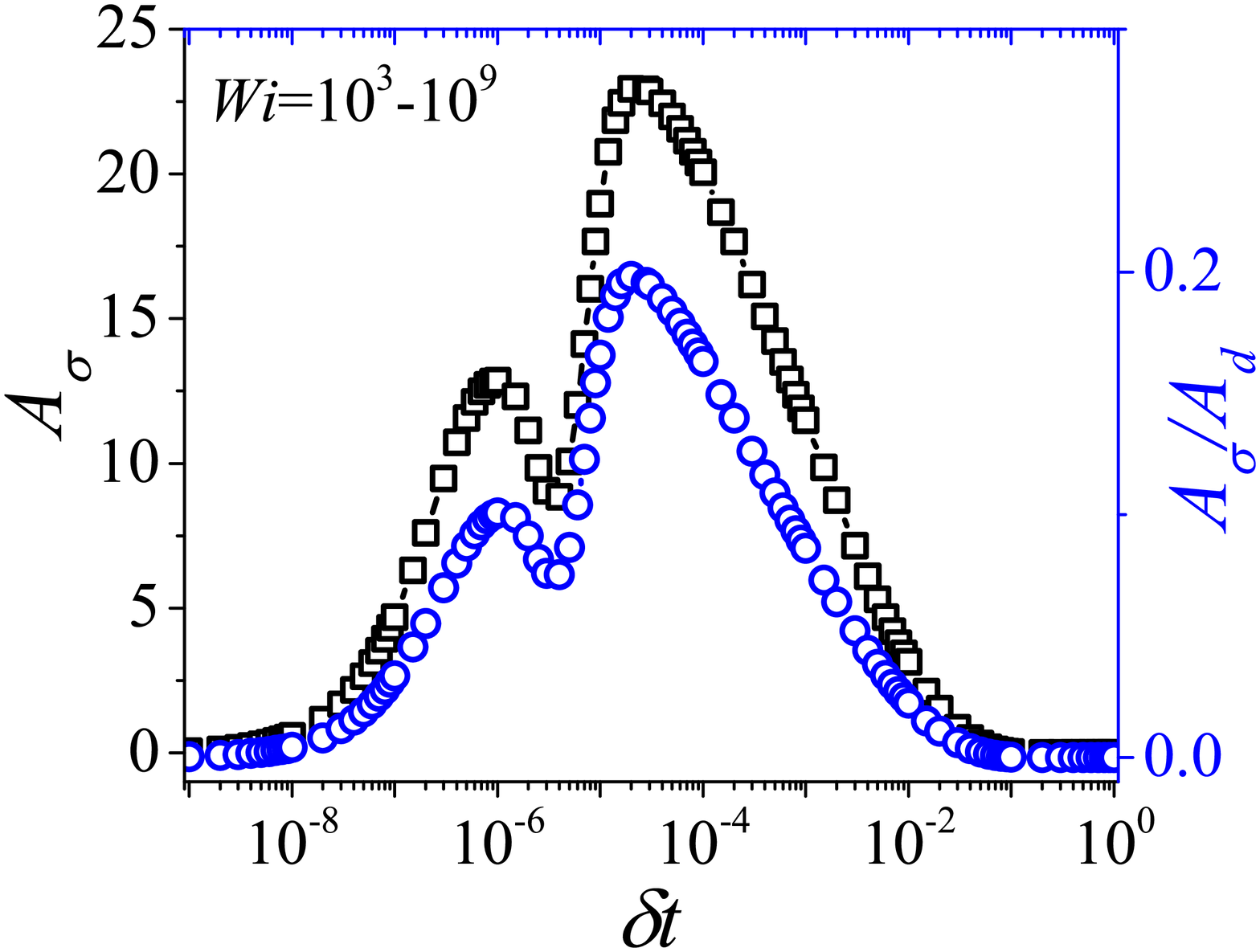}\label{area_1e3}}
\caption{Variation of area of hysteresis loop $(A_{\sigma})$ (on left y-axis)  and normalised area of hysteresis loop $(A_{\sigma}/A_d)$ (on right y-axis) is plotted with respect to $\delta t$. (a) $A_{\sigma}$ is calculated for $Wi$ varying in the range $10^{-5}-10^{-1}$ for $n=10$, $\bar{\eta_s}=10^{-3}$ (i.e., $\tau=10^{-2}$ s, $(\eta_s+\eta_p)=1$ Pa s). (b) $Wi$ is varying in the range $10^{-4}-10^{2}$ for $n=10$, $\bar{\eta_s}=10^{-3}$ (i.e., $\tau=10^{-1}$ s, $(\eta_s+\eta_p)=1$ Pa s) and $A_{\sigma}$ is calculated corresponding to the results plotted in Fig.\,\ref{fig:overall}. (c) $A_{\sigma}$ is calculated for $Wi$ varying in the range $10^{-3}-10^{3}$ for $n=10$, $\bar{\eta_s}=10^{-7}$ (i.e., $\tau=1$ s, $(\eta_s+\eta_p)=10^4$ Pa s). (d) $Wi$ is varying in the range $10^{5}-10^{9}$ for $n=10$, $\bar{\eta_s}=10^{-7}$ (i.e., $\tau=10^{6}$ s, $(\eta_s+\eta_p)=10^4$ Pa s) and area is calculated corresponding to the results plotted in Fig.\,\ref{fig:hw_overall}. In Fig.\,(a), the hysteresis loops obtained are open for all the values of $\delta t$. In Fig.\,(c) hysteresis loops are open for $\delta t<10^{-1}$. Square symbols shows $A_{\sigma}$ and circle symbol shows $A_{\sigma}/A_d$ and lines are guide to eyes in all the figures. In these figures, loop area shows bi-modal curves, which are qualitatively similar to bell shaped and bi-modal curve obtained using thixotropic materials in literature [\onlinecite{divoux2013rheological,radhakrishnan2017understanding,jamali2019multiscale,mckinley2022mneymosymearxiv}]. (All the variables in this figure are dimensionless as mentioned in section \ref{section_model}.)} 
\label{area_dt}
\end{figure}

In addition to the qualitative characteristics, hysteresis loops have also been characterised quantitatively using area of hysteresis loop in the literature. The area of hysteresis loop, $A_{\sigma}$ is obtained as mentioned in section \ref{section_model}. The plot of $A_{\sigma}$ as a function of $\delta t$ showed a bell-shaped curve for thixotropic materials and models [\onlinecite{divoux2013rheological,radhakrishnan2017understanding,jamali2019multiscale}]. In these studies, the plot of $A_{\sigma}$ as a function of $\delta t$ showed a monotonic decrease of $A_{\sigma}$ with $\delta t$ for simple yield stress materials and models. The presence of bell-shaped curve in this plot has been treated as a signature of thixotropic nature of these materials and the value of $\delta t$ corresponding to the peak in the bell-shaped curve has been related to thixotropic time-scale of the material. In this subsection, we study the area of hysteresis loop obtained using a non-thixotropic viscoelastic model during the down-sweep and up-sweep shear flow. We measure the variation of area of hysteresis loop as a function of $\delta t$ and also plot the loop area that is normalised with the area under the down-sweep curve $(A_{d})$. We show results for four cases (i) $\bar{\eta_s}=10^{-3}$, and $Wi$ is varying from $10^{-5}$ to $10^{-1}$ in Fig.\,\ref{area_1e_5}, (ii) $\bar{\eta_s}=10^{-3}$, and $Wi$ is varying from $10^{-4}$ to $10^{2}$, in Fig.\,\ref{area_1e_4}, (iii) $\bar{\eta_s}=10^{-7}$, and $Wi$ is varying from $10^{-3}$ to $10^{3}$, in Fig.\,\ref{area_1e_3}, and (iv) $\bar{\eta_s}=10^{-7}$, and $Wi$ is varying from $10^{3}$ to $10^{9}$, in Fig.\,\ref{area_1e3}. We find that loop area shows bimodal dependence for (ii), (iii), and (iv), and bell shaped dependence for (i). In all four cases, the loop area tends to zero if $\delta t$ is significantly low. As explained above, this is due to no change in stress during down-sweep and up-sweep because of insufficient time of shearing at each step. The first increase in area as a function of $\delta t$ in all four cases is due to incomplete stress relaxation at the end of down-sweep, that continues in up-sweep and drives the up-sweep stress away from down-sweep stress. The incomplete decrease in stress during down-sweep continues to increase the loop area untill the up-sweep stress shows a significant increase by getting more time of shearing in up-sweep, which results in decreasing the loop area. These three processes lead to bell shaped curve in (i). The presence of stress overshoot in the up-sweep causes second increase in the loop area and its absence on increasing $\delta t$ causes its final decay. The addition of second peak results into a bi-modal distribution of loop area when plotted against $\delta t$.

We also find that the second peak of $(A_{\sigma})$ is less significant for case (ii) and more significant for case (iii) and case (iv). This is due to two reasons, one is decrease in $\bar{\eta_s}$ value, which causes amplification of underlying viscoelastic effects (these effects might get diminished at high $Wi$ if $\bar{\eta_s}$ value is also high), and second is high value of $Wi$ that results in more prominent overshoot in stress. The variation of $A_{\sigma}/A_d$ in (ii), (iii) and (iv) show that the first peak of bi-modal graph is lower and the second peak is higher. In case (ii), the order of height of the first and the second peak of bi-modal graph is completely opposite to the order of first and second peak of $A_{\sigma}$ bi-modal graph. This is due to the fact that for small values of $\delta t$, the change in down-sweep stress is minimal as shown in Fig.\,\ref{fig:overall}, which results in significantly high value of $A_d$ for low values of $\delta t$ and changes the height of first and second peak. In cases (iii) and (iv), $~A_{\sigma}/A_d$ is qualitatively similar to $A_{\sigma}$ plot as $A_d$ remains almost same for small values of $\delta t$ (corresponding to first peak of $A_{\sigma}/A_d$ plot). The value of $A_d$ decreases for intermediate values of $\delta t$ causing the amplification of second peak of $A_{\sigma}/A_d$ plot. 

\begin{figure}[htbp]
    \centering
    \includegraphics[scale=0.4]{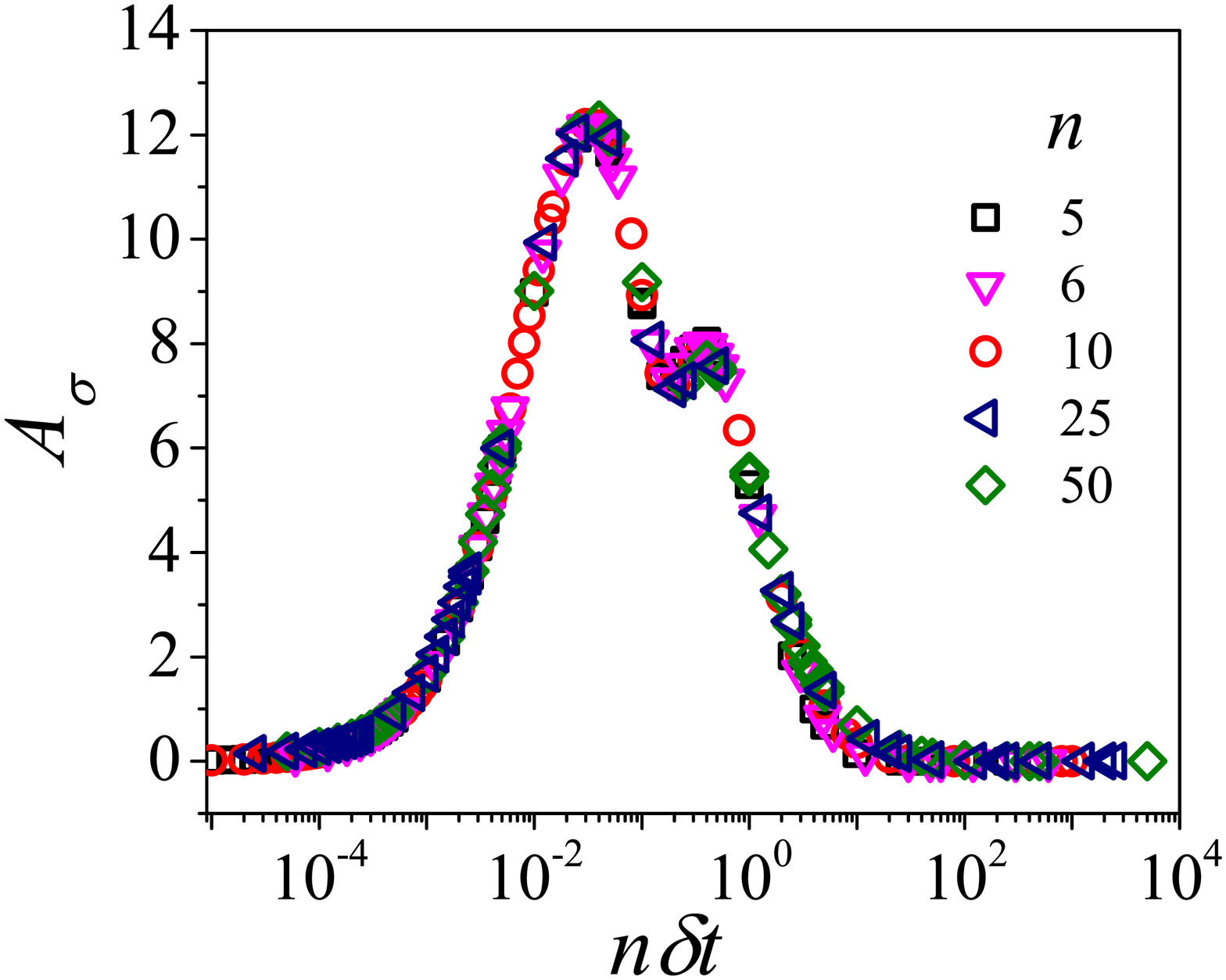}
    \caption{ Master curve of $A_{\sigma}$ is plotted as a function of $n\delta t$ for different values of number of steps per decade, $n$. This result is obtained for $Wi$ varying in the range $10^{-4}-10^{2}$ and $\bar{\eta_s}=10^{-3}$. (All the variables in this figure are dimensionless as mentioned in section \ref{section_model}.)} 
    \label{area_master}
\end{figure}

The bell-shaped/bi-modal dependence of $A_{\sigma}$ and $A_{\sigma}/A_d$ on $\delta t$ obtained using a non-thixotropic viscoelastic model is qualitatively similar to that obtained using a thixotropic material and models in literature [\onlinecite{divoux2013rheological,radhakrishnan2017understanding,jamali2019multiscale,mckinley2022mneymosymearxiv}]. Furthermore, we examine the effect of number of steps taken per decade $(n)$ and plot $A_{\sigma}$ as a function of $n\delta t$ in Fig.\,\ref{area_master}. This plot also shows a bi-modal master curve, which is independent of the value of $n$ as shown in Fig.\,\ref{area_master}. This result corresponds to results presented in Fig.\,\ref{area_1e_4}. This master curve can be compared qualitatively with the master curve of the thixotropic material obtained experimentally and by simulations [\onlinecite{divoux2013rheological,radhakrishnan2017understanding,jamali2019multiscale}]. This result shows the underlying similarity in successfully getting a master curve using a non-thixotropic viscoelastic model and results of thixotropic models and materials reported in literature. Therefore, a master curve of $A_{\sigma}$ as a function of $n\delta t$ also does not guarantee the exclusive identification of thixotropic material. This result also shows that a bi-modal curve is independent of the type of flow protocol used to study hysteresis. For example, a combination of extremely high value of $n$ and a very small value of $\delta t$ can show results for continuous ramp down and ramp up flow.

\subsection{Analysis}

Figures \ref{fig:marsh_paper_figure}-\ref{area_master} clearly show that the presence of (i) hysteresis loop, and (ii) bell-shaped/bi-modal dependence of $A_{\sigma}$ on $\delta t$ do not necessarily guarantee a material to be thixotropic. In this subsection, we analyze these two aspects in further details and discuss implications of the same.

\subsubsection{Comparison of type of hysteresis loops}
%\st{that are restricted only to a thixotropic material. They studied rheological hysteresis using a constitutive model and material that shows simple yield stress and viscoelastic bifurcation. The real materials were also viscoelastic in nature. Authors performed down-sweep followed by up-sweep shear rate experiments in a step-wise manner. Radhakrishnan et al. categorised hysteresis loops obtained using simple yield stress model with added viscoelasticity and viscosity bifurcating models (both models have a structure buildup term.). Authors mentioned that for a simple yield stress fluid with added viscoelasticity, stress during up-sweep shall be lesser (for the complete or some part of the range of shear rates) than stress during down-sweep shear flow. The authors also stated that a thixotropic material shall always show a stress higher in the up-sweep shear flow as compared to stress during down-sweep shear flow. This is due to a fact that a thixotropic material shows structure buildup at lower shear rate, and hence show a higher stress during up-sweep shear flow. The authors also explain that in the case of simple yield stress fluids, lower stress during up-sweep is due to the time taken by a material to reach the steady state because of added viscoelasticity. However, they added that stress during up-sweep may become higher than stress during down-sweep for a small range of shear rates because of overshoot in shear stress.}

Radhakrishnan et al.\,[\onlinecite{radhakrishnan2017understanding}] proposed the qualitative features of the hysteresis loops of a thixotropic structural kinetic model with a single mode Maxwell model as constitutive equation. They proposed two kinds of thixotropic hysteresis loops. In the first kind, up-sweep stress is lower than down-sweep stress for some shear rate values and in the second kind, up-sweep stress is higher than down-sweep stress. We find that the first kind of hysteresis loop proposed by Radhakrishnan et al.\, [\onlinecite{radhakrishnan2017understanding}] is qualitatively similar to loops obtained for a non-thixotropic viscoelastic model studied in the present work. Both the models (structural kinetic model by Radhakrishnan et al.\, and Giesekus model in this work) can show open and closed hysteresis loops. The loops presented in their paper (Figs. 5 (a) and 5 (c) of Radhakrishnan et al.\, [\onlinecite{radhakrishnan2017understanding}]) are qualitatively similar to loops shown in Figs. \ref{1e_1}-\ref{3e_1}. The hysteresis loops obtained for low relaxation time viscoelastic material also show similarity with hysteresis loops obtained for other thixotropic materials in the literature. We mention here a few examples. (i) Hysteresis loop obtained for waxy potato starch in paper by Krystyjan et al.\, [\onlinecite{krystyjan2016thixotropic}] is similar to the hysteresis loop in Figs. \ref{1e_1}-\ref{1}. (ii) Hysteresis loop of stress as a function of Mason number obtained using DPD simulation for an attractive colloidal particulate system by Jamali et. al.\, [\onlinecite{jamali2019multiscale}] in their Fig.\,1(e) is similar to Figs. \ref{3e_1} and \ref{1} of this study using a viscoelastic model. 

\begin{figure}[htbp]
    \centering
    \includegraphics[scale=0.3]{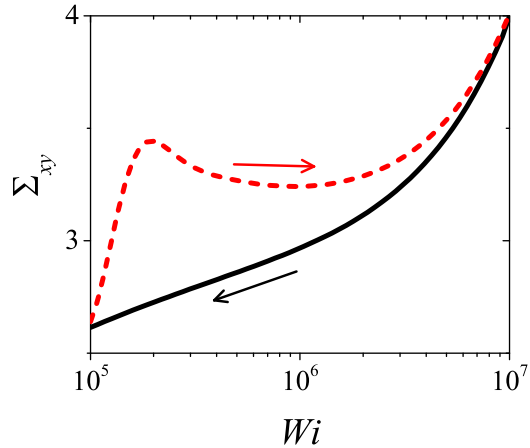}
    \caption{Shear stress is plotted as a function of $Wi$ for down-sweep and up-sweep shear flow by varying the $Wi$ in the range of $10^5-10^7$. The value of $\bar{\eta_s}$ used to calculate this result is $10^{-7}$ with $n=10$. The down-sweep and up-sweep shear flow result for this case shows a hysteresis loop, which is qualitatively similar to hysteresis loop observed for a thixotropic materials. Solid line shows down-sweep and dashed line shows the up-sweep shear flow result.  (All the variables in this figure are dimensionless as mentioned in section \ref{section_model}.)} 
    \label{fig:thixotropic_type_loop}
\end{figure}

However, the second kind of hysteresis loops proposed by Radhakrishnan et al.\, [\onlinecite{radhakrishnan2017understanding}] for thixotropic materials, wherein down-sweep stress is below the up-sweep stress is not observed for the low relaxation time viscoelastic material. Nonetheless, these type of loops can be obtained using a high relaxation time viscoelastic material as shown in Figs. \ref{hw45e_3}-\ref{hw1e_2}. These loops are similar to many loops observed for thixotropic materials. A few examples are listed as follows. (i) Figure 2 (b) of Kurokawa et al.\, [\onlinecite{kurokawa2015avalanche}] using Ludox gel is similar to Figs. \ref{hw1e_4} and \ref{hw1e_3}. (ii) Figure 8 (e) and 8 (f) of Fazilati et al.\, [\onlinecite{fazilati2021thixotropy}] using CNC suspension is similar to Fig.\,\ref{hw45e_3}. (iii) Figure 8 of Divoux et al.\, [\onlinecite{divoux2011stress}] obtained using Carbopol microgel is similar to Figs. \ref{hw45e_3}-\ref{hw1e_2}. (iv) Figure 5(a) of Jamali and McKinley [\onlinecite{mckinley2022mneymosymearxiv}] obtained using structure kinetic model for a thixoviscous fluid is similar to Fig.\,\ref{hw45e_3}-\ref{hw1e_2}. Figure \ref{fig:thixotropic_type_loop} shows a hysteresis loop in which shear stress is higher in the up-sweep shear flow than the stress during down-sweep shear flow for the complete range of $Wi$. This is obtained by varying $Wi$ only for two decades i.e., $10^5-10^7$. Even though this type of loop is obtained by varying $Wi$ for two decades, this plot is presented here only to show that a viscoelastic model is capable to show this type of loop. This loop is similar to Fig.\,1(b) of Divoux et al.\, [\onlinecite{divoux2013rheological}] obtained using Laponite dispersion. Interestingly, we would like to highlight that various closed hysteresis loops as shown in Figs. \ref{fig:hw_overall}, and \ref{fig:thixotropic_type_loop} wherein up-sweep stress is higher than down-sweep stress can also be obtained using the Giesekus model with $\bar{\eta_s}=0$ for $Wi$ varying in the range $10-10^7$. Here selection of $\bar{\eta_s}=0$ and range of $Wi$ typically corresponds to a polymer melt with relaxation time of the order of $10^4$ s and experimentally explorable shear rate range as discussed above (data not shown).

We also compare the type of hysteresis loops obtained using an inelastic thixotropic material with a characteristic time scale of the order of $10^6$ s. The details of the model are discussed in the appendix section. We have used the same shear rate range $(10^{-3}$ to $10^3$ s$^{-1})$ and performed the shear rate down-up sweep using the flow protocol discussed in section \ref{section_model}. Therefore, the results obtained using this inelastic thixotropic model can be compared with the results of Giesekus model with a relaxation time of $10^6$ s that are shown in Fig.\,\ref{fig:hw_overall}. The results of the shear rate downsweep and upsweep path show that stress in both direction do not overlap and form a hysteresis loop for some values of $\delta t$. These results are shown in Fig.\,S1. The hysteresis loops shown in Fig.\,S1 are similar to hysteresis loops obtained for a thixotropic material in the literature [\onlinecite{radhakrishnan2017understanding,divoux2013rheological}] in which upsweep stress is higher than downsweep stress. Interestingly, these hysteresis loops are also qualitatively similar to some of the hysteresis loops obtained using a viscoelastic material with relaxation time of $10^6$ s i.e., Figs. \ref{hw45e_3}-\ref{hw5e_2} in which upsweep stress is higher than downsweep stress.

 The hysteresis loops presented in Figs.\,\ref{fig:marsh_paper_figure}-\ref{fig:hw_overall}, \ref{fig:thixotropic_type_loop} for the Giesekus model and also the hysteresis loops for polymer melts (high relaxation time and $\bar{\eta_s}=0$) covers various types of qualitatively similar loops presented in literature for thixotropic materials/models including that shown in Fig.\,S1 for inelastic thixotropic model. This similarity between the results of non-thixotropic viscoelastic model and thixotropic materials raises an important question as to whether it is possible to differentiate between a viscoelastic material and thixotropic material on the basis of hysteresis loops. Therefore, if a completely unknown material is to be identified as thixotropic or viscoelastic, then rheological hysteresis may not be an ideal method to serve the purpose. This is because, as demonstrated in this study, just based on the hysteresis loop, the material could be either non-thixotropic viscoelastic material or a thixotropic material.
 
 %\textcolor{blue}{It must be pointed out that the hysteresis loops shown in Figs. \ref{fig:overall}, \ref{fig:hw_overall}, and \ref{fig:thixotropic_type_loop} are affected by the value of $\bar{\eta_s}$. However, a closed hysteresis loop with up-sweep stress higher than down-sweep stress can also be obtained using Giesekus model for $Wi$ varying in the range $10-10^7$ (i.e., a relaxation time of the order of $10^4$ s and $\bar{\eta_s}=0$ (results are not shown here). Therefore, a typical thixotropic loop can also be obtained for a polymer melt which generally have a very high relaxation time (i.e., usually greater than or equal to $O(10^2)$ s) [\onlinecite{povolo1996stress}] .}
 
 \subsubsection{Comparison of area of hysteresis loop}
 
As mentioned above, the loop area has been reported to show a bell-shaped/bi-modal dependence for thixotropic materials in the literature [\onlinecite{divoux2013rheological,radhakrishnan2017understanding,jamali2019multiscale,mckinley2022mneymosymearxiv}]. We also observe a bell-shaped curve for normalised loop area as a function of $\delta t$ for an inealstic thixotropic model with thixotropic characteristic timescale of the order of $10^6$ s that is shown in Fig.\,S2. We compare the bell-shaped/bi-modal dependence of thixotropic materials and inelastic thixotropic model with bell-shaped/bi-modal dependence of non-thixotropic viscoelastic model. We find that the first peak of bi-modal dependence of $A_{\sigma}/A_d$ is observed for very low values of $\delta t$ which might not be realisable experimentally. Therefore, one may end up getting only the second peak in $A_{\sigma}/A_d$ that is observed due to stress overshoot. For example, in case (iv) if $\tau=10^5$ s, the first peak is observed for $\delta t\approx10^{-6}$, which is around $0.1$\,s of shearing time at each step. Most of the experimental studies use shearing time in the range $1-10^3$ s as also studied by Divoux et al.\, [\onlinecite{divoux2013rheological}]. In this case, only second peak can be obtained for a viscoelastic fluid with very high relaxation time, which also may erroneously lead one to identify the material as thixotropic. Similarly, the width of bell-shaped/bimodal distribution in Fig.\,\ref{area_master} is also comparable with results of Divoux et al.\, [\onlinecite{divoux2013rheological}].

Interestingly, in the case of inelastic thixotropic model, we find that the value of $\delta t^*$ and $A_{\sigma}/A_d$ at which the peak is observed in the bell-shaped curve of normalised loop area (shown in Fig.\,S2), is of the same order as for a viscoelastic material with relaxation time of the order of $10^6$ s (Fig.\ref{area_1e3}).

%\textcolor{blue}{The maximum value of dimensionless area in Figs. \ref{area_1e_4}-\ref{area_1e3} is of the order of $(10)$. However, the dimensional value in each case depends on the value of zero-shear viscosity and the relaxation time of the solution. The dimensional value of loop area will be higher for each case (cases shown in Figs. \ref{area_1e_4}-\ref{area_1e3}) that has a lower relaxation time and higher zero shear viscosity. Therefore, the dimensional value of area of hysteresis loop for case shown in Fig.\,\ref{area_1e_5} will lie in range   

Many studies also employ characteristic features of hysteresis loop to obtain thixotropic timescale. In various studies [\onlinecite{radhakrishnan2017understanding,divoux2013rheological,jamali2019multiscale,mckinley2022mneymosymearxiv,javadi2022thixotropy}] the thixotropic timescale (or characteristic timescale) is related to the value of $\delta t$ corresponding to the peak in the $A_{\sigma}-\delta t$ plot as this plot has been reported to show a bell-shaped type for thixotropic materials. This technique assumes that an increase and decrease of $A_{\sigma}$ is essentially due to thixotropy in the material. In the present work, we clearly show that viscoelasticity solely by itself, leads to a bell-shaped/bi-modal $A_{\sigma}-\delta t$ plot. 

We have also discussed before that the thixotropic materials undergo structural build-up under no or weak flow conditions. On the other hand, structural break-down takes place in the same when subjected to strong deformation fields. Consequently, subsequent to shear melting, which is a strong deformation field, many thixotropic materials attain a liquid state with predominantly viscous response. Under such conditions, viscoelastic relaxation time of these materials is very small and may lead to a broad separation of viscoelastic and thixotropic timescale. Therefore, there could be several thixotropic materials, where there is a natural separation of timescale. However, as the shear rate is decreased, structure formation takes place in the same causing relaxation time to increase. Depending upon nature of a material, such increase may cause viscoelastic relaxation time to be of the same order as thixotropic timescale since there is nothing fundamental that prevents two timescales to be of the same order. For instance, it is possible that relaxation time of a material does not decrease significantly upon shear melting (may take place for materials that have high volume fraction of the dispersed phase, e.g. colloidal glasses, composites, etc.). Furthermore, it may happen that relaxation time does decrease upon shear melting, but increases very rapidly when the structure formation takes place (may occur in certain kinds of colloidal gels, such as dispersions of clay, silica nanoparticles, etc.). Consequently, not all materials are expected to show such separation for every state of a material during the shear rate sweep irrespective of the direction of the sweep. Therefore, if the relaxation timescale and characteristic timescale of a thixotropic material is of the same order at any state during the sweep, it may lead to a possibility that the nature of bell-shaped curve, in general, and the position of peak in the bell-shaped curve, in particular, of $A_{\sigma}-\delta t$ plot (thixotropic timescale) will be influenced/contaminated by the viscoelastic relaxation times.

%In addition, very importantly, most of the experimentally probed thixotropic materials are not inelastic, and have a finite non-zero relaxation time (in reality high relaxation time), and hence are strongly viscoelastic. Therefore, if the relaxation time is significantly high in a thixotropic material, which is usually the case, then there may be a case that peak will get shifted due to viscoelastic effects. As a result, quantification of thixotropic timescale using hysteresis loop method will always have a contribution from viscoelastic nature of a material. This contamination of thixotropic timescale by viscoelasticity may, therefore, lead to misleading or even erroneous results. Hence, hysteresis loop method may not be used to determine thixotropic timescale as well as any other thixotropic feature of the material if a material is also viscoelastic.

Our results depicted in Fig.\,\ref{fig:overall}-\ref{fig:thixotropic_type_loop} show that an area enclosed by a hysteresis loop strongly depends on values of $\delta t$ and the minimum and maximum value of $Wi$. In real viscoelastic materials, with large relaxation time and broad relaxation time distribution, the hysteresis is, therefore expected to be significant over a wide range of $Wi$ and $\delta t$. Consequently, in a thixotropic material with broad viscoelastic relaxation time distribution, utmost care needs to be exercised while quantifying any thixotropic attributes of the same.

The analysis of hysteresis behavior of non-thixotropic viscoelastic, inelastic thixotropic as well as viscoelastic thixotropic materials suggests that the shear rate must be applied for a time-scale greater than a specific value in order to obtain the intrinsic flow curve, which is invariant of the direction of shear rate sweep. In case of viscoelastic material, such time-scale is greater than that associated with relaxation time of the same. If relaxation time is large, time to obtain steady state may exceed practically achievable timescales in a laboratory.  In case of inelastic thixotropic material, time-scale to attain steady state is that associated with microstructural reorganization when subjected to change in strain rate/stress. When viscoelasticity happens to be important in a thixotropic material, both the timescales, relaxation and characteristic thixotropic timescale of the material, influence the behavior. Furthermore, the thixotropic nature causes viscoelastic timescale to increase as a function of time, making such distinction exceedingly difficult. Therefore, for any unknown material, in order to determine whether it is viscoelastic or thixotropic it is necessary to know the relaxation dynamics of the same. By monitoring the time variation of viscosity after application/removal of strain rate/stress, the magnitude of relaxation time (compared to the observation timescale) can be identified irrespective of whether the material is thixotropic or not. However if the material is viscoelastic with high relaxation time, material's behavior can be discerned by applying step stress and/or step strain at different waiting times after pre-shear, with applied strain/stress in the linear region. Such experiment will not just allow determination of relaxation time of a material but will also determine whether material is thixotropic or not as per the recently proposed criterion [\onlinecite{agarwal2021distinguishing}].

\subsection{Viscoelastic hysteresis and shear banding} \label{subsection_shear_banding}

As mentioned in the Introduction, the rheological hysteresis has been linked with shear banding in the literature for thixotropic systems [\onlinecite{divoux2013rheological,radhakrishnan2017understanding}]. Divoux et al.\, [\onlinecite{divoux2013rheological}] performed experiments on aqueous solution of Laponite with different concentrations and showed the presence of shear banding during up-sweep shear flow because of stress overshoot. The authors also showed the same for 1 wt. \% aqueous solution of Carbopol. Radhakrishnan et al.\, [\onlinecite{radhakrishnan2017understanding}] showed the presence of shear banding using structure kinetic model for simple yield stress fluids and viscosity bifurcating fluids. However, Puisto et. al.\, [\onlinecite{puisto2015dynamic}] showed that the presence of shear banding is only associated with a small range of shear rate in the up-sweep shear flow. Jamali et al.\, [\onlinecite{jamali2019multiscale}] noted that rheological hysteresis is more closely related to microstructure level dynamics than the flow heterogenity or mescoscale dynamics.

The above discussion is focused on the correlation between hysteresis and shear banding in thixotropic and simple yield stress fluids. However, in this work we have studied hysteresis using a non-linear viscoelastic model. In our recent study [\onlinecite{sharma2021onset}], we have proposed that stress overshoot does not necessarily lead to transient shear banding in a viscoelastic material using Johnson-Segalman, non-stretching Rolie-Poly, and the Giesekus model. Non-linear simulation results [\onlinecite{moorcroft14}] also showed that stress overshoot does not lead to transient shear banding for Giesekus model. We, therefore, suggest that there may not be any correlation between viscoelastic hysteresis and shear banding in a viscoelastic material. This conclusion, naturally, is applicable only if the underlying constitutive curve of the material is monotonic.

%\newpage
\section{Conclusions} \label{section_conclusions}

In this work, we have studied rheological hysteresis using a nonlinear viscoelastic model. We have followed the protocol suggested by Divoux et al.\, [\onlinecite{divoux2013rheological}] wherein shear rate is varied in a step-wise manner from a high value to a low value (down-sweep) followed by a step-wise increase (up-sweep) to the original high value. We model the behaviour of two kinds of materials using Giesekus model, which is a standard nonlinear viscoelastic constitutive equation. We use systems with low $O$(0.1 s) and high relaxation times $O$($10^6$ s) over a range of realistic shear rates available in conventional rheometers ($10^{-3}$-$10^3$ s$^{-1}$). We observe that both the kind of systems show pronounced hysteresis for practically realizable intermediate step time $\delta t$, with diminishing area enclosed by loop at high and low values of $\delta t$. This observation leads to a bell shaped or bimodal curve of area enclosed by hysteresis loop when plotted as a function of $\delta t$. The increasing part of the bell-shaped curve is due to incomplete stress relaxation during down-sweep shear flow. On the other hand, the decreasing part of the bell-shaped curve is because of up-sweep stress approaching its steady state value corresponding to each shear rate. The second peak in the bi-modal distribution curve, when present, is because of stress overshoot during up-sweep shear flow. It is observed that the bell-shaped curve is obtained even when the shear rates are in the linear viscoelastic region, while the non-linear viscoelastic effects lead to bi-modal shaped curve. The hysteresis loops observed in the case of low relaxation time viscoelastic material can be open and closed but the hysteresis loops obtained using a high relaxation time viscoelastic materials are observed to be closed for all the values of $\delta t$ explored in this work.

Thixotropic materials also show a hysteresis loop for the flow protocol adopted in this study. In the literature, characteristics of the hysteresis loop are often employed to interpret behavior of thixotropic materials. Thixotropic materials also show a bell shaped when area enclosed by hysteresis loop is plotted against $\delta t$. In this context an important question that we address is: in a down-up shear rate sweep experiment, can an unknown material be guaranteed to be thixotropic if it exhibits hysteresis loop and other associated features? Interestingly, the present work clearly shows that a viscoelastic material shows various qualitative features of hysteresis that are usually attributed to the thixotropic materials (i.e, viscoelastic thixotropic models and materials reported in the literature and inelastic thixotropic model with comparable characteristic timescale studied in this work). 
%For an inelastic thixtropic material with characteristic timescale of $O(10^6 s)$, the type of hysteresis loop and the normalised area of hysteresis loop shows qualitatively similar features as compared to a viscoelastic material with a relaxation time of $10^6$ s. Since both thixotropic and viscoelastic materials exhibit the hysteresis loop and associated characteristics, the most crucial question that arises is whether it is possible to identify an unknown material using the hysteresis approach. The most crucial question that arises is whether it is possible to identify a completely unknown material using the method of hysteresis since both thixotropic and viscoelastic materials exhibit hysteresis loop and the associated features. But the most important query is: in a down-up shear rate sweep experiment, can a material be guaranteed to be thixotropic if it exhibits hysteresis loop and other associated features?
However, it is known that various real thixotropic materials employed in the literature may have finite relaxation time and hence could also be viscoelastic.  Furthermore, for an unknown soft material we cannot have a prior knowledge of whether there is a broad separation of viscoelastic and thixotropic timescales over all the shear rate domains to be employed in the sweep. Therefore, in general, the present work suggests that while characterizing the hysteresis loop for a thixotropic material, caution needs to be exercised in order to avoid the possibility of influence of viscoelastic effects on the hysteresis behavior. We also note that while in some cases rheological hysteresis in thixotropic materials is also associated with a transient shear banding instability, rheological hysteresis in a viscoelastic material may not have any correlation with the transient shear banding instability if the underlying constitutive curve is monotonic.

\section*{Acknowledgment}
We acknowledge financial support from the Science and Engineering Research Board, Government of India. 

\appendix

\renewcommand\thefigure{\thesection.\arabic{figure}}  

%\section{Appendix}\label{section_appendix}

\setcounter{figure}{0}   

\section{Thixoviscous model}

In this section, we study hysteresis in a shear rate sweep experiment for an inelastic thixotropic material using a purely viscous thixotropic model that was first proposed by Goodeve [\onlinecite{goodeve1939general}] and has been studied by many researchers in the literature [\onlinecite{mujumdar2002transient,larson2019review, mckinley2022mneymosymearxiv,larson2015constitutive}]. According to this model, the microstructure of a thixotropic material can be represented by a scalar structure parameter $\lambda$. If $\lambda=0$, it shows that the material is completely rejuvenated and if $\lambda=1$, then material is completely arrested. The generalised Newtonian model can be used to represent the constitutive relationship between stress $(\sigma^*)$ and shear rate $(\dot{\gamma}^*)$ for this inelastic thixotropic material, which is given by:
\begin{equation}
    \sigma^*=\mu\dot{\gamma^{*}},
\end{equation}
where $\mu$ is the viscosity of the solution. The variation of $\mu$ with respect to $\lambda$ is given by [\onlinecite{larson2015constitutive,moore1959rheology,mckinley2022mneymosymearxiv,mewis2012colloidal}]:
\begin{equation} \label{mu_lambda}
    \mu=\mu_0(1+C\lambda)
\end{equation}
 where $\mu_0$ is the viscosity of completely rejuvenated solution and $C$ is a model parameter. Finally, the evolution of structure parameter $\lambda$ is given by [\onlinecite{goodeve1939general,mujumdar2002transient,larson2019review,mckinley2022mneymosymearxiv,larson2015constitutive}]:
 \begin{equation}
     \frac{d\lambda}{dt^{*}}=k_{+}(1-\lambda)-k_{-}\lambda\dot{\gamma^{*}}.
 \end{equation}
In this equation, the first term shows the build-up of structure with a time constant $\displaystyle\frac{1}{k_{+}}$ and the second term contributes to breakdown of structure with a constant $k_-$. Here, $\displaystyle\frac{1}{k_{+}}$ has been interpreted as a thixotropic timescale in Refs. [\onlinecite{mujumdar2002transient,larson2019review,mckinley2022mneymosymearxiv}]. We solve the thixoviscous model for a cyclic shear rate sweep protocol. We assume that the inertial effects are negligible and use the flow protocol mentioned in section II of the paper. Also, the variables with $*$ superscript are dimensional and the variables without $*$ superscript are dimensionless. The non-dimensional scheme used to obtain results using thixoviscous model is as follows: $t=k_+t^*$, $\sigma=\sigma^*/\mu_0k_+$. We use the same expressions of $A_{\sigma}$ and $A_{\sigma}/A_d$ is as given in section II of the paper (Eqs. 7-8).

\section{Results}

\begin{figure}[h]
\centering
    \subfigure[]{
    \includegraphics[scale=0.19]{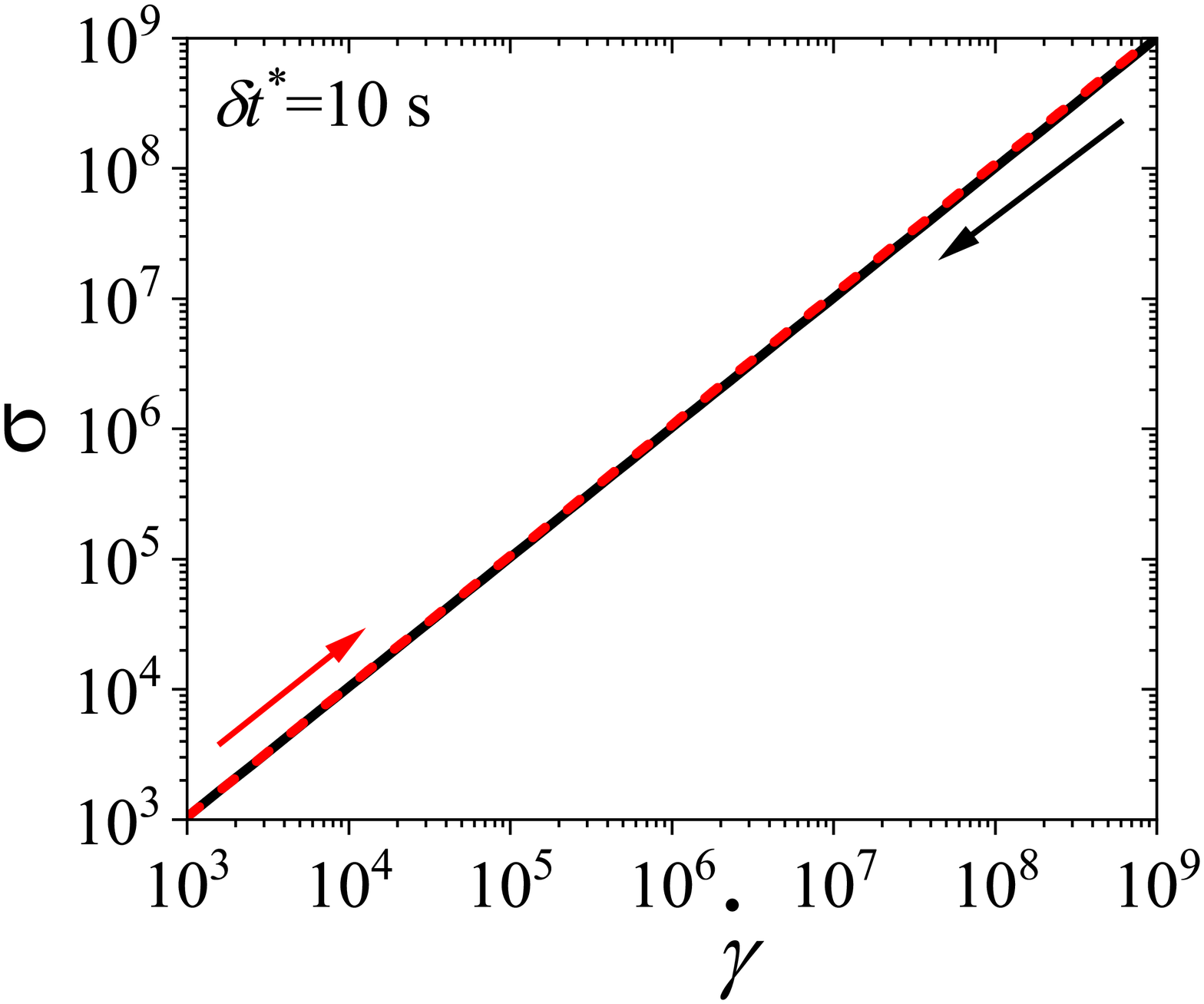}
    \label{th_1}
  }
     \subfigure[]{
\includegraphics[scale=0.19]{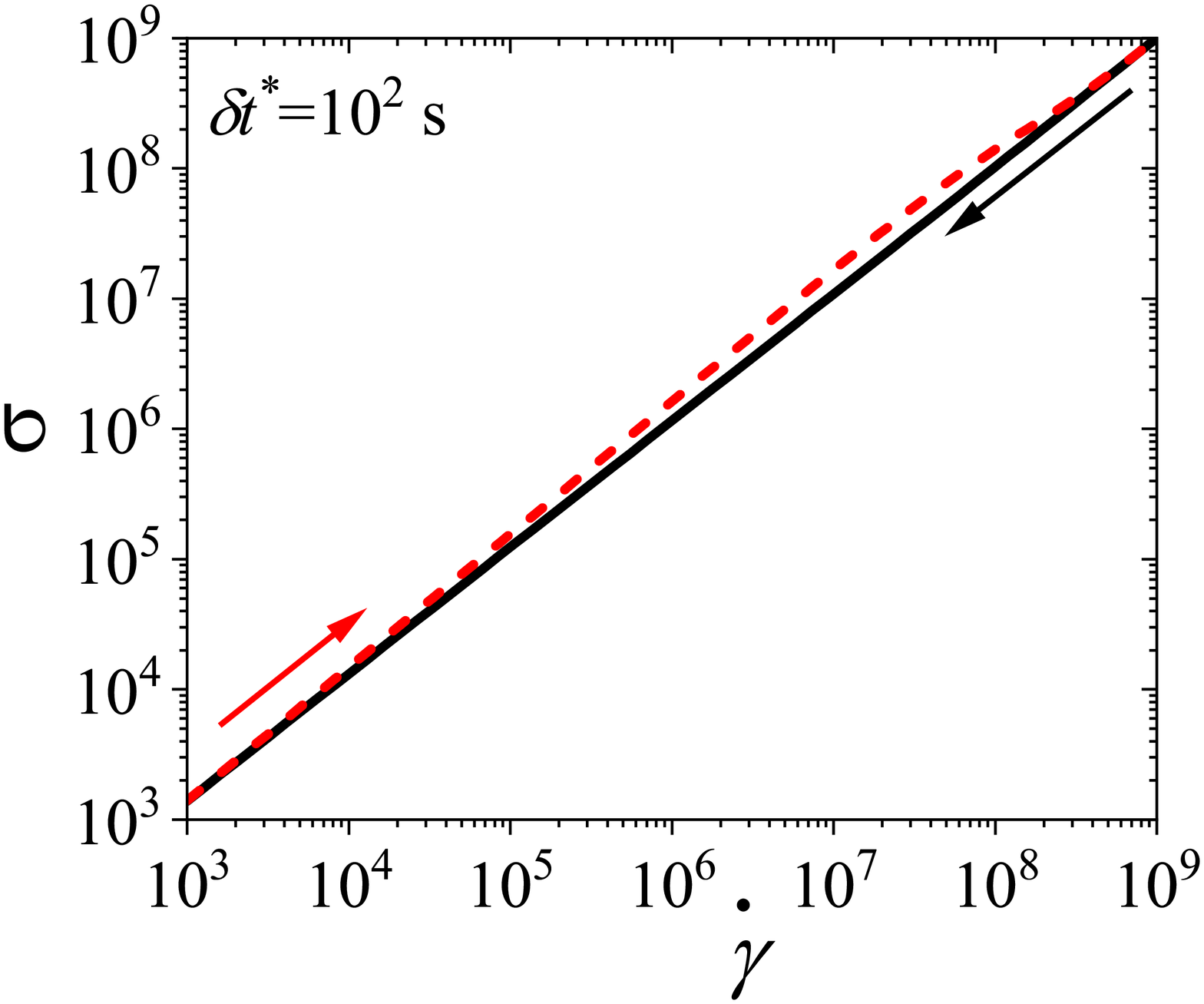}
    \label{th_2}
  }
    \subfigure[]{
    \includegraphics[scale=0.19]{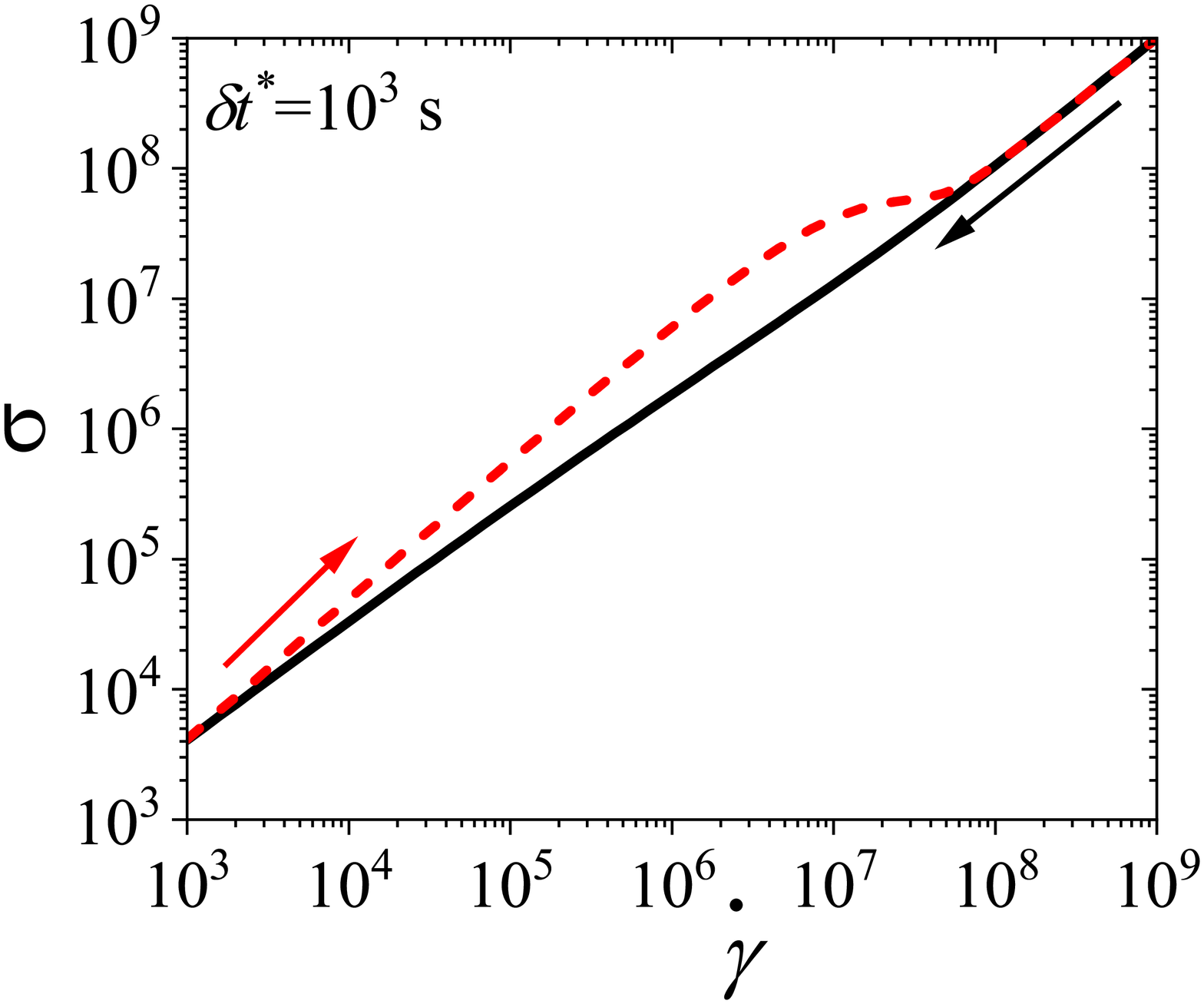}
    \label{th_3}
  }
         \subfigure[]{
\includegraphics[scale=0.19]{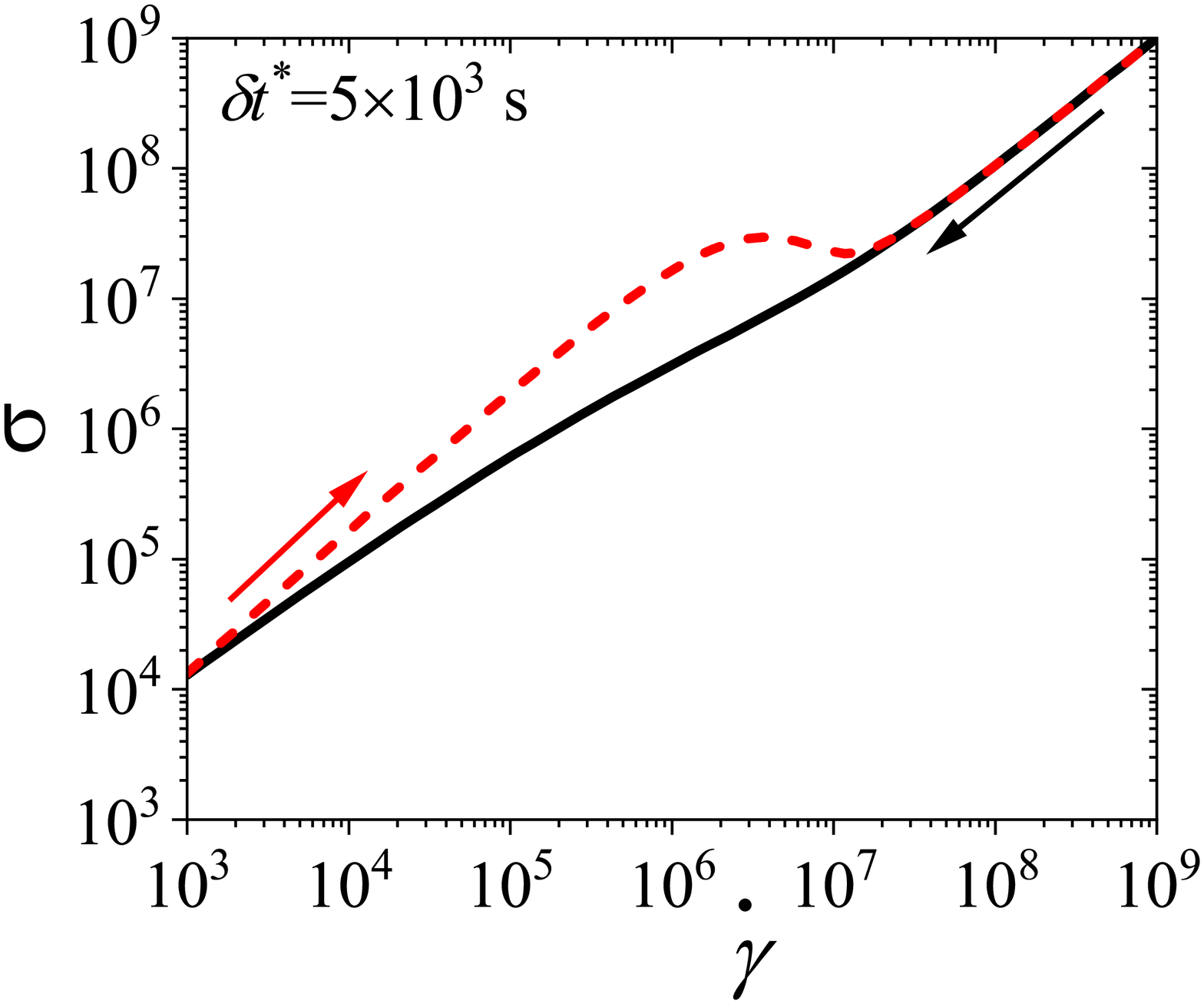}
    \label{th_4}
  }
    \subfigure[]{
    \includegraphics[scale=0.19]{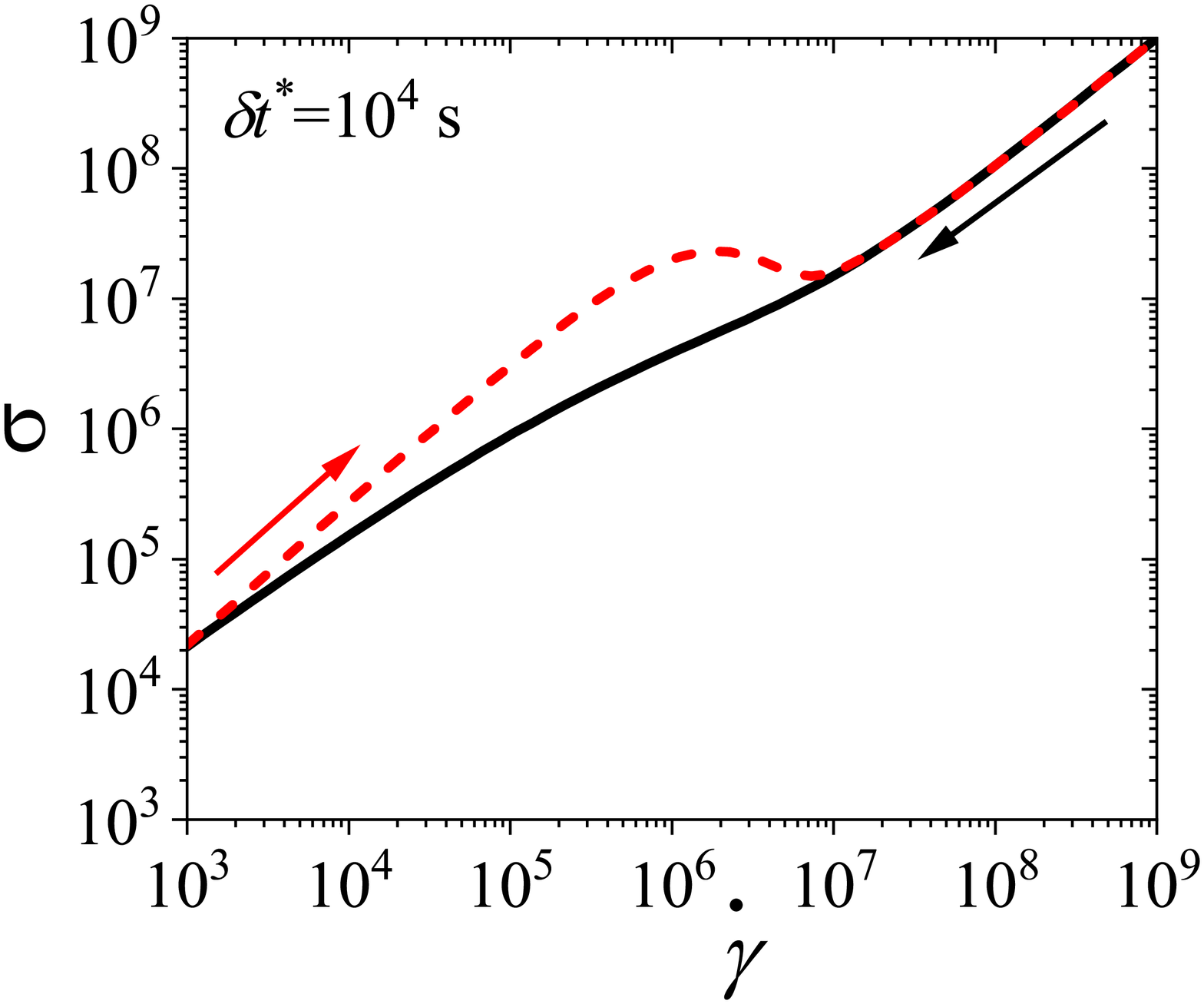}
    \label{th_5}
  }
     \subfigure[]{
\includegraphics[scale=0.19]{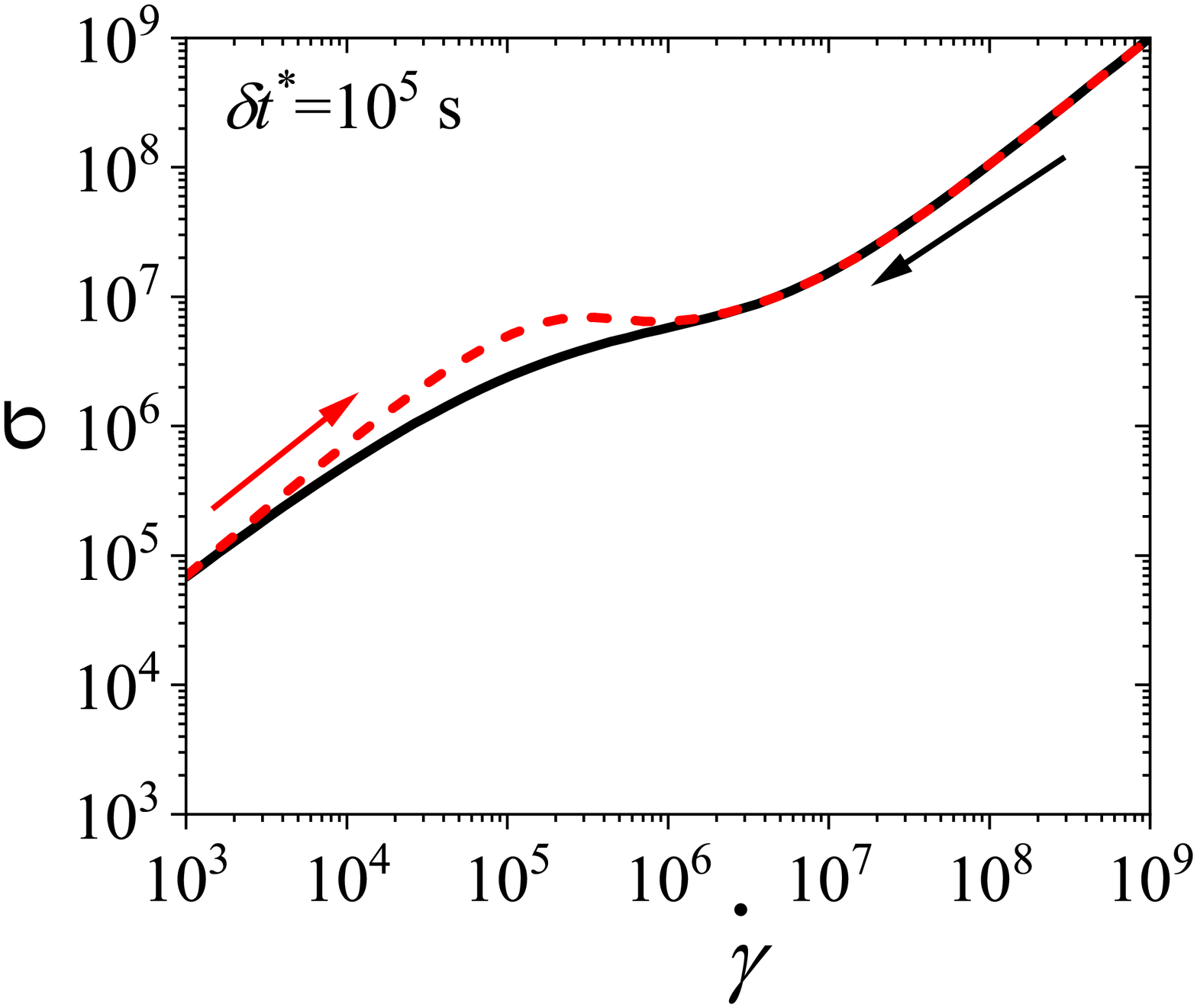}
    \label{th_6}
  }
   \subfigure[]{
\includegraphics[scale=0.19]{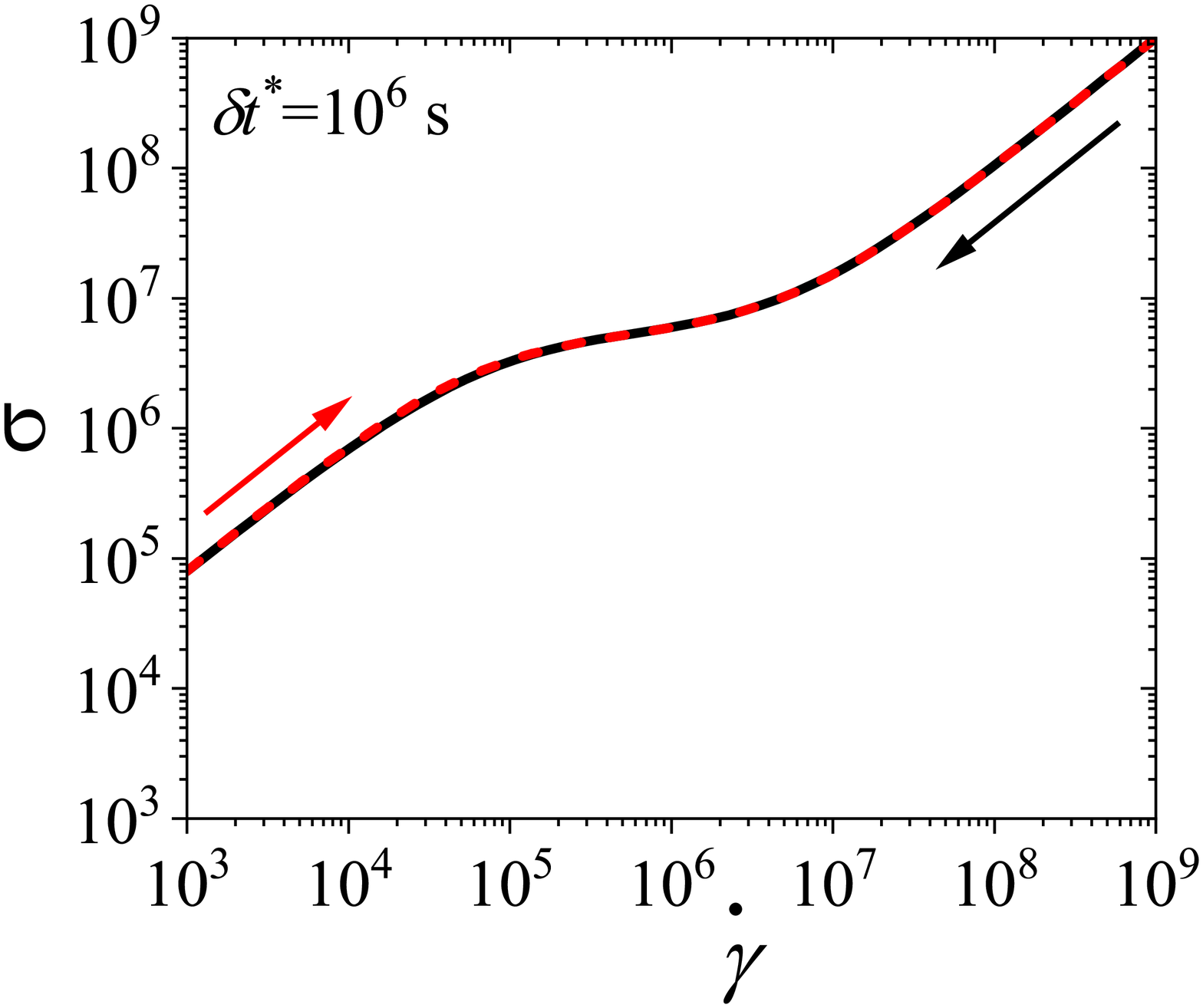}
    \label{th_7}
  }
\caption{\footnotesize Shear stress $(\sigma=\sigma^*/\mu_0k_+)$ is plotted as a function of shear rate $(\dot{\gamma}=\dot{\gamma}^*k_+)$ with $1/k_+=10^{6}$ s for different values of $\delta t^{*}$ (s). These plots predict the behavior of purely viscous thixotropic material in a shear rate sweep experiment with $1/k_+=10^{6}$ s and shear rate is varying from $10^{-3} $s$^{-1}$ to $10^{3}$s$^{-1}$. The value of model parameters $C$ and $k_{-}$ are $80$ and $1.5\times10^{-5}$, respectively.  }
\label{fig:thixo_loops}
\end{figure}

\begin{figure}[h]
    \centering
    \includegraphics[scale=0.35]{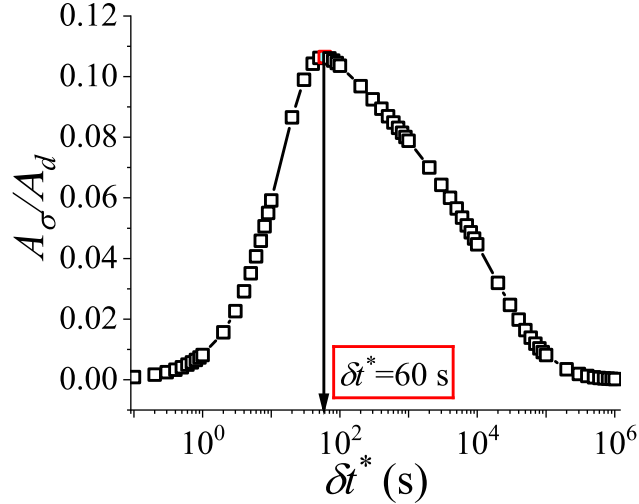}
    \caption{Variation of normalized area ($A_{\sigma}/A_d$) of hysteresis loops as a function of $\delta t^{*}$ (s) for a  thixotropic material in a shear rate sweep experiment and $1/k_+=10^6$ s. These results are corresponding to Fig. \ref{fig:thixo_loops}.}
    \label{fig:area_thixo_loops}
\end{figure}

We first present the results for cyclic shear rate sweep flow in Fig. \ref{fig:thixo_loops} for different values of $\delta t^*$ (dimensional). We use $1/k_+=10^6$ s to compare results of the thixoviscous model with results of the Giesekus model with $\tau=10^6$ s.
 We use the identical shear rate range ($10^{-3}$ s$^{-1}$ to $10^{3}$ s$^{-1}$) that was employed for the Giesekus model in the paper. It can be seen that for low values of $\delta t^*$, stress during down sweep and up sweep stress overlaps and no hysteresis is observed (Fig. S\ref{th_1}). As the value of $\delta t^*$ increases, shear stress during up sweep increases and shear stress in the down sweep shear flow decreases. Consequently, shear stress during down sweep and up sweep does not overlap and show hysteresis (Fig. S\ref{th_2}). We observe that area enclosed by up sweep and down sweep stress first increases with $\delta t^*$ as also shown in Fig. \ref{fig:area_thixo_loops}. As the value of $\delta t^*$ is further increased, down sweep and up sweep stress approach steady state (Figs. S\ref{th_3}-S\ref{th_6}) and consequently, the area between down sweep and up sweep stress curves decreases with further increase in value of $\delta t^*$ as shown in Fig. \ref{fig:area_thixo_loops}. The generic thixo-viscous model used in the present work clearly shows a bell-shaped dependence $A_{\sigma}/A_d$ and $\delta t^*$.

This observation of a bell shaped curve of area of hysteresis loop as a function of $\delta t^*$ is similar to results of various thixotropic models and materials reported in the literature [\onlinecite{divoux2013rheological,radhakrishnan2017understanding,mckinley2022mneymosymearxiv}]. The hysteresis loops shown in Figs. S\ref{th_2}-S\ref{th_6} are qualitatively similar to hysteresis loops shown in Figs. 4(g)-4(i) and Fig. 8 of the paper in which up sweep stress is above the down sweep stress. Also, the value of $\delta t^*$ at which peak is observed in the $A_{\sigma}/A_d-\delta t^*$ plot in Fig. \ref{fig:area_thixo_loops} is 60 s. This value is comparable with the value of $\delta t^*$ at which the peak (second and larger) is observed ($\approx 23 s$) in Fig. 7(d) of the paper. Also, the value of $A_{\sigma}/A_d$ corresponding to the peak is comparable in both cases. These results show that the results of a high relaxation time ($\tau$) viscoelastic material can also be qualitatively similar to a thixotropic material with a high value of characteristic thixotropic timescale $(1/k_+)$.

It must be noted that we get the similar value of $\delta t^*$ and $A_{\sigma}/A_d$ corresponding to peak in the $A_{\sigma}/A_d$ vs. $\delta t^*$ curve for different values of characteristic timescale $(1/k_+)$ varying from $10-10^6$ s, when other parameters are suitably varied. The corresponding peak values are of the same order as that observed for Giesekus model with $\tau=10^6$ s.

%\nocite{*}
\bibliography{draft_mybibfile.bib}% Produces the bibliography via BibTeX.
%\bibliographystyle{aipnum4-1}
%\bibliography{draft_mybibfile.bib}

\end{document}